\documentclass[a4paper,12pt]{article}
\usepackage[left=2cm,right=2cm,top=2.5cm,bottom=2.5cm,]{geometry}
\usepackage[T1]{fontenc}

\usepackage{amsmath}
\usepackage{
    accents, adjustbox, afterpage, amssymb, amsfonts, amsthm, array, booktabs, calc, cancel, caption, color, comment, datetime, dcolumn, dsfont, eurosym, float, geometry, graphicx, lipsum, longtable, mathtools, mathbbol, multirow, pdflscape, pdfpages, placeins, ragged2e, rotating, setspace, subcaption, subfiles, textpos, threeparttable, soul, xcolor, zref-savepos
}

\usepackage[hang, flushmargin]{footmisc}

\usepackage{enumitem}
\usepackage{tikz}
\usetikzlibrary{calc,arrows.meta}

\usepackage[authoryear]{natbib}
\setcitestyle{authoryear,aysep={},open={(},close={)}}
\bibpunct{(}{)}{;}{a}{,}{,}

\usepackage[hypertexnames=false,colorlinks]{hyperref}
\hypersetup{
  pdffitwindow=false,
  colorlinks=true,
  linkcolor={red!50!black},
  citecolor={blue!50!black},
  urlcolor={black},
  linkbordercolor={white},
}

\usepackage{etoolbox}
\input{citepos.tex}

\usepackage{xparse}
\NewDocumentCommand{\hyref}{m O{}O{}}{\hyperref[#1]{#2 \ref{#1}#3}}
\newdateformat{DDMonthYYYY}{%
  \THEDAY\, \monthname[\THEMONTH] \THEYEAR
}

\DeclareMathOperator*{\argmax}{arg\,max}

\DeclareMathOperator*{\sign}{sign}

\DeclareMathOperator*{\logit}{logit}
\newcommand*\diff{\mathop{}\!\mathrm{d}}

\theoremstyle{plain}
\newtheorem{theorem}{Theorem}
\newtheorem{proposition}{Proposition}
\newtheorem{lemma}{Lemma}

\newtheorem{corollary}{Corollary}

\theoremstyle{definition}

\newtheorem{remark}{Remark}

\usepackage{dsfont}
\usepackage{fouriernc}
\usepackage{charter}
\usepackage{libertine}

\newcommand\blfootnote[1]{
  \begingroup
  \renewcommand\thefootnote{}\footnote{#1}
  \addtocounter{footnote}{-1}
  \endgroup
}

\usepackage[ruled,vlined]{algorithm2e}
\usepackage{bookmark}

\usepackage[sf,sl,outermarks]{titlesec}
\newcommand{\addperiod}[1]{#1.}

\titleformat{\section}[block]
{\normalfont\Large\bfseries}{\thesection.}{.5em}{\Large\bfseries}
\titlespacing*{\section}{0pt}{*1.3}{*0.2}

\titleformat{\subsection}[block]
{\normalfont\large\bfseries}{\thesubsection.}{.5em}{\large\bfseries}
\titlespacing*{\subsection}{0pt}{*1}{*0}

\titleformat{\subsubsection}[runin]
{\normalfont\bfseries}{}{0em}{\normalsize\bfseries\addperiod}
\titlespacing*{\subsubsection}{0pt}{*1}{*1}

\definecolor{mathematica1}{rgb}{0.368417, 0.506779, 0.709798}
\definecolor{mathematica2}{rgb}{0.880722, 0.611041, 0.142051}

\setlist[enumerate]{leftmargin=*,wide=0pt,itemsep=0pt,topsep=2pt}
\AtBeginDocument{
  \setlength\abovedisplayskip{4pt}
  \setlength\belowdisplayskip{4pt}
}
\begin{document}
\thispagestyle{empty}
\setcounter{page}{0}

\setcounter{footnote}{0}
\renewcommand{\thefootnote}{\fnsymbol{footnote}}
\begin{center}\Large
    {\noindent\bfseries
    Speed, Accuracy, and Complexity
    }
\end{center}
\vspace*{1em}

\makebox[\textwidth][c]{
    \begin{minipage}{1.2\linewidth}
        \Large\centering
        Duarte Gonçalves\footnotemark
    \end{minipage}
}
\setcounter{footnote}{1}\footnotetext{
    \setstretch{1} Department of Economics, University College London; \hyperlink{mailto:duarte.goncalves@ucl.ac.uk}{\color{black}duarte.goncalves@ucl.ac.uk}.
}
\blfootnote{
    I am very grateful to the editor and to four referees, whose comments and suggestions greatly improved this paper. 
    I also thank 
	Carlos Al\'{o}s-Ferrer, 
    C\'{e}sar Barilla,  
    Doruk Cetemen, 
    Teresa Esteban-Casanelles, 
    Philippe Jehiel, 
    Deniz Kattwinkel, 
    Ian Krajbich, 
    Ryan Oprea, 
    Pietro Ortoleva, 
    Antonio Penta, 
    Antonio Rangel, 
    Ludvig Sinander, 
    Jakub Steiner,  
    Ran Spiegler, 
    William Zhang, 
	and the participants at 
    Glasgow, UBC, USC, UCSD, UCSB, Caltech, RHUL, Essex, Columbia, UPenn, Princeton, Surrey, Vienna, HEC Paris, LMU, Berlin Behavioural Economics, 
    MSU, Maastricht, Zurich, Cambridge, Stanford GSB, PSE, Pittsburgh and Carnegie Mellon, EHU, 
    as well as 
    the Southeast Theory Festival (Oxford),
    UCL-LSE Theory Workshop, 
    Junior Theory Workshop (Bonn),
    EAYE Annual Meeting, 
    Barcelona Summer Forum, 
    MathPsych, 
    AMASES, and the
    Symposium on Cognitive Economics, 
	for valuable feedback.
    Lastly, I am indebted to Nick Netzer for many long conversations about this project.
	\\
	\emph{First posted draft}: 12 February 2024. 
    \emph{This draft}: \DDMonthYYYY\today.
}
\setcounter{footnote}{0} \renewcommand{\thefootnote}{\arabic{footnote}}

\begin{center} {\bfseries\large Abstract} \vspace*{1em} \end{center}

\noindent\makebox[\textwidth][c]{
    \begin{minipage}{.85\textwidth}
        \noindent
        This paper studies when response time is informative about problem complexity. It revisits a canonical sequential-sampling model in which a decision-maker chooses when to stop acquiring costly information. Problem complexity is measured by the noise-to-signal ratio of the evidence process. Under exogenous stopping rules—as when the decision-maker does not optimally adjust to problem complexity—response time increases with complexity. By contrast, this monotonicity breaks down when the decision-maker observes problem complexity ex ante and optimally adjusts to it. Expected stopping time is then inverse-U-shaped in complexity, so choices are fast in both very simple and very complex problems. Ability and response time are similarly ambiguously related: more able decision-makers are faster on simple problems but slower on complex ones. Finally, this paper shows that complexity and ability can be inferred from the sensitivity of choices to subsidies, which is greater in more complex problems and for less able decision-makers.

        \vspace*{1em}
    \end{minipage}
}
\newpage

\section{Introduction}
\label{section:introduction}


It is almost definitional that more complex decision problems entail more mistakes and poor decisions. 
This has been widely documented both in experimental settings and in real-world contexts with significant financial implications.\footnote{
    Examples include 
    lottery choice \citep{RabinWeizsacker2009AER,Puri2023WP},
    inference and failure of contingent reasoning \citep{Martinez-MarquinaNiederleVespa2019AER,EnkeGraeber2023QJE}, 
    bandit problems \citep{BanovetzOprea2023AEJMicro}, 
    as well as choosing between health insurance plans \citep{SinaikoHirth2011JHE,BhargavaLoewensteinSydnor2017QJE}, 
    pension plans \citep{ChoiLaibsonMadrian2011REStat}, 
    and mortgages \citep{KeysPopePope2016JFE,AgarwalRosenYao2016MnSc}.
}
When these generate complex decision problems, tax schedules, allocation mechanisms, and various other incentive structures can then lead to inefficient outcomes and unreliable predictions, even when optimal. 
In contrast, deliberately complex pricing schemes, obfuscation of product characteristics, and other business practices can be used to exploit less sophisticated decision-makers.\footnote{
    This observation is at the root of ongoing work on complexity and simplicity in mechanism design and industrial organisation.
    Recent theoretical work includes \citet{Li2017AER}, \citet{BorgersLi2019Ecta}, \citet{PyciaTroyan2023Ecta}, and \citet{LiDworczak2021EC}; an example of empirical research in the same domain is \citet{Rees-JonesTaubinsky2020REStud}.
    On obfuscation by principals more broadly, see \citet{EllisonWolitzky2012RAND}, \citet{Jehiel2011TE}, and \citet{Spiegler2016ARE}.
} 
Understanding what makes some decision problems more complex than others is therefore important both for positive and normative reasons.


Response time is often used as a behavioural indicator of problem complexity, that is, the intrinsic features of a problem that affect how difficult it is for a decision-maker to understand, process, or solve it.\footnote{
    This paper focuses on ``problem complexity'' as a property intrinsic to the problem rather than the decision-maker. 
    This is distinct from more subjective or resource-based notions of behavioural or procedural complexity. 
    We refer to \citet{Oprea2025WP} for a comprehensive discussion on different uses of the term complexity.
} 
This intuition is deeply connected to sequential sampling models of decision-making, in which a decision-maker gradually accumulates noisy information before choosing between alternatives. 
Depending on the application, information acquisition in these models has been interpreted as acquiring or searching through existing data, cognitive reasoning, experiencing physical stimuli, or recalling past experiences. 
When the problem is more complex, evidence is less informative, learning is slower and decisions are expected to take longer. 
This logic underlies a large empirical literature relating response times and complexity to study decision-making and strategic interaction, as well as cognition and inference.\footnote{ 
    This comprises settings such as 
    lottery choice \citep{Wilcox1993EJ,Alos-FerrerFehrFehr-DudaGaragnani2024WP}, 
    loan choice \citep{KalayciSerra-Garcia2016EE}, 
    portfolio choice \citep{CarvalhoSilverman2017WP}, 
    knapsack problems \citep{FrancoYadavBossaertsMurawski2021JMathPsy}, 
    dominance-solvable and global games \citep{Esteban-CasanellesGoncalves2020WP,SchotterTrevino2021EE}, and 
    extensive-form games \citep{HorensteinGrabiszewski2022JBEE}; 
    more examples are surveyed in \citet{Clithero2018JEPsy}, \citet{SpiliopoulosOrtmann2018EE}, and \citet{GillProwse2023EJ}. 
    A large literature in psychology and neuroscience also looks at response time as an indicator of problem complexity; 
    see \citet{BogaczWagenmakersForstmannNieuwenhuis2010TINS}, \citet{Heitz2014FrontiersNeuro}, and \citet{ForstmannRatcliffWagenmakers2016ARPsy} for surveys. 
    Many of these studies \emph{show} a monotonic relation between response time and problem complexity rather than assuming it, while others 
    \emph{assume} it.
}


Whilst response time may be a compelling indicator of problem complexity, both existing evidence and introspection suggest that the relationship between complexity and response time is more nuanced than this standard intuition implies. 
Simple problems are often dealt with quickly because it is easy to identify the best course of action. 
However, very complex problems may also lead to fast responses when decision-makers quickly realise that further deliberation is not worthwhile.
Empirical evidence similarly points in conflicting directions.\footnote{
    For instance, both monotonic relationships and non-monotonic ones have been documented in psychology, in instances of visual discrimination \citep{PalmerHukShadlen2005JVision,Swensson1972PercepPsy}, 
    numerical problems \citep{BuckleyGillman1974JExpPsy,AshcraftBattaglia1978JExpPsy}, and choice with multiple attributes \citep{OnkenHastieRevelle1985JExpPsy}.
}
Some environments exhibit the conventional pattern in which more difficult problems generate slower and less accurate responses, whereas others exhibit non-monotone relationships between response time and difficulty.
This raises two basic questions: 
When should longer response times be interpreted as evidence of greater problem complexity? 
And, if not with response time, how can we identify problem complexity?


In this paper, we reconcile these contrasting intuitions within a canonical sequential sampling framework (\citealt{Wald1945AnnStats}, \citeyear{Wald1947Ecta}; \citealt{DvoretzkyKieferWolfowitz1953AMStats}).
A central theme of the paper is the role of optimality and meta-cognition: whether decision-makers recognise how complex a problem is and can adjust their behaviour accordingly. 
We show that when stopping is exogenous relative to problem complexity --- be it because problem complexity is not recognised or due to suboptimal exogenous conventions --- response time is monotone in complexity under mild conditions. 
In contrast, when decision-makers recognise complexity ex ante and optimally adapt stopping to realised complexity, this monotonicity breaks down and an inverse-$U$-shaped relationship emerges: 
Simple problems generate fast and accurate choices; moderately more complex problems foster slower and less accurate choices; but very complex problems lead to lower accuracy while again inducing \emph{faster} responses. 
Extending the model to dynamic effort control, we uncover that this non-monotonicity also suggests restraint in using response times to infer the ability of decision-makers, as high ability individuals always choose better, but do so faster in simple problems and slower in complex ones.
Finally, we return to the question of identifying problem complexity and show that choices are more sensitive to subsidies in more complex problems, providing an alternative way to infer problem complexity from simple incentive manipulations.


We study a standard drift-diffusion model in which a decision-maker gradually accumulates noisy information before choosing between two alternatives. 
Problem complexity is measured by the noise-to-signal ratio of the information process. 
A higher noise-to-signal ratio makes evidence less informative and therefore slows learning. 
The decision-maker stops gathering information at a stopping time and then chooses the alternative with the highest posterior expected payoff.
Throughout the paper, we use ``response time'' and ``stopping time'' interchangeably.\footnote{
    Depending on the literature, related concepts --- even if reflecting subtle differences --- are also referred to as reaction time or decision time. 
    Our focus is on the duration of deliberation before a choice is made.
} 
Our results focus on how speed (expected stopping time) and accuracy (probability of choosing the ex-post optimal alternative) depend on problem complexity.


The paper delivers four main sets of results. 
First, we study environments in which stopping behaviour is exogenous or invariant to realised complexity. 
\hyref{theorem:exogenous-stopping:speed-accuracy}[Theorem] shows that, under mild conditions on the stopping rule, more complex problems lead both to longer expected response times and to lower choice accuracy. 
When stopping depends on achieving a target level of certainty about which alternative is optimal --- as given by fixed belief thresholds --- greater complexity simply slows down the evolution of beliefs, so it takes longer for beliefs to reach the target level. 
For time-dependent stopping given by evolving stopping thresholds, as discussed in the literature \citep[e.g.,][]{HawkinsForstmannWagenmakersRatcliffBrown2015JNeuro}, the comparison is subtler,\footnote{
    In this case, a more complex problem is equivalent to running the belief process on a slower clock, which means that the same effective belief path is evaluated against later stopping thresholds. 
    When stopping thresholds shrink over time, these later thresholds are tighter and can push towards earlier stopping. 
    We show that, for a broad class of exogenous stopping rules, the slower-learning effect dominates.
} but, for a broad class of exogenous stopping rules, the same is true: 
more complex problems generate longer response times and lower accuracy. 
This benchmark also captures situations in which the decision-maker cannot separately adjust behaviour to complexity because learning about complexity is inseparable from learning which action is correct \citep[as in][]{FudenbergStrackStrzalecki2018AER}. 
Examples include tasks in which the feature that determines which option is optimal is intrinsically tied to what makes the problem easier or harder, such as the absolute difference in value when identifying the item with the highest value.


Second, we study optimal stopping, where the decision-maker optimally adjusts to problem complexity.
Some problems are recognisably harder than others even before deliberation begins, and decision-makers may therefore optimally adjust how much information they acquire.
\hyref{theorem:optimal-stopping:speed-accuracy}[Theorem] shows that the comparative statics change fundamentally once stopping optimally adapts to problem complexity. 
Accuracy also decreases with problem complexity, but expected response time becomes non-monotone and inverse-$U$ shaped. 
Greater complexity slows learning, which tends to increase response time, but it also narrows the optimal stopping thresholds as additional information becomes less valuable. 
For sufficiently complex problems, this second force dominates, leading decision-makers to stop earlier despite learning more slowly.
Hence, very simple and very complex problems may both generate fast responses, albeit for entirely different reasons.
We also show in \hyref{appendix:general} that the contrast between exogenous and optimal stopping extends to more general informational environments, allowing the decision-maker to be uncertain about problem complexity.


Third, we extend the model to incorporate endogenous effort and heterogeneous ability. 
The motivation is twofold. 
On the one hand, time spent on a task need not coincide with effort expended on it: a decision-maker may spend a long time on a simple problem while exerting little effort, or spend little time on a difficult problem while exerting substantial effort. 
On the other hand, the empirical literature documents conflicting relationships between ability and response time.
In some settings, slower responses are associated with greater sophistication \citep{Rubinstein2016QJE,AgranovCaplinTergiman2015JESA}; in others, faster responses are interpreted as evidence of higher ability \citep[see, e.g.,][]{Goldhammer2015Meas}.
In the two cases mentioned above, where the only decision affecting learning is whether to stop or not, we then allow decision-makers to choose not only when to stop but also how much effort to exert while gathering information, with higher ability lowering the cost of effort.


The extension yields two main implications.
As we prove in \hyref{theorem:ability:speed-accuracy}[Theorem], higher-ability decision-makers always make more accurate choices because they optimally exert greater effort and therefore obtain more informative signals.
However, the relationship between ability and response time is itself non-monotone.
In relatively simple problems, higher-ability decision-makers respond faster because the problem is effectively easier for them.
In sufficiently complex problems, however, higher-ability decision-makers respond more slowly because their greater ability makes continued deliberation worthwhile.
Thus short response times may reflect either high ability or low ability.
The model therefore reconciles conflicting empirical findings on the relationship between speed and ability.


Finally, we return to the fundamental question of how an analyst can infer which problem is more complex. 
Naturally, when the analyst knows which choice is optimal, they can infer problem complexity from accuracy directly, where the simpler problem is the one in which decision-makers are more likely to choose the best alternative. 
In many cases, however, which alternative is best is subjective and depends on unobserved parameters or idiosyncratic preferences, making it impossible to infer complexity directly from choice accuracy. 
At the same time, our earlier results imply that response time is not generally monotone in problem complexity.

We propose an alternative behavioural measure based on the sensitivity of choices to subsidies. 
Subsidising one alternative makes it more appealing and, therefore, chosen more often. 
\hyref{theorem:identification}[Theorem] shows that the effect of subsidies on choices is greater in more complex problems. 
Put differently: for the same subsidy, the probability with which the subsidised alternative is chosen increases more in more complex problems than in easier ones. 
Intuitively, in complex problems the stopping thresholds are narrower, so small changes in stopping thresholds have larger effects on behaviour.
In the heterogeneous-ability extension, the same logic implies that subsidy responsiveness is decreasing in ability.
This provides a practical monotone behavioural measure of problem complexity and ability even when the analyst cannot observe the correct action. 
This approach, however, is not free from assumptions: decision-makers must recognise problem complexity, at least imperfectly, and must respond to the subsidy.\footnote{
    Indeed, existing studies have documented insensitivity to financial incentives \citep[see, e.g.,][]{EnkeGneezyHallMartinNelidovOffermanvandeVan2023REStat}.
} 
In comparing very simple and very complex problems, ceiling effects may lead this incentive manipulation to produce insignificant differences and limit the ability to infer complexity ordering.

The remainder of the paper proceeds as follows. 
\hyref{section:setup}[Section] presents the setup of our baseline model and discusses modelling assumptions. 
\hyref{section:exogenous-stopping}[Sections] and \hyref{section:optimal-stopping} present the main results regarding speed and accuracy under exogenous and optimal stopping, respectively. 
\hyref{section:ability}[Section] examines the extension to heterogeneous ability and \hyref{section:identification}[Section] returns to how problem complexity impacts the effect of subsidies on choices. 
\hyref{section:conclusion}[Section] concludes.
Proofs are provided in \hyref{appendix:proofs}. 
\hyref{appendix:general} and \hyref{appendix:variants} extend results to more general environments and discuss variants and related models.

\section{Setup and Problem Complexity}
\label{section:setup}

The understanding of the relationship between speed and accuracy has been indelibly shaped by the widespread use of sequential sampling models in psychology, neuroscience, and economics.
In this section we introduce a simple version of the drift-diffusion model, a particularly influential stylised model of decision-making in which evidence in favour of one of two alternatives evolves as a Brownian motion with drift.

\subsection{Setup}
\label{section:setup:setup}

\subsubsection{Alternatives and Payoffs}
A decision-maker is choosing from a set $A=\{a,b\}$, with generic element $\alpha \in A$. 
Payoffs depend on a parameter $\theta\in\Theta=\{-1,1\}$ determining which alternative is best, and are given by $u:A\times\Theta\to\mathbb R$.
Specifically, $u(a,1)>u(b,1)$ and $u(a,-1)<u(b,-1)$.

Let $(\Omega, \mathcal F, \mathbb P)$ denote a probability space, where $\omega \in \Omega$ denotes the state of the world, $\mathcal F$ a Borel $\sigma$-algebra, and $\mathbb P$ a probability measure. 
Denote by $p_0:=\mathbb P(\theta>0)$ the prior probability that $a$ is optimal, and extend the utility function to beliefs over $\Theta$ by writing
$u(\alpha,p):=\mathbb E_{\theta\sim p}[u(\alpha,\theta)]$.
For simplicity, we assume each option is equally likely. 
This simplifies the exposition and the proofs, but does not affect the substance of the results.\footnote{
    Instead of understanding variation in $p_0$ as changing the measure $\mathbb P$, one can equivalently understand it as varying the random variable $\theta:\Omega\to\Theta$.
}

\subsubsection{Sequential Sampling}
Prior to choosing, the decision-maker can acquire information about $\theta$ paying a flow cost per unit of time $c>0$. 
The decision-maker observes a continuous-time process $\theta \, \mu \, t + \sigma\, B_t$, where $B_t$ is a standard Brownian motion defined on $(\Omega, \mathcal F, \mathbb P)$ independent of $\theta$, $\mu>0$ is the drift rate, and $\sigma>0$ scales the instantaneous volatility. 
Equivalently, this can be understood as the continuous-time limit of observing Gaussian signals $z_{\diff t}\sim N(\diff t \, \theta \, \mu,\, \diff t \,\sigma^2)$ every instant of time $\diff t$. 
For notational convenience, we focus on the observationally equivalent process 
$X_t:=\theta \, \kappa^{-1} \, t + \, B_t$, where $\kappa:=\sigma/\mu>0$ corresponds to the noise-to-signal ratio.

Let $\mathcal F_t^X:=\sigma(X_s:0\leq s\leq t)$ denote the filtration generated by the signal process. 
The decision-maker's posterior belief $p_t:=\mathbb P(\theta >0\mid \mathcal F_t^X)$ evolves depending on the information obtained. 
It is conveniently expressed in log-odds, $\logit(p):=\ln(p/(1-p))$, since $\logit(p_t)=
2\kappa^{-1} X_t = 2 \theta \kappa^{-2} \, t + 2\kappa^{-1}\,B_t$.\footnote{
    Any diffusion process  
    $\diff X_t = \hat\mu(X_t,t,\theta) \diff t + \hat\sigma(X_t,t) \diff B_t$ with Lipschitz continuous $\hat\mu,\hat\sigma$ satisfying $\hat\sigma(X_t,t)/(\hat\mu(X_t,t,a)-\hat\mu(X_t,t,b)) = \kappa/2$ will give rise to the same law of motion for the decision-maker's beliefs, 
    $\diff p_t = 2 \kappa^{-1} p_t(1-p_t)\diff \tilde B_t$, where $\tilde B_t$ describes the innovation process for beliefs relative to the observed filtration $\mathcal F_t^X$. 
    This accommodates not only Brownian motion with unknown drift, but also, for instance, Ornstein-Uhlenbeck and geometric Brownian motion with hidden drift.
}

The model is often interpreted as a stylised representation of deliberation or reasoning, with the information process representing accumulated evidence in favour of or against the alternatives.
Whilst in psychology and neuroscience it became a benchmark model of the neural basis of decision-making,\footnote{
    For instance, sequential sampling models have been used to study memory retrieval and sampling from memory \citep{Ratcliff1978PsyRev,ShadlenShohamy2016Neuron}, sensory perception \citep{GoldShadlen2001TRENDSCogSci,RatcliffSmithBrownMcKoon2016TrendsCognSci}, and neural activation during decision-making \citep{BogaczWagenmakersForstmannNieuwenhuis2010TINS,SummerfieldTsetsos2012FrontiersNeuro,ShushruthZylberbergShadlen2022CBio}. 
    Variants have also been used to interpret eye-gaze dynamics in product choice and purchasing decisions \citep{KrajbichLuCamererRangel2012FrontiersPsy,CallawayRangelGriffiths2021PlosCompBio}.
    See \citet{KrajbichFernandezYang2026Ch} for an excellent overview of the literature.
} 
in economics it has been used to model a wide range of situations and behavioural patterns, including randomness in consumer choice \citep{Webb2018MnSc,Clithero2018JEBO,Alos-FerrerFehrNetzer2022JPE}, consumer search \citep{BrancoSunVillas-Boas2012MnSc}, group deliberation and social learning \citep{FrydmanKrajbich2022MnSc,ReshidiLizzeriYarivChanSuen2024WP}, brand recognition and advertising \citep{Alos-Ferrer2018JBDM,ChiongShumWebbChen2020WP}, strategic choice \citep{SchotterTrevino2021EE}, persuasion \citep{OrlovSkrzypaczZryumov2020JPE,EscudeSinander2023TE}, and policy experimentation \citep{CallanderHummel2014Ecta,McClellanEcta2022,Wong2025TE}.

\subsubsection{Stopping Problem and Time Costs}

The decision-maker then faces a trade-off between taking longer to have a better understanding of which alternative is better, and the cost of time involved in doing so. 
Formally, the decision-maker chooses a stopping time $\tau$ adapted to $\mathcal F_t^X$. 
Upon stopping at $\tau$, the decision-maker chooses an alternative $\alpha_\tau\in\arg\max_{\alpha\in A}u(\alpha,p_\tau)$.
Different stopping times will lead to different expected payoffs, $\mathbb E[\max_\alpha \mathbb E[u(\alpha,\theta)|\mathcal F_\tau^X]-c\tau]$. 
We will consider both exogenously given and optimal stopping times. 

Our main focus is in characterising the behavioural implications of changes in $\kappa$, namely, 
the stopping time $\tau$ and 
the accuracy of choice $\mathbb P(\alpha_\tau \in \alpha^\theta)$, where $\alpha^\theta = \argmax_\alpha u(\alpha,\theta)$.

\subsection{Problem Complexity}
\label{section:setup:problem-complexity}

We take the noise-to-signal ratio $\kappa$ as the baseline index of \emph{problem complexity}, with higher values of $\kappa$ corresponding to more complex problems.
This is a natural index of how difficult it is to learn which alternative is better from the available evidence: for a fixed amount of time, higher $\kappa$ implies less informative posterior beliefs, and, under fixed decision criteria, this in turn implies that more time is required to attain a given level of confidence.

Naturally, many features may make a choice problem more or less complex.
These include the intrinsic properties of the alternatives involved --- for instance, deciding between savings products is arguably easier than understanding which structured financial product is best for oneself --- as well as how information is presented or how easy it is to obtain and compare across alternatives, for example through attribute-by-attribute comparison tools.
We nevertheless believe that the noise-to-signal ratio captures the central intuition that simple problems are those in which the decision-maker quickly realises which alternative is best, whereas in more complex problems it takes longer to become similarly confident.
This is in line with related interpretations in economics, psychology, and cognitive science \citep[e.g.,][]{Callander2011AER,FehrRangel2011JEP,ForstmannRatcliffWagenmakers2016ARPsy}.

We will obtain comparative statics in two settings: settings in which the decision-maker ex-ante recognises and adjusts to problem complexity and those in which they do not.
The former correspond to cases in which the decision-maker knows ex ante whether a problem is more or less complex than another, just not which alternative is best. 
We make the simplifying assumption that the level of problem complexity is known; in \hyref{appendix:general}, we show that our comparative statics generalise to cases in which the decision-maker faces uncertainty about the exact level of complexity of the problems, knowing just how to compare them ex ante in a probabilistic sense.
The latter settings include cases in which the decision-maker is not able to distinguish whether a problem is more or less complex than another.
The distinction between recognised and unrecognised problem complexity will be central in \hyref{section:exogenous-stopping}[Sections] and \hyref{section:optimal-stopping}, which reveal that the two settings induce strikingly different comparative statics.

\subsubsection{Relation to Other Notions of Complexity}
Our notion of problem complexity is closely related to existing notions of complexity in computer science, economics, and cognitive science, recently surveyed by \citet{Oprea2025WP}. 
For instance, it is related to computational complexity, namely the minimum amount of resources (such as time, memory, or energy) required to solve a problem,\footnote{
    See \citet{Sipser2013Book} for a textbook reference in computer science;
    \citet{EcheniqueGolovinWierman2011WP,Camara2025WP,Sanjurjo2025WP} for applications to consumer theory; and
    \citet{Gilboa1988JET,AbreuRubinstein1988Ecta} for early applications to repeated games.
} 
and to sample complexity, the sample size required to attain a specified level of accuracy with high probability.\footnote{
    See \citet[ch. 3]{Shalev-ShwartzBen-David2014Book} in machine learning, \citet{Roughgarden2020Book} in algorithmic game theory, and \citet{HuangMansourRoughgarden2018SIAMJComputing} in algorithmic mechanism design.
    Closely related ideas also arise in statistical decision theory \citep{Wald1945AnnStats,Wald1947Ecta}, dynamic information acquisition \citep{BrancoSunVillas-Boas2012MnSc,SteinerStewartMatejka2017Ecta,FudenbergStrackStrzalecki2018AER,CheMierendorff2019AER}, rational inattention \citep{Sims2003JME,MatejkaMcKay2015AER,CaplinDean2015AER,CaplinDeanLeahy2019REStud,CaplinCsabaLeahyNov2020REStud}, signal detection theory in psychology \citep{GreenSwets1966Book}, and drift-diffusion modelling \citep{RatcliffSmithBrownMcKoon2016TrendsCognSci,ForstmannRatcliffWagenmakers2016ARPsy}.
}

Problem complexity is related to, but distinct from, cost complexity \citep{ShenhavLiederMusslickBotvinick2017ARNeuro,Oprea2020AER,Schulzeetal2025TrendsCogSci}. 
Cost complexity refers to the costs or resource burden associated with specific choice procedures that a specific decision-maker experiences when faced with a specific problem, whether due to implementation, computational, or cognitive costs and limitations. 
It therefore requires considering not just the complexity of the problem, but also how the decision-maker approaches it. 
In some cases, these costs may be proportional to time --- as in our baseline model above. 
More generally, the cost incurred or effort exerted under a given choice procedure also depends on features of the decision-maker's behaviour rather than of the environment itself.
We return to this distinction in \hyref{section:ability}[Section], when discussing endogenous effort and heterogeneous ability.

\section{Speed and Accuracy under Exogenous Stopping}
\label{section:exogenous-stopping}

Before turning to optimal stopping, it is useful to isolate the mechanical effect of changing problem complexity whilst holding the stopping rule fixed. 
This benchmark serves two purposes. 
First, it separates the effect of changing the information process from the effect operating through endogenous adjustment of stopping behaviour. 
Second, it clarifies the connection between our framework and other standard sequential-sampling models in which stopping boundaries are fixed, or otherwise do not adjust to realised problem complexity.

\subsection{Comparative Statics in Problem Complexity}
\label{section:exogenous-stopping:speed-accuracy}

Throughout this section, we consider cases in which stopping is exogenously given. 
We focus on stopping rules that depend only on the decision-maker's posterior belief and the time spent engaging with the problem thus far. 
This captures, for instance, stopping upon reaching a pre-determined level of certainty about which alternative is best, or upon securing a satisficing level of expected payoff. 
More generally, we write the stopping time as $\tau^{\kappa}:=\inf\{t>0\mid p_t^{\kappa}\notin (\underline p_t,\overline p_t)\}$, 
where $(p_t^{\kappa})_{t\geq 0}$ denotes the posterior belief process for a problem with complexity $\kappa$, and $\underline p,\overline p:\mathbb R_+ \to (0,1)$ are potentially time-varying and $\mathcal C^1$ stopping thresholds such that $\overline p\geq \underline p$.\footnote{
    There is an active discussion in psychology and cognitive science as to which shape of the stopping boundaries best captures behaviour --- see, for instance, \citet{HawkinsForstmannWagenmakersRatcliffBrown2015JNeuro} and \citet{Bhui2019CBB}. 
    We allow for a general class of stopping thresholds, under the mild restriction that they depend only on the current belief and calendar time, capturing many variations used in the existing literature.
} 
To avoid overburdening notation, we suppress the dependence on problem complexity $\kappa$ when no confusion can arise.

At every time $t$, the thresholds define a continuation region $C_t:=(\underline p_t,\overline p_t)\subseteq (0,1)$, corresponding to the set of beliefs at which the decision-maker continues acquiring information. 
The decision-maker chooses $\alpha_t=a$ whenever their belief exceeds the upper threshold, $p_t\geq \overline p_t$, and chooses $\alpha_t=b$ if it exceeds lower threshold, $p_t\leq \underline p_t$.

We assume that $C_t$ is weakly decreasing in $t$ with respect to the subset order so that stopping occurs in finite time almost surely. 
In addition, we require one of the following conditions to be satisfied:
\begin{enumerate}[label=(A\arabic*)]
    \item $-\overline p_t'\overline \rho_t\leq 1$ and $\underline p_t'\underline \rho_t\leq 1$ for all $t\geq 0$ such that $C_t\neq \emptyset$, where, for some $K\geq \kappa$ 
    $\overline \rho_t:=\frac{K^2}{\overline p_t-\underline p_t}\int_{\underline p_t}^{\overline p_t}\frac{q-\underline p_t}{2q^2(1-q)^2}\diff q$ and $\underline \rho_t:=\frac{K^2}{\overline p_t-\underline p_t}\int_{\underline p_t}^{\overline p_t}\frac{\overline p_t-q}{2q^2(1-q)^2}\diff q$; 
    \label{assumption:continuation-region:belief-rate}
    or
    \item $\exists$ $\overline x,\underline x:\mathbb R_+\to \mathbb R$ such that 
    $\overline p_t={\logit}^{-1}(2\kappa^{-1}\overline x_t)$ and $\underline p_t={\logit}^{-1}(2\kappa^{-1}\underline x_t)$,
    or 
    $\overline p_t={\logit}^{-1}(\overline x_t)$ and $\underline p_t={\logit}^{-1}(\underline x_t)$, where (a) $\overline p_t=1-\underline p_t$ or (b) $\overline p,\underline p$ are constant in $t$.
    \label{assumption:continuation-region:symmetry}
\end{enumerate}
Condition \hyref{assumption:continuation-region:belief-rate} imposes a bound on how fast the continuation region can shrink. 
Condition \hyref{assumption:continuation-region:symmetry} instead allows us to accommodate cases in which the stopping thresholds depend on the evidence process directly, noting that $\{p_t \in ({\logit}^{-1}(2\kappa^{-1}\underline x_t),{\logit}^{-1}(2\kappa^{-1}\overline x_t))\}=\{X_t \in (\underline x_t\overline x_t)\}$. 
These are technical but substantive assumptions; in \hyref{appendix:proofs:theorem:exogenous-stopping:speed-accuracy}[Appendix] we discuss how these ensure the results in this section.
Although substantive, these conditions are satisfied by a large class of stopping rules of general interest.

Our main result for this section is the following:
\begin{theorem}
    \label{theorem:exogenous-stopping:speed-accuracy}
    Under the standing assumptions above, (1) the expected stopping time $\mathbb E[\tau]$ increases in problem complexity $\kappa$, and (2) accuracy $\mathbb P(\alpha_\tau = \alpha^\theta)$ is decreasing in problem complexity $\kappa$.
\end{theorem}

\begin{figure}[t!]
    \centering\small\singlespacing
    \begin{subfigure}{.485\linewidth}
        \includegraphics[width=\linewidth]{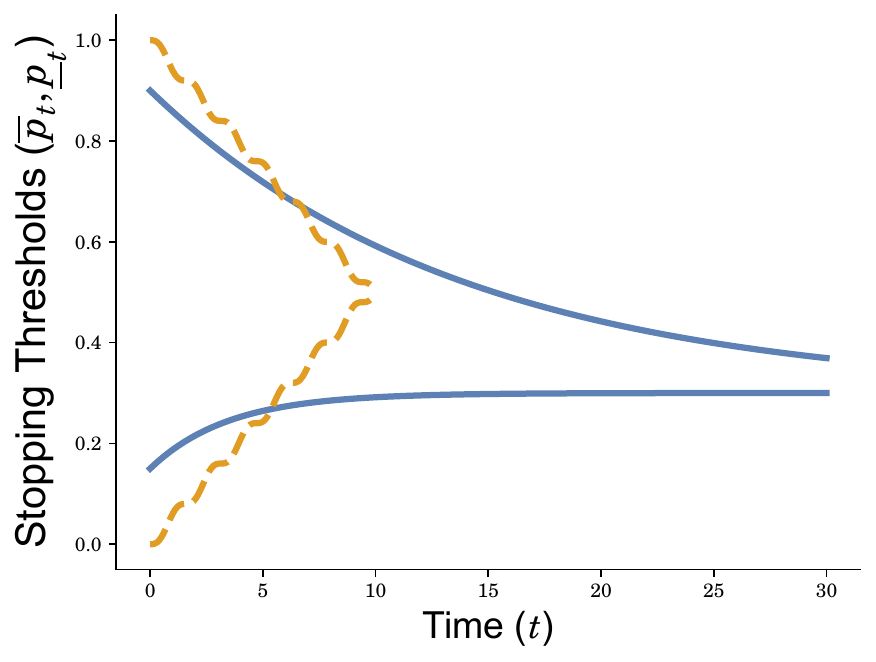}
        \caption{Stopping Thresholds}
        \label{figure:optimal-stopping:thresholds}
    \end{subfigure}

    \begin{subfigure}{.485\linewidth}
        \includegraphics[width=\linewidth]{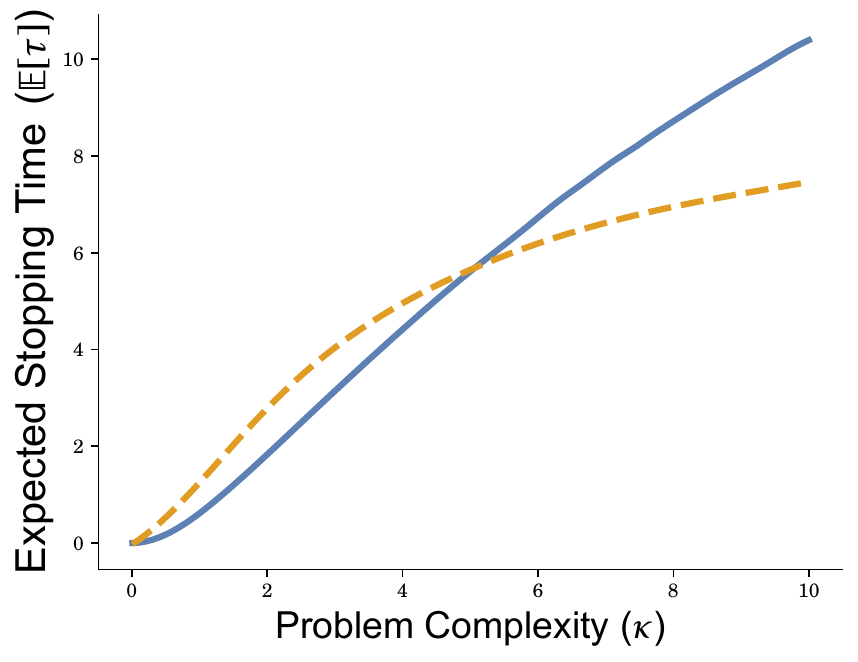}
        \caption{Expected Stopping Time}
        \label{figure:optimal-stopping:time}
    \end{subfigure}
    \begin{subfigure}{.485\linewidth}
        \includegraphics[width=\linewidth]{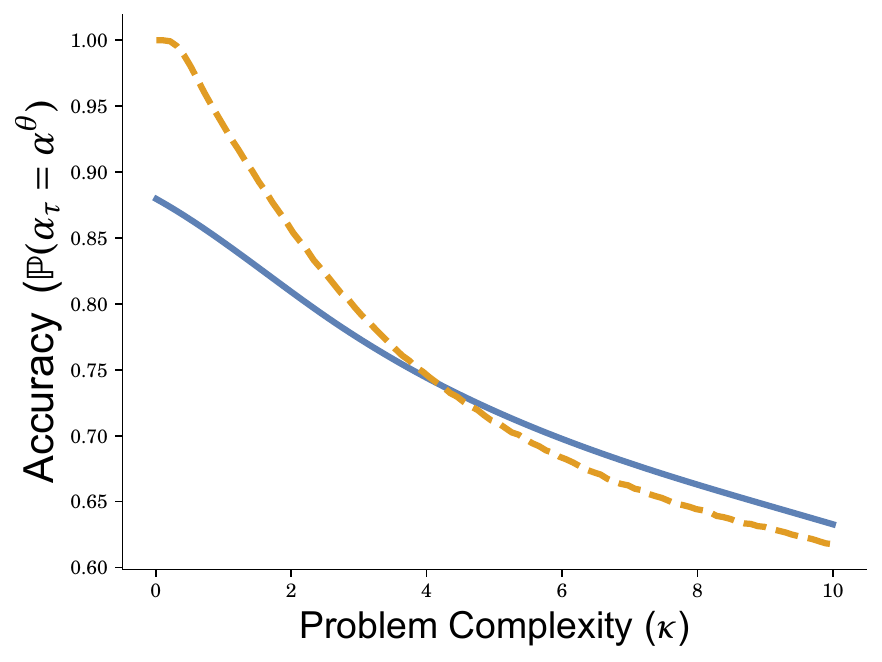}
        \caption{Accuracy}
        \label{figure:optimal-stopping:accuracy}
    \end{subfigure}
    \begin{minipage}{1\linewidth}
        \caption{Exogenous Stopping: Speed and Accuracy}
        \label{figure:exogenous-stopping:speed-accuracy}
        \emph{Notes}: This figure shows, in panel (a), the stopping thresholds for two continuation regions, and, in panels (b) and (c), the associated expected stopping time and accuracy as a function of $\kappa$, for $\kappa \in [0,1]$.

        The blue solid lines refer to a continuation region given by $C_t^1:=(\underline p_t^1,\overline p_t^1)$, 
        where, for some $q \in (0, 1)$, $\overline d \in (0,1-q)$, and $\underline d \in (0, q)$, 
        $\overline p_t := q + \overline d \exp(-{2 t}/({K^{2} m \overline d(\overline d +\underline d)}))$ and 
        $\underline p_t := q - \underline d \exp(-{2 t}/({K^{2} m \underline d(\overline d +\underline d)}))$, 
        with $m := \max_{p \in [q-\underline d, q+\overline d]}(p(1-p))^{-2}/2$.
        We set $q=1/3$, $\overline d=1/2$, $\underline d = 1/10$, and $K=1$.
        By construction, $C^1$ satisfies \ref{assumption:continuation-region:belief-rate} $\forall \kappa\leq K$. 
        
        The dashed orange lines refer to a continuation region given by $C_t^2:=(1-\overline p_t^2,\overline p_t^2)$, where
        $\overline p_t^2 := 1- \frac{1}{2}\frac{4t-\sin(4 t)}{4 T -\sin(4 T)}$ for $t<T$ and $\overline p_t^2=1/2$ for $t\geq T:=10$.
        By construction, $C^2$ satisfies \ref{assumption:continuation-region:symmetry}.
        The continuation region in panel (a) is depicted for $C^2$ with $\kappa=1/20$.
    \end{minipage}
\end{figure}

\hyref{theorem:exogenous-stopping:speed-accuracy}[Theorem] recovers the classical `faster is easier' logic exactly in the environments where stopping does not adjust to complexity.\footnote{
    Although one might conjecture that greater problem complexity should always increase expected stopping time under any fixed time-dependent boundary, \hyref{appendix:proofs:theorem:exogenous-stopping:speed-accuracy}[Appendix] shows that this can fail. 
    In particular, with asymmetric continuation regions that shrink sufficiently abruptly, greater problem complexity can reduce expected stopping time. 
    These conditions are intended to capture economically relevant cases in which the stopping rule does not itself create a selection effect strong enough to overturn the slower-learning effect. 
} 
It is stated for a restricted but practically important class of exogenous stopping rules, including fixed boundaries, symmetric process-based boundaries, and other shrinking continuation regions satisfying the conditions above. 
\hyref{figure:exogenous-stopping:speed-accuracy}[Figure] provides an illustration. 
Panel (a) depicts the corresponding continuation region; panels (b) and (c) show the associated expected stopping time and accuracy as a function of $\kappa$. 

The key observation behind \hyref{theorem:exogenous-stopping:speed-accuracy}[Theorem] is that a more complex problem is observationally equivalent to running the same belief process on a slower clock. 
Since the decision-maker stops by time $t$ only if beliefs have exited the continuation region at some point prior to $t$, greater problem complexity affects stopping through two channels. 
First, the belief process evolves more slowly, which tends to delay exit. 
Second, at a given calendar time the same realised belief path is effectively compared with thresholds from a later point in time.

If the continuation region were constant or weakly expanding over time, both forces point towards slower stopping in more complex problems: beliefs are compared with the continuation region and, at the same time, have had less effective time to diffuse. 
In that case, a simple coupling argument yields the result. 
When the continuation region shrinks over time, the two forces work in opposite directions: beliefs feature slower learning, but later thresholds are tighter. 
Part (1) of \hyref{theorem:exogenous-stopping:speed-accuracy}[Theorem] shows that, under the standing assumptions, the former effect dominates. 
The proof exploits the built-in symmetry when the continuation region is symmetric about 1/2 (i.e., $\overline p_t=1-\underline p_t$), and otherwise uses the differential characterisation of the expected stopping time combined with an optional stopping argument.
It is important to note that the comparative statics for expected stopping time can fail when the standing assumptions are violated. 
In particular, when the continuation region is asymmetric and shrinks abruptly, lower complexity can actually lead to a selection into continuation being more likely, and induce a higher expected stopping time for the less complex problem.

Part (2) of \hyref{theorem:exogenous-stopping:speed-accuracy}[Theorem] follows from the fact that beliefs change faster. 
The intuition is clearest when the stopping thresholds are symmetric about $1/2$, so that $\overline p_t=1-\underline p_t$. 
In that case, the posterior probability of making the correct choice at stopping is simply $\mathbb P(\alpha_\tau = \alpha^\theta) = \overline p_\tau$. 
Hence, when $\overline p_t$ is decreasing in $t$, later stopping implies lower accuracy, whereas when $\overline p_t$ is increasing in $t$, later stopping implies higher accuracy. 
The proof of the general case extends this logic by combining coupling with optional stopping.
In short, in contrast to what occurs with expected stopping time, the comparative statics on accuracy rely only on the continuation region shrinking over time.

\subsection{Speed-Accuracy Complementarity in Related Models}
\label{section:exogenous-stopping:related-models}

\subsubsection{Uncertain Complexity}

The finding that, for a broad class of stopping rules, more complex problems are associated with slower and less accurate decisions bears resemblance to existing findings in the literature, in particular those in \citet{FudenbergStrackStrzalecki2018AER}.\footnote{
    \citet{DrugowitschMoreno-BoteChurchlandShadlenPouget2012JNeuro} discuss related findings using numerical solutions. 
    \citet{Bhui2019CBB} provides recent experimental evidence.
}
They study accuracy \emph{conditional on stopping time} under different informational environments about the drift rate, corresponding to the absolute difference in payoffs across alternatives. 
They show that, in a symmetric environment, when the drift rate is known, accuracy is constant over time, whereas when the drift rate is unknown and normally distributed, accuracy decreases over time. 

Whilst our notion of speed-accuracy complementarity is different --- we compare expected stopping time and accuracy across problems of different complexity --- the same broad mechanism is present in both settings. 
First, the decision-maker does not optimally adjust the stopping rule to realised problem complexity; here because we consider exogenous stopping rules, there because the decision-maker is uncertain about problem complexity. 
Second, the continuation region is shrinking over time.\footnote{
    \citet{FudenbergStrackStrzalecki2018AER} formulate the continuation region in the space of beliefs over the drift rate.
    However, it can equivalently be expressed in terms of time-varying boundaries for the posterior probability assigned to a given alternative being optimal, yielding a shrinking continuation region which is symmetric about $1/2$.
}

To relate the two problems more tightly, we show that analogous comparative statics apply to \citet{FudenbergStrackStrzalecki2018AER}: there too, greater \emph{realised} problem complexity is associated with longer stopping times and lower accuracy. 
As we discuss in the proof (\hyref{appendix:proofs:theorem:exogenous-stopping:speed-accuracy}[Appendix]) and in \hyref{appendix:variants:FSS}[Appendix], \hyref{theorem:exogenous-stopping:speed-accuracy}[Theorem] extends to more general belief processes in which decision-maker is uncertain about problem complexity, including that arising in \citet{FudenbergStrackStrzalecki2018AER}.

\subsubsection{Robustness and Misspecification}

Our results are also related to settings in which the stopping rule is optimal under a robust maxmin criterion. 
Drawing on \citet{Wald1947Ecta}, consider a symmetric problem, $u(a,1)=u(b,-1)>u(a,-1)=u(b,1)$, in which the decision-maker entertains a set of possible complexity levels $[\underline \kappa,\overline \kappa]$. 
A maxmin decision-maker then chooses a stopping rule that is optimal for the worst-case complexity level $\overline\kappa$ which, anticipating the discussion in the next section, correspond to constant stopping thresholds. 
The stopping rule is therefore invariant to realised problem complexity and, as in \hyref{theorem:exogenous-stopping:speed-accuracy}[Theorem], problems with greater realised complexity again induce slower stopping and strictly lower accuracy. 
Similar comparative statics also hold if the decision-maker behaves optimally, but is misspecified regarding the actual problem complexity.\footnote{
    A large recent literature draws on \citet{Berk1966AnnMathStats} to study learning in misspecified settings \citep[e.g.,][]{EspondaPouzo2016Ecta,FrickIijimaIshii2023REStud,FudenbergStrackLanzani2021Ecta} --- see \citet{BohrenHauser2025ARE} for a review.
    Although this literature mainly focuses on issues of convergence, long-run dynamics, and stability in active learning settings, our results provide useful tools for comparative statics in settings in which decision-makers stop learning in finite time as in cases of misspecified costly information acquisition.
}

\section{Speed and Accuracy under Optimal Stopping}
\label{section:optimal-stopping}
The main focus of this section is to characterise how problem complexity affects stopping time and choice accuracy when the decision-maker recognises and optimally adjusts to problem complexity. 
This contrasts with \hyref{section:exogenous-stopping}[Section], where stopping rules were invariant to realised complexity and more complex problems were therefore slower and, under shrinking continuation regions, also less accurate.

Once the decision-maker can adjust the stopping rule to the realised complexity of the problem, the relation between speed and complexity changes fundamentally. 
Two forces pull in opposite directions. 
Simpler problems entail more informative signals, which raises the marginal value of continued deliberation. 
At the same time, because time is costly, the decision-maker may instead choose to stop earlier and save on the cost of further information acquisition. 
We show that the effect is ambiguous: expected stopping time is generally non-monotone in problem complexity, even though accuracy remains monotone. 
The ability of the decision-maker to recognise and optimally adjust to problem complexity therefore challenges the use of response time as a proxy for complexity. 
The section concludes by reviewing related models and empirical evidence consistent with this observation.

\subsection{Characterising Optimal Stopping}
\label{section:optimal-stopping:characterisation}

The decision-maker chooses a stopping time $\tau^*$ to maximise their expected payoff net of the cost of time. 
This induces the value function 
$V(p_0):=\sup_\tau\mathbb E[\max_\alpha \mathbb E[u(\alpha,\theta)|\mathcal F_\tau^X]-c\tau]$. 

The solution to this optimal stopping problem is known; see \citet[Theorem 21.1]{PeskirShiryaev2006Book} and the earlier discussion in \citet[Ch. 4.2]{Shiryaev2007Book}.\footnote{
    \citet{Shiryaev2007Book} studies the same problem but incorrectly assumed the value function is twice continuously differentiable. 
} 
We nevertheless briefly go through the main steps of the characterisation, both because they clarify the economic structure of the problem and because they are used directly in the comparative statics below.

First, we separate relative from absolute incentives. 
Let $\tilde p\in (0,1)$ denote the belief that makes the decision-maker indifferent between choosing either alternative $\tilde p: u(a,\tilde p)=u(b,\tilde p)$, and define $\delta: = (u(a,a)-u(b,a))+(u(b,b)-u(a,b))$. 
The parameter $\tilde p$ captures relative incentives between the two alternatives, whereas $\delta$ captures the overall stakes of the problem. 
For instance, adding a constant to $u(a,\cdot)$ lowers the belief threshold $\tilde p$ at which alternative $a$ is optimal, but does not affect $\delta$; conversely, scaling up $u$ linearly increases $\delta$ but does not change $\tilde p$. 



The earliest optimal stopping time is characterised by getting a higher expected payoff by stopping as by continuing, that is, $\tau^*:=\inf\{t>0 | V(p_t) = v(p_t)\}$, where $v(p):=\max_\alpha u(\alpha,p)$. 
Since, by standard arguments, $V$ is convex, and equals $v$ on $\{0,1\}$, where beliefs are degenerate, then, since $v$ is continuous and piecewise linear, 
then the earliest optimal stopping time is characterised by two constant thresholds in beliefs, $\underline p, \overline p$ such that $\tau^*=\inf\{t>0\mid p_t\notin(\underline p,\overline p)\}$.

The optimal stopping thresholds follow from the differential characterisation of the value function. 
By standard arguments \citep[Proposition 3.1]{OksendalReikvam1998SSR}, the value function $V$ is the unique viscosity\footnote{
    The value function $V$ is not a classical solution: it fails to be $\mathcal C^2$ at the stopping thresholds.
} solution to the free-boundary problem
\begin{align*}
    0&=-c +2 \kappa^{-2}\left(p(1-p)\right)^2V''(p) &\text{ on }p \in(\underline p, \overline p) \tag{HJB} \\
    0&=V(p)-\max_\alpha u(\alpha,p) &\text{ on }p \notin (\underline p, \overline p). \tag{BC}
\end{align*}

We therefore obtain the following result.

\begin{proposition}
    \label{proposition:optimal-stopping}
    The optimal stopping time $\tau^*$ is given by $\tau^* =\inf\{t\geq 0: p_t \notin (\underline p, \overline p)\}$, where $\underline p$ and $\overline p$ satisfy $0<\underline p<\tilde p<\overline p<1$ and are the unique solution to 
    \begin{align}
        2\kappa^{-2}\delta/c
        &=(A(\overline p)+B(\overline p))-(A(\underline p)+B(\underline p))
        \label{equation:boundaries:mu}
        \\
        \tilde p
        &=[C(\overline p)-C(\underline p)]/[(A(\overline p)+B(\overline p))-(A(\underline p)+B(\underline p))]
        \label{equation:boundaries:ptilde}
    \end{align}
    with $A(p):=\exp(\logit(p))-\exp(-\logit(p))$, $B(p):=2\logit(p)$, and $C(p):=\logit(p)+\exp(\logit(p))$. 
\end{proposition}

\subsection{Comparative Statics under Optimal Stopping}
\label{section:optimal-stopping:speed-accuracy}

Recognising and optimally adjusting to problem complexity leads to a markedly different set of comparative statics from those in \hyref{section:exogenous-stopping}[Section].

\begin{theorem}
    \label{theorem:optimal-stopping:speed-accuracy}
    Under the optimal stopping time $\tau^*$, (1) the expected stopping time $\mathbb E[\tau^*]$ is non-monotone and quasi-concave in problem complexity $\kappa$; and (2) accuracy, $\mathbb P(\alpha_{\tau^*} = \alpha^\theta)$, is decreasing in problem complexity $\kappa$.
\end{theorem}

The proof combines two ingredients. 
First, by optional stopping, expected stopping time and choice accuracy admit closed-form expressions as functions of the stopping thresholds alone. 
Second, the comparative statics of the thresholds with respect to problem complexity follow from the implicit function theorem applied to the system in \eqref{equation:boundaries:mu}--\eqref{equation:boundaries:ptilde}. 
Substituting these threshold comparative statics into the closed-form expressions yields the result.

A noteworthy feature of \hyref{theorem:optimal-stopping:speed-accuracy}[Theorem] is that the qualitative comparative statics do not depend on either relative or absolute incentive intensity. 
In particular, the comparative statics remain the same regardless of whether or not the decision-maker cares more about choosing $a$ when $b$ is better than the converse --- as given by a high $\tilde p$ --- or that that the problem at hand has high or low stakes --- as reflected by $\delta$.
In short, \hyref{theorem:optimal-stopping:speed-accuracy}[Theorem] complements existing related results \citep[e.g.,][]{BogaczBrownMoehlisHolmesCohen2006PsyRev}.\footnote{
    Although we focus on continuous evidence processes as given by diffusion processes, recent literature has also indicated that in binary decision problem with one-sided Poisson process, the expected decision time could be non-monotone in the arrival rate \citep{BobtcheffLevy2017AEJMicro,HalacKartikLiu2016REStud}. 
}

A crucial observation is that greater problem complexity induces narrower stopping thresholds. 
This has two implications. 
First, the decision-maker acquires less information in the Blackwell sense, which lowers accuracy. 
Second, it reveals the two opposing forces behind the non-monotonicity of the stopping time: beliefs take longer to exit a fixed continuation region, but the continuation region itself becomes narrower.

The contrast with \hyref{section:exogenous-stopping}[Section] is immediate. 
Under exogenous stopping, greater problem complexity slows down the belief process whilst leaving the stopping rule fixed, so more complex problems are slower. 
Under optimal stopping, greater complexity still slows down learning, but it also narrows the optimal continuation region. 
These two effects pull in opposite directions. 
Narrower stopping thresholds reduce the amount of information acquired and therefore lower accuracy, but they may either shorten or lengthen expected stopping time depending on which force dominates. 
\hyref{theorem:optimal-stopping:speed-accuracy}[Theorem] shows that the threshold effect dominates for sufficiently complex problems, whereas the slower-learning effect dominates for sufficiently simple ones. 
This is what gives rise to the inverse-$U$ relationship between stopping time and problem complexity.

\begin{figure}[t!]
    \centering\small\singlespacing
    \begin{subfigure}{.485\linewidth}
        \includegraphics[width=\linewidth]{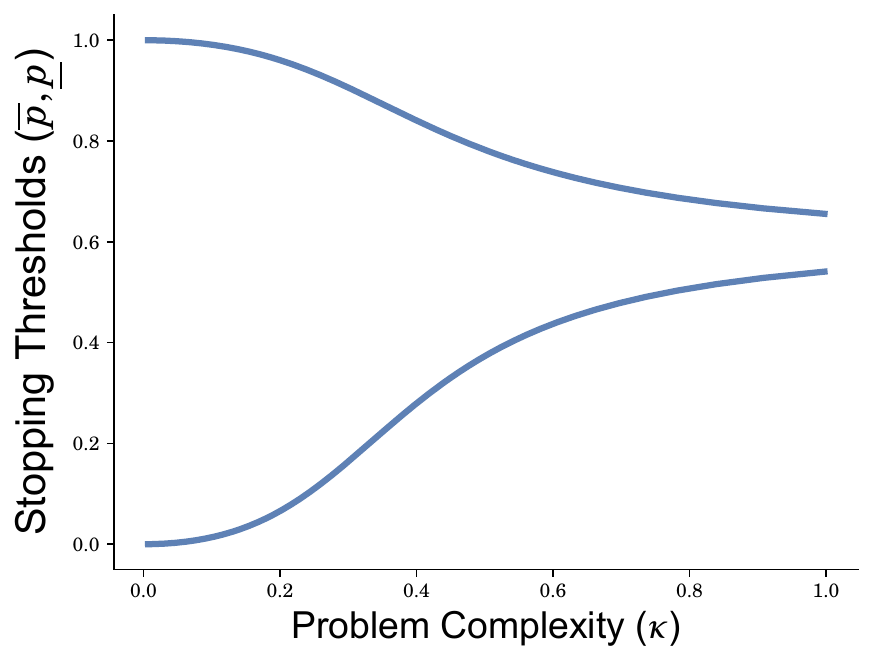}
        \caption{Stopping Thresholds}
        \label{figure:optimal-stopping:thresholds}
    \end{subfigure}

    \begin{subfigure}{.485\linewidth}
        \includegraphics[width=\linewidth]{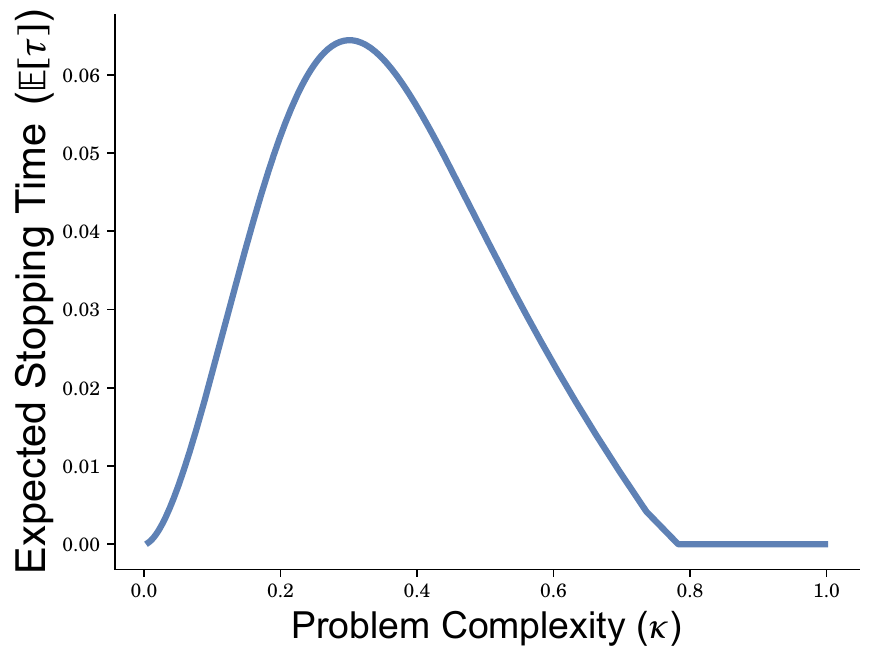}
        \caption{Expected Stopping Time}
        \label{figure:optimal-stopping:speed}
    \end{subfigure}
    \begin{subfigure}{.485\linewidth}
        \includegraphics[width=\linewidth]{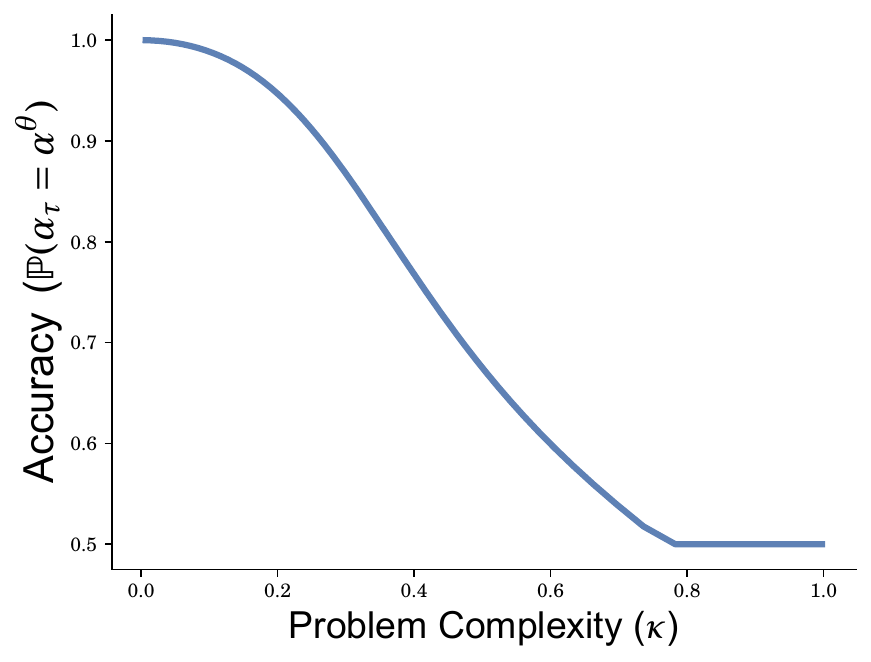}
        \caption{Accuracy}
        \label{figure:optimal-stopping:accuracy}
    \end{subfigure}
    \begin{minipage}{1\linewidth}
        \caption{Optimal Stopping: Speed and Accuracy}
        \label{figure:optimal-stopping:speed-accuracy}
        \emph{Notes}: This figure shows, in panels (a), (b), and (c), the optimal stopping thresholds, the expected optimal stopping time, and the optimal level of accuracy as a function of problem complexity $\kappa$.
        We set $\delta=c=1$ and $\tilde p=3/5$.
    \end{minipage}
\end{figure}

\hyref{figure:optimal-stopping:speed-accuracy}[Figure] exhibits how speed and accuracy depend on problem complexity and the resulting inverse-$U$ relationship between stopping time and complexity. 
The figure also illustrates another implication of the model. 
As problem complexity rises, the optimal stopping thresholds move towards the indifference point $\tilde p$. 
Generically, when $\tilde p\neq p_0$, this implies that for sufficiently high problem complexity the prior no longer lies in the continuation region. 
In such cases, the decision-maker stops immediately and chooses the prior-favoured alternative without acquiring additional information.

\subsection{Speed-Accuracy Non-Monotonicity in Related Models}
\label{section:optimal-stopping:related-models}

\subsubsection{Preference Intensity and Discounting}

The same qualitative comparative statics arise in several common variants of the model. 
In many applications, the speed of learning is proportional to preference intensity \citep[e.g.,][]{FehrRangel2011JEP}. 
In our setting, this is observationally equivalent to replacing problem complexity $\kappa$ with $\kappa\delta^{-1}$. 
The qualitative conclusions of \hyref{theorem:optimal-stopping:speed-accuracy}[Theorem] therefore continue to hold.\footnote{ 
    The non-monotonicity of the expected optimal stopping time also applies to cases in which preference intensity is unknown, such as \citepos{FudenbergStrackStrzalecki2018AER}, where the learning rate is nevertheless also affected by an exogenous parameter analogous to $\kappa$. 
    We discuss this case in \hyref{appendix:variants:FSS}[Appendix].
} 
Another common variant replaces the flow cost of time with exponential discounting. 
In \hyref{appendix:variants:discounting}[Appendix], we show that the qualitative relation between problem complexity, accuracy, and expected stopping time remains the same in that setting.

\subsubsection{Costly Information Acquisition}

In \hyref{appendix:variants:cia}[Appendix], we discuss closely related non-monotone comparative statics in standard static costly information acquisition models, highlighting that the non-monotonicity is not an artefact of the Brownian formulation. 
In the baseline model, time determines the amount of information acquired and problem complexity acts as a time-scaling parameter, making the cost of information is therefore directly tied to time. 
We then consider a general rational inattention environment in which the decision-maker may choose arbitrary information structures \citep{MatejkaMcKay2015AER,CaplinDean2015AER,CaplinDeanLeahy2022JPE}, whilst maintaining two assumptions: problem complexity acts as a time-scaling parameter, and the cost of information is monotone in the amount of time-equivalent information acquired. 
The same non-monotone relation between time and problem complexity emerges there as well, although single-peakedness requires additional assumptions.

\subsubsection{Uncertain versus Recognised Complexity}

It is the capacity of the decision-maker to optimally adjust their behaviour to the complexity of the problem that underlies the differences between \hyref{theorem:exogenous-stopping:speed-accuracy}[Theorem] and \hyref{theorem:optimal-stopping:speed-accuracy}[Theorem]. 
The inability to adjust optimally to realised complexity may be reasonable when the decision-maker is unsure about how complex the problem is. 
For a given prior about problem complexity, the stopping rule is fixed against different realised values of $\kappa$, effectively taking us to the previous section's setup. 

However, uncertainty about complexity does not by itself eliminate the non-monotonicity from \hyref{theorem:optimal-stopping:speed-accuracy}[Theorem].
If the decision-maker is uncertain about $\kappa$ and $\theta$, as they learn jointly about $\theta \kappa^{-1}$, learning which alternative is best is inseparable from learning problem complexity, as in \citet{FudenbergStrackStrzalecki2018AER}.
In these settings, one analogue of our comparative static is with respect to the prior regarding problem complexity rather than the realised value. 
Take the case of choosing between two computers and two computer cables. 
It may be that each of the choice problems may be more or less complex, depending on the specific items we are comparing, but despite the complexity of each being uncertain, the difficulty in choosing between computers is arguably stochastically higher than that in choosing between computer cables. 
As we show in \hyref{appendix:general}, expected optimal stopping time will here too be non-monotone with respect to the distribution of problem complexity.\footnote{
    In \hyref{appendix:general} we consider only cases in which the decision-maker faces richer uncertainty about $\theta$ but the problem is otherwise the same. 
    We discuss how our results relate to the setting in \citet{FudenbergStrackStrzalecki2018AER} in more detail in \hyref{appendix:variants:FSS}[Appendix].
}

\subsection{Reconciling Empirical Evidence on Speed and Problem Complexity}

\hyref{theorem:exogenous-stopping:speed-accuracy}[Theorems] and \hyref{theorem:optimal-stopping:speed-accuracy} suggest that the relation between response times and problem complexity depends on how complexity enters the problem. 
In some environments, the feature that makes the problem more or less complex also determines which action is optimal. 
In others, complexity varies independently of which action is best. 
The former case is more closely aligned with the exogenous-stopping benchmark, whereas the latter is where the non-monotonicity from \hyref{theorem:optimal-stopping:speed-accuracy}[Theorem] is most likely to arise.

In the first case, when the optimal action depends on what makes the problem more or less complex, the complexity of the problem and its solution are indissociable. 
There, decision-makers cannot adjust to problem complexity independently of learning the optimal solution and, in line with \hyref{theorem:exogenous-stopping:speed-accuracy}[Theorem], longer stopping times are more likely to reflect greater problem complexity. 
This is the case, for instance, in visual perception tasks where the goal is to identify the dominant colour or direction of movement and the intensity of that dominance is what varies \citep[e.g.,][]{LindePaivio1979MemCog,RatcliffRouder1998PsyRev,VossRothermundVoss2004MemCog,MaanenGrasmanForstmannWagenmakers2011FrontiersNeuro}. 
Another example is numerical comparison tasks in which the feature being varied is precisely the magnitude of the difference between the two values \citep[e.g.,][]{MoyerLandauer1967Nature,BuckleyGillman1974JExpPsy,KrajcsiLengyelKojouharova2016FrontiersPsy,GoncalvesNunnariZarate-PinaWP}.

In the second case, problem complexity is affected by a feature that is separable from the determination of the optimal action. 
If decision-makers recognise and optimally respond to that feature, stopping times need not vary monotonically with problem complexity, as in \hyref{theorem:optimal-stopping:speed-accuracy}[Theorem]. 
This is the case, for instance, when choosing products with more or fewer attributes \citep{OnkenHastieRevelle1985JExpPsy}, identifying the dominant colour of a larger or smaller set of items \citep{GoncalvesNunnariZarate-PinaWP}, or comparing two values obtained from sums with more or fewer summands \citep{GoncalvesNunnariZarate-PinaWP}. 
\citet{WrightAyton1988MemCog} also document a non-monotone relationship between stopping time and accuracy in true-or-false questions of arguably varying difficulty, and \citet{AgranovSchotterTrevino2025WP} provide evidence of a non-monotone relation across several environments, including lottery choice, contingent reasoning, public-good games, auctions, and belief-updating tasks.

\section{Speed and Accuracy with Heterogeneous Ability}
\label{section:ability}

We extend the model to incorporate effort and heterogeneous ability. 
This serves two purposes. 

First, it makes explicit that time spent on a task need not be a good measure of the effort expended on it. 
This is precisely the distinction between problem complexity and a more individual-specific notion of cost complexity \citep{Oprea2020AER}, which depends not only on the problem but also on how a decision maker chooses to approach it: a decision-maker may spend a long time on a task whilst exerting little effort, or spend little time on a difficult task and exert substantial effort. 
We therefore allow the decision-maker to control how much effort they exert at every point in time, positing that more effort is costly but increases how much progress they make per unit of time.

Second, it allows the model to speak to the ambiguous relation between ability and response time documented in the literature.
In many settings, slower response times are taken to be indicative of greater ability or sophistication. 
For instance, more financially literate individuals make slower financial decisions \citep{DarrietGuilleVergnaudShimizu2020JEPsy} and, in strategic environments, individuals who spend more time on a problem often tend to be more successful or sophisticated \citep{Rubinstein2016QJE,AgranovCaplinTergiman2015JESA,Alos-FerrerBruckenmaier2021EE,Esteban-CasanellesGoncalves2020WP,GillProwse2023EJ}. 
By contrast, work on educational testing and intelligence measurement has long stressed speed as an indicator of ability \citep{ThorndikeBregmanCobbWoodyard1926,Thurstone1937Psyta}.\footnote{
    \citet{Goldhammer2015Meas} provides an extensive discussion of this literature and the tensions created by the conflicting evidence.
}
Although these studies typically measure ability independently of response time, one might still be tempted to infer ability directly from how fast a decision-maker responds when performance is ambiguous or unobserved. 
To show how our model can formalise this ambiguity between time and ability, we incorporate heterogeneous ability, with decision-makers of higher ability bearing a lower cost to effort.

\subsection{Comparative Statics under Optimal Control}
\label{section:ability:optimal-control}

We revisit the model in \hyref{section:setup}[Section] and allow the decision-maker to control instantaneous effort $e_t\geq 0$, borrowing from \citet{MoscariniSmith2001Ecta}. 
Greater effort makes signals more informative, so the information process becomes 
$\diff X_t=\theta \kappa^{-1}\sqrt{e_t}\,\diff t+\,\diff B_t$,
where $\kappa>0$ is as before and $B_t$ is a standard Brownian motion independent of $\theta$.

Exerting effort is costly. 
A decision-maker of ability $\lambda>0$ incurs flow cost $c(e_t/\lambda)$, where $c> 0$ is increasing and convex, with $c',c''>0$ and $c'''\geq 0$. 
Higher values of $\lambda$ therefore correspond to lower effort costs and hence greater ability.

A decision-maker of ability $\lambda$ chooses both a stopping time $\tau$ and an effort path $(e_t)_{t\geq 0}$, adapted to $\mathcal F_t^X$, to maximise
\begin{align*}
    \mathbb E\left[\max_\alpha\mathbb E[u(\alpha,\theta)\mid \mathcal F_t^X]-\int_0^\tau c(e_{t}/\lambda)\diff t \right].
\end{align*}
Ability does not affect stopping time or accuracy directly. 
It matters only through the endogenous choice of effort. 
Accordingly, we focus on the relation between ability, stopping time, accuracy, and problem complexity when the decision-maker recognises and optimally adjusts to problem complexity, as in \hyref{section:optimal-stopping}[Section].

The main implication is that response time alone is not a reliable measure of ability.

\begin{theorem}
    \label{theorem:ability:speed-accuracy}
    Under optimal effort choice $(e_t^*)_{t\geq 0}$ and the corresponding optimal stopping time $\tau^*$,
    (1) expected stopping time $\mathbb E[\tau^*]$ is non-monotone and quasi-concave in ability $\lambda$; 
    (2) accuracy, $\mathbb P(\alpha_{\tau^*}=\theta)$, is increasing in ability $\lambda$; and
    (3) when $\mathbb E[\tau^*]>0$, expected stopping time is quasi-supermodular in ability and problem complexity, $(\lambda;\kappa)$.
\end{theorem}

The proof proceeds by observing that the Hamilton--Jacobi--Bellman equation implies that optimal effort is constant over time. 
Moreover, if $e_\lambda^*$ denotes the optimal effort of a decision-maker of ability $\lambda$, then $e_\lambda^*=\lambda e^*$, where $e^*$ solves
$e^*c'(e^*)=c(e^*)$. 
Hence higher-ability decision-makers exert more effort. 
At the same time, the flow cost of optimal effort is identical across ability levels, since $c(e_\lambda^*/\lambda)=c(e^*)=:c^*$. 
Ability therefore affects behaviour entirely through the effective informativeness of the signal process: for a decision-maker of ability $\lambda$, the effective problem complexity becomes $\kappa e_\lambda^{*-1/2}$. 
The problem is thus equivalent to the optimal-stopping problem from \hyref{section:optimal-stopping}[Section], with complexity rescaled by ability, so the result then follows from \hyref{theorem:optimal-stopping:speed-accuracy}[Theorem]. 
Furthermore, the comparative statics with respect to problem complexity $\kappa$ from \hyref{theorem:optimal-stopping:speed-accuracy}[Theorem] are also immediately preserved. 

The interpretation is simple. 
Greater ability is tantamount to the same problem being effectively simpler. 
Equivalently, the same underlying problem is more complex for a low-ability decision-maker than for a high-ability one. 
As a result, higher-ability decision-makers always make more accurate choices, but the relation between ability and stopping time is nuanced. 
Short response times may reflect either high ability or low ability: a highly able decision-maker may stop early because the problem is very simple for them, whereas a low-ability decision-maker may stop early because further deliberation is not worthwhile. 
For example, students who hand in an examination early may do so because additional time would be unproductive, but this may reflect either strong performance or a lack of progress.

The quasi-supermodularity implies that faster decisions are more reflective of higher ability in simpler than in more complex problems.
In particular, whether higher-ability decision-makers stop sooner or later than lower-ability ones depends on the complexity of the problem. 
In relatively simple problems, higher-ability decision-makers stop sooner than lower-ability ones. 
In sufficiently complex problems, the opposite holds: higher-ability decision-makers stop later because greater effort makes continued deliberation worthwhile.

This is also where the distinction between problem complexity and cost complexity becomes particularly useful. 
Problem complexity is an ex ante property of the environment, indexed here by $\kappa$. 
Cost complexity, by contrast, depends on how demanding it is for a particular decision-maker to deploy a given procedure. 
Ability operates through this second channel: it changes how costly it is to generate informativeness, and therefore changes the effective complexity of the same underlying problem.

\subsection{Speed and Ability in Related Models}
\label{section:ability:related}

\subsubsection{Discounting} 
The same qualitative results continue to hold under exponential discounting. 
The non-monotonicity of expected optimal stopping time with respect to problem complexity extends immediately, whereas the accuracy remains monotone in problem complexity. 
Moreover, there exist problem complexity levels $\kappa>\kappa'$ and ability levels $\lambda>\lambda'$ such that in the more complex problem the higher-ability decision-maker stops strictly later than the lower-ability one, whereas in the simpler problem, $\kappa'$, the higher-ability decision-maker stops strictly earlier. 
We discuss this extension in \hyref{appendix:variants:discounting-ability}[Appendix].

\subsubsection{Exogenous Stopping} 
The results above rely on the decision-maker being able to adjust both effort and stopping to realised problem complexity and their ability. 
Ability does not affect stopping time or accuracy directly; it matters only through the effort margin. 
If higher ability translates pointwise into higher effort, but stopping is exogenous or is not adapted to either problem complexity or individual ability, then results analogous to those in \hyref{section:exogenous-stopping}[Section] apply: shorter stopping times are then more likely to reflect higher ability. 
This would be the case when decision-makers are unaware of how easy a problem is or of their own ability to deal with it.

\subsection{Reconciling Empirical Evidence on Speed and Ability}
\label{section:ability:evidence}

\hyref{theorem:ability:speed-accuracy}[Theorem] helps reconcile conflicting evidence indicating either a positive or a negative relationship between response time and ability. 
The main implication is that this relationship cannot be interpreted without taking problem complexity into account. 
Higher ability always improves accuracy, but it need not make response times uniformly shorter or uniformly longer.

Existing evidence is consistent with this prediction. 
\citet{LindleyWilsonSmithBathurst1995PID} report a visual categorisation task in which measured intelligence, using the Wonderlic Personnel Test, is related to response time in an inverted-$U$ shaped way. 
\citet{GoldhammerNaumannStelterTothRolkeKlieme2014JEducPsy} find heterogeneous relationships between accuracy and response time modulated by both task difficulty and skill level in standardised assessments. 
\citet{DodonovaDodonov2013Intelligence} and \citet{GoldhammerNaumannGreiff2015JIntell} show that faster responses are more indicative of higher ability in easier tasks. 

Taken together, the evidence is consistent with the model's prediction that, in simple problems, faster responses are more likely to indicate greater ability, whereas in complex problems the opposite may be true. 
Our framework provides a simple rationale for this pattern: once decision-makers optimally adjust effort to effective problem complexity, ability changes both how informative signals become and whether continued deliberation is worthwhile.

\section{Identifying Problem Complexity (and Ability)}
\label{section:identification}

We now turn to the question that often motivates the use of response times in the first place: how to identify problem complexity. 
Complexity is often closely associated with mistakes and suboptimal choices in a variety of settings, from inference and misinformation,\footnote{
    See, e.g., \citet{EnkeZimmermann2019REStud}, \citet{NiederleVespa2023ARE}, \citet{GoncalvesLibgoberWillis2023WP}, and \citet{GoncalvesNunnariZarate-PinaWP}.
} to lottery choice and financial markets,\footnote{
    There is a large literature examining issues tied to complexity in choice under risk, including 
    \citet{Wilcox1993EJ}, \citet{FrydmanJin2022QJE}, \citet{EnkeGraeber2023QJE}, \citet{Oprea2023WP}, \citet{Puri2023WP}, \citet{AgranovSchotterTrevino2025WP}, and \citet{EnkeShubatt2023NBERWP}.
    On financial choices, see \citet{AgarwalRosenYao2016MnSc}, \citet{BhargavaLoewensteinSydnor2017QJE}, \citet{Rees-JonesTaubinsky2020REStud}, and \citet{CarvalhoSilverman2017WP}.
} to market institutions and consumer choice,\footnote{
    Much attention in mechanism design has been devoted to the problem of mechanism simplicity \citep{Li2017AER,BorgersLi2019Ecta,PyciaTroyan2023Ecta,LiDworczak2021EC}. 
    Recent work on consumer choice and market institutions includes 
    \citet{Spiegler2016ARE}, \citet{BossaertsMurawski2017TrendsCogSci}, \citet{Martinez-MarquinaNiederleVespa2019AER}, \citet{Jakobsen2020AER}, \citet{GilboaPostlewaiteSchmeidler2021RE}, and \citet{Camara2025WP}. 
} 
or strategic choice and voting.\footnote{
    See, among others, \citet{AlaouiPenta2016REStud,AlaouiPenta2022JPE} and \citet{EspondaVespa2024REStud}.
}
In order to understand how to simplify problems, or even to study what makes some problems harder than others, it is first necessary to find a way to rank problems by complexity. 

Throughout the paper, we have shown that accuracy is monotone in problem complexity --- even under exogenous stopping, under mild conditions --- whereas response time need not be. 
Accordingly, if an analyst knows which choice is optimal, then accuracy can be used directly to infer complexity: more complex problems generate lower choice accuracy. 
This is often feasible in laboratory settings, where the true state is observed by the analyst and mistakes can be directly measured.

In many environments, however, the analyst does not know which action is optimal. 
For instance, if one wishes to assess which features of a mechanism or interface make it more complex, then the optimal choice typically depends on the very private information the analyst is trying to elicit. 
Our earlier results also show that expected response time is not generally a reliable proxy for problem complexity.\footnote{
    This non-monotonicity is not confined to the mean: it persists for the variance of the stopping time.
    With heterogeneous populations, these need not be even quasiconcave with respect to problem complexity.
}
This raises a natural question: is there some observable behavioural response that remains monotone in problem complexity and can therefore be used to identify it?

\subsection{An Identification Result}
\label{section:identification:result}

We propose an alternative approach based on the effect of subsidies on choice. 
Consider two problems, $A$ and $\hat A$, in which the decision-maker chooses between alternatives $\{a,b\}$. 
Within our model, problem $A$ is more complex than problem $\hat A$ if it has a higher noise-to-signal ratio $\kappa$.

Consider offering a subsidy $s$ for choosing alternative $b$ in problem $A$ and alternative $\hat b$ in problem $\hat A$. 
In each problem, this shifts payoffs to choosing $b$ by a constant, $u^s(b,\theta):=u(b,\theta)+s$, whilst keeping payoffs to choosing $a$ the same, $u^s(a,\theta):=u(a,\theta)$. 
This raises the relative attractiveness of the subsidised alternative, and therefore increases the indifference threshold $\tilde p$, leaving the overall stakes $\delta$ unchanged. 
Hence the subsidised alternative should be chosen more often. 
Under optimal stopping, this is in fact the case: if $\tau^*$ denotes the optimal stopping time from \hyref{proposition:optimal-stopping}[Proposition], then $\frac{\partial}{\partial s}\mathbb P(\alpha_{\tau^*}=b)>0$.
We say that subsidies are more effective when their effect on choices, $\frac{\partial}{\partial s}\mathbb P(\alpha_{\tau^*}=b)>0$, is greater. 

Our main result in this section is that subsidy effectiveness is itself increasing in problem complexity.

\begin{theorem}
    \label{theorem:identification}
    Let $\mathbb P(\alpha_{\tau^*}=\alpha)>0$, $\alpha=a,b$. 
    Under the optimal stopping time $\tau^*$, the effect of subsidies on choices, $\frac{\partial}{\partial s}\mathbb P(\alpha_{\tau^*}=b)$, is increasing in problem complexity $\kappa$.
\end{theorem}

The intuition for \hyref{theorem:identification}[Theorem] comes from the interaction between the marginal effect of subsidies on stopping thresholds and the level effect of problem complexity. 
Subsidising alternative $b$ raises the indifference threshold $\tilde p$, which shifts both optimal stopping thresholds upward. 
As a result, the decision-maker becomes more willing to stop and choose the subsidised alternative, requiring now stronger evidence in order to stop and choose the non-subsidised one. 
Thus the subsidised alternative is chosen more often and more quickly, whereas the non-subsidised alternative is chosen less often and more slowly --- a comparative static which holds in our model as well as in sequential sampling models with optimal stopping more generally.\footnote{
    See \hyref{proposition:optimal:choice:shift:general}[Proposition] in \hyref{appendix:general:optimal}[Appendix]. 
    This comparative static also holds when for general information structures, and misspecified priors, as shown in \citet{Goncalves2023WP}.
}

\begin{figure}[t!]
    \centering\small\singlespacing
    \begin{subfigure}{.485\linewidth}
        \includegraphics[width=\linewidth]{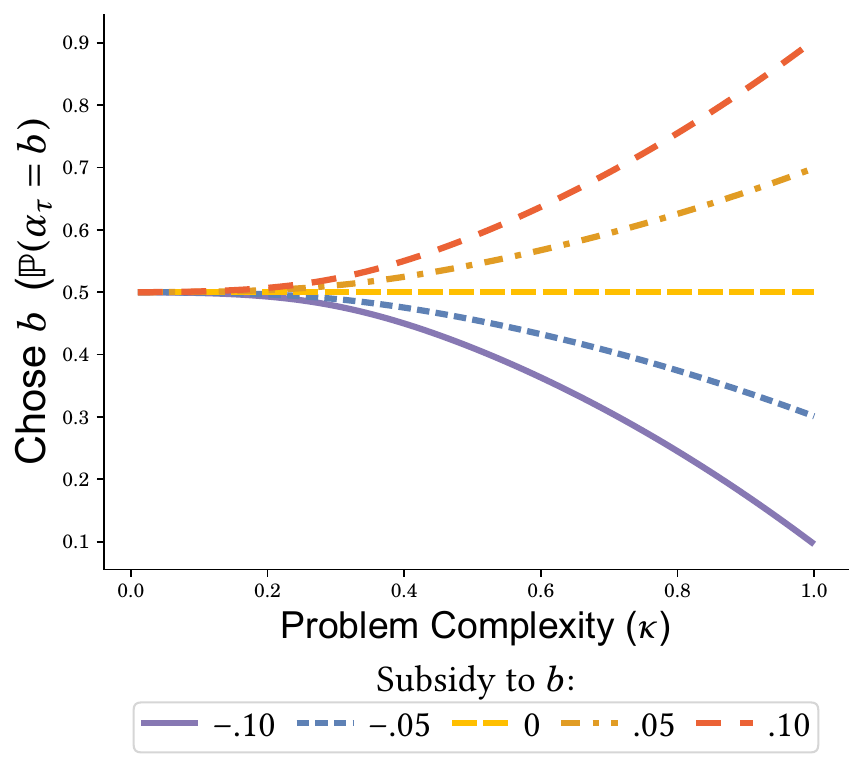}
    \end{subfigure}
    \begin{minipage}{1\linewidth}
        \caption{Optimal Stopping: Sensitivity to Subsidies}
        \label{figure:optimal-stopping:subsidy}
        \emph{Notes}: This figure shows the probability with which alternative $b$ is chosen under optimal stopping as a function of problem complexity $\kappa$, for different levels of subsidies to that alternative $b$.
        We set the baseline payoff to $u(a,1)=u(b,-1)=1$, $u(a,-1)=u(b,1)=0$, and $\delta=c=1$.
    \end{minipage}
\end{figure}

The essential point is that the same shift in thresholds has a larger effect when the problem is more complex. 
In simple problems, stopping occurs only at beliefs already close to certainty, so small changes in thresholds have little effect on choice probabilities. 
In complex problems, by contrast, the continuation region is narrow, which amplifies the impact that the threshold shifts have on choices. 
Although the magnitude of the threshold shifts also varies with problem complexity, this turns out to be a second-order effect. 
The proof formalises this intuition by deriving a closed-form expression for the derivative and studying its dependence on problem complexity.

\hyref{figure:optimal-stopping:subsidy}[Figure] provides a visualisation of how the effect of a subsidy on choice probability depends on the level of problem complexity.
In easy problems, the decision-maker chooses the best alternative with probability close to one. 
Since, in our baseline, alternative $b$ is optimal with probability $p_0=1/2$, in easy problems the decision-maker chooses it with probability close to 1/2, and subsidies have a limited effect on choices.
In contrast, when problem complexity is high, even a small subsidy has a markedly significant effect on choices.

\hyref{theorem:identification}[Theorem] therefore provides a monotone behavioural measure of problem complexity. 
More precisely, it shows that interior choice probabilities satisfy increasing differences in subsidies and problem complexity. 
This yields a practical way of identifying which of two problems is more complex, even when the analyst does not observe the correct action.

The same logic extends immediately to heterogeneous ability in the model of \hyref{section:ability}[Section]. 
Under optimal effort $e_t^*$ and stopping $\tau^*$, the decision-maker's belief process has an effective noise-to-signal given by $ (e^* \lambda)^{-1/2} \kappa$. 
Hence ability has the opposite effect on subsidy effectiveness to problem complexity: all else equal, higher-ability decision-makers respond less to the same subsidy. 
This gives the following corollary.

\begin{corollary}
    \label{corollary:identification}
    Let $\mathbb P(\alpha_{\tau^*}=\alpha)>0$, $\alpha=a,b$. 
    Under optimal effort choice $(e_t^*)$ and the corresponding optimal stopping time $\tau^*$, the effect of subsidies on choices, $\frac{\partial}{\partial s}\mathbb P(\alpha_{\tau^*}=b)$, is decreasing in ability $\lambda$.
\end{corollary}

\subsection{Scope and Limitations}
\label{section:identification:limitations}

\subsubsection{Optimal versus Exogenous Stopping}
The identification result crucially depends on optimal stopping and does not extend mechanically to exogenous stopping rules. 
Suppose that stopping thresholds remain consistent with the decision-maker's preferences, so that $\overline p_t>\tilde p>\underline p_t$ and the stopping decision always leads to the posterior-optimal action. 
If the stopping thresholds do not vary with $\tilde p$, then $\frac{\partial}{\partial s}\mathbb P(\alpha_\tau=b)=0$. 
If they vary with $\tilde p$ but remain constant and invariant to $\kappa$, then the martingale property still implies that $\mathbb P(\alpha_\tau=b)$ does not depend on $\kappa$. 
More generally, even when stopping boundaries shrink over time and move upward with $\tilde p$ --- and, hence, with a subsidy $s$ --- monotonicity of $\frac{\partial}{\partial s}\mathbb P(\alpha_\tau=b)$ with respect to $\kappa$ may fail.\footnote{
    For example, let $(\underline p_t,\overline p_t)=(1/2-1/20-1/10\exp(-t)+1/10h,\, 1/2+1/5+1/20\exp(-t)+h)$. 
    Let $q(\kappa,h)$ denote the probability that $p_t$ exits through the upper boundary for complexity level $\kappa$ and upward shift $h$. 
    An analogue of \hyref{theorem:identification}[Theorem] would require
    $\Delta(h):=(q(\kappa,h)-q(\kappa,0))-(q(\kappa',h)-q(\kappa',0))>0$
    for $\kappa>\kappa'$ and $h>0$. 
    Yet, letting $\kappa\to\infty$ and $\kappa'\to 0$ yields
    $\Delta(h)=\frac{23(1-54h)h}{40(4+9h)(5+18h)}$, 
    which is positive for $h<1/54$ and negative for $h>1/54$.
} 
Thus the comparative statics in \hyref{theorem:identification}[Theorem] are specific to optimal stopping.

\subsubsection{Other Confounding Factors}
Since \hyref{theorem:identification}[Theorem] provides a monotone relation between problem complexity and the effect of subsidies on choice, it is robust to aggregation over heterogeneous populations, provided the distribution of other relevant characteristics is the same across the problems being compared. 
As with any comparative-static exercise, the result is an all-else-equal statement, and this distributional qualification is essential. 
It is also important to guard against false negatives: very simple problems, very complex problems, or very small subsidies may all generate responses that are too small to be detected reliably in finite samples. 
This is all the more relevant since behavioural responses to incentives are oftentimes more muted than expected.\footnote{
    See, e.g., \citet{EnkeGneezyHallMartinNelidovOffermanvandeVan2023REStat}, who scale up absolute incentives (stakes) significantly and find at best muted behavioural effects in choices.
    Whilst our proposed approach pertains to manipulating relative incentives (subsidies), muted reactions to incentive manipulations will exacerbate false negatives.
}

\subsubsection{Subsidy Sensitivity and Behavioural Attenuation}
Concurrent experimental work by \citet{EnkeGraeberOpreaYang2026WP} shows that, in more complex problems, decision-makers become less sensitive to problem fundamentals. 
This is fully consistent with \hyref{theorem:identification}[Theorem].
Here, subsidies are orthogonal to $\theta$, which corresponds to the fundamentals of the problem determining the optimal action. 
Hence, a greater sensitivity to subsidies implies behavioural attenuation with respect to $\theta$. 
The result can then be seen as providing a sequential sampling rationale for behavioural attenuation.

\subsubsection{Costly Information Acquisition}
Given the discussion in \hyref{section:optimal-stopping:related-models}[Section] of static costly information acquisition models, it is natural to ask whether an analogous identification result might hold there as well. 
A related monotonicity result does obtain under mutual-information costs, but it does not extend generally.\footnote{
    \citet{AmbuehlOckenfelsStewart2025REStat} study whether higher participation fees can disproportionately attract individuals with lower costs of learning, and derive a closely related result in a rational inattention model with mutual-information costs. 
    However, the result fails more generally under uniformly posterior-separable costs \citep{CaplinDeanLeahy2022JPE}. 

    More broadly, whilst sequential sampling with additive flow costs can be embedded in rational inattention, the two frameworks differ in important comparative statics. 
    For instance, subsidising an alternative need not increase the probability with which it is chosen under rational inattention \citep{PeaseWhitmayer2025WP}, whereas it always weakly does so under sequential sampling with optimal stopping \citep{Goncalves2023WP}.
}
Hence the identification strategy proposed here is not a generic property of all costly-information models, but rather a distinctive implication of optimal sequential sampling.

\subsection{Empirical Evidence on Subsidy Effectiveness, Problem Complexity, and Ability}
The general effects of subsidies on choices had previously been documented in \citet{Goncalves2023WP}. 
Concurrent work by \citet{GoncalvesNunnariZarate-PinaWP} provides direct evidence consistent with \hyref{theorem:identification}[Theorem] and \hyref{corollary:identification}[Corollary] across a range of domains, including perception, computation, logic, prediction, and inference. 
That paper also distinguishes empirically between exogenous and endogenous stopping by varying different dimensions of problem complexity. 

For example, in the context of choosing which of two sums has the greater value, they replicate the familiar result that smaller absolute differences in sums lower accuracy, suggesting greater complexity. 
However, in that setting, learning the absolute difference is inseparable from learning which sum is larger, so the stopping rule should be invariant to that feature. 
Consistent with this, smaller absolute differences also induce longer response times and an insignificant reaction to subsidies. 
They also vary the number of summands in each expression, a feature that plausibly affects problem complexity whilst remaining orthogonal to which sum is larger. 
There, accuracy decreases with the number of summands, response time is non-monotone, and choices respond more strongly to subsidies as the number of summands rises. 
This is exactly the pattern predicted by \hyref{theorem:identification}[Theorem]. 

More broadly, our results highlight the importance of metacognition. 
The identification results in \hyref{theorem:identification}[Theorem] and \hyref{corollary:identification}[Corollary] rely on decision-makers recognising, at least imperfectly, both problem complexity and their own ability. 
Consistent with this, \citet{GoncalvesNunnariZarate-PinaWP} show that decision-makers learn over the course of the experiment to react more strongly to subsidies precisely in those environments that also induce lower accuracy.

\section{Final Remarks}
\label{section:conclusion}

We believe that our results have important implications for the use of response times in the social sciences. 
Whilst response times provide an important source of information that is able to shed light on the cognitive processes underlying decision-making, our results support existing evidence that, on their own, they may not be a reliable indicator of problem complexity or ability. 
Instead, we propose to use small incentive distortions to infer problem complexity.
These are easy to implement in real-life settings and could be used to study which features make specific problems more or less complex, and evaluate if particular policies improve them. 
In current ongoing work, \citet{GoncalvesNunnariZarate-PinaWP} not only corroborate the results in \hyref{theorem:exogenous-stopping:speed-accuracy}[Theorems], \hyref{theorem:optimal-stopping:speed-accuracy}, and \hyref{theorem:ability:speed-accuracy} --- already evidenced in the above-referenced literature --- but also provide support for the empirical relevance of \hyref{theorem:identification}[Theorem], demonstrating that subsidies can indeed be used to reveal problem complexity.

Our results also highlight the importance of recognising how complex a problem is, as well as one's own ability in tackling problems of a particular nature.
From consumer choice and labour supply decisions, to contract and regulation design, gaining a better understanding of problem complexity can help policy-makers and practitioners. 
Several questions emerge naturally. 
Which features make a problem of a particular class simpler or more complex? 
Can uncertainty about problem complexity lead individuals to exert too much or too little effort, be able to rationalise phenomena akin to over- and underconfidence? 
Are agents with different abilities able to learn to self-sort into tackling problems (activities or jobs) for which they hold a comparative advantage? 
We find these to be promising avenues for future research.

\FloatBarrier

\section{References}
{   
    \setstretch{1}
    \bibliographystyle{econ-aea}
    \bibliography{speed-accuracy-complexity.bib}
}

\FloatBarrier \newpage

\phantomsection
\addcontentsline{toc}{section}{Appendices}

\setcounter{section}{0}
\renewcommand{\thesection}{Appendix \Alph{section}}
\renewcommand{\thesubsection}{\Alph{section}.\arabic{subsection}}

\section{Proofs of the Main Results}
\label{appendix:proofs}

\subsection{Proof of \hyref{theorem:exogenous-stopping:speed-accuracy}[Theorem]}
\label{appendix:proofs:theorem:exogenous-stopping:speed-accuracy}

Assume \ref{assumption:continuation-region:belief-rate} holds. 
Then, \hyref{theorem:exogenous-stopping:speed-accuracy}[Theorem] is a special case of \hyref{proposition:exogenous:speed:shrinking:general}[Propositions] and \ref{proposition:exogenous:accuracy:shrinking:general}, with $\sigma(t,y,m)=m/2 (1-y^2)$, where $y=2p-1$ and $m=\kappa^{-1}$.
Indeed, there we consider a more general class of belief diffusions.
As shown in \hyref{appendix:general:belief-process}[Appendix], this can accommodate cases in which the decision-maker has richer uncertainty about $\theta$ and $\kappa$, including the setting in which $\theta \sim N(0,1)$ with a known $\kappa$, or where $\theta=\pm 1$ with equal probability and $\kappa^{-1}$ is distributed according to a folded standard normal.

Assume now that \hyref{assumption:continuation-region:symmetry} holds with $C_t^\kappa:=(\logit^{-1}(2\kappa^{-1}\underline x_t),\logit^{-1}(2\kappa^{-1}\overline x_t))$. 
Since $C_t^\kappa$ weakly decreases in $t$ and strictly decreases in $\kappa$ in the subset order, then 
\hyref{theorem:exogenous-stopping:speed-accuracy}[Theorem][(2)] follows immediately from \hyref{proposition:exogenous:accuracy:shrinking:general}[Propositions] and \hyref{proposition:exogenous:speed-accuracy:nested:general}, with $\sigma(t,y,m)=m/2 (1-y^2)$, where $y=2p-1$ and $m=\kappa^{-1}$.
When instead $C_t^\kappa:=(\logit^{-1}(\underline x_t),\logit^{-1}(\overline x_t))$, $C_t^\kappa$ still weakly decreases in $t$ in the subset order and is constant in $\kappa$, hence the claim follows from \hyref{proposition:exogenous:accuracy:shrinking:general}[Propositions].
It then only remains to show \hyref{theorem:exogenous-stopping:speed-accuracy}[Theorem][(1)].

We proceed to analyse the case when \hyref{assumption:continuation-region:symmetry}[][(a)] holds.
We prove the following:
\begin{proposition}
    \label{proposition:exogenous-stopping:symmetric-thresholds:speed}
    Fix $\theta \ne 0$ and let $X_t^{\kappa}:=\theta\kappa^{-1} t + B_t$, where $\kappa>0$ and $B_t$ is a standard Brownian motion independent of $\theta$.
    Let $\overline x_t:\mathbb R_+ \to \mathbb R_+$ be continuous but at finitely many points and $\hat\tau^{\kappa}:=\inf\{t>0|X_t^\kappa \notin (-\overline x_t,\overline x_t)\}$.
    Then, $\hat\tau^\kappa$ increases in the first-order stochastic dominance sense in $\kappa$.
    Furthermore, if $\hat\tau^\kappa\ne 0$ and finite almost surely, then $\hat\tau^\kappa$ strictly increases in the first-order stochastic dominance sense in $\kappa$.
\end{proposition}
\begin{proof}
    We first discretise time and show that the continuation probability at finitely many points in time increases in $\kappa>0$.
    Fix any $T>0$ and partition $\pi_n:0=t_0<t_1<\cdots<t_n=T$, and set $T^{\pi_n}:={({t_i})}_{i=1,...,n}$.
    Define the event $E_t^{\kappa}:=\{X^\kappa_t \in (-\overline x_t,\overline x_t)\}$ and $E^{\kappa,\pi_n}:=\cap_{t_i \in T^{\pi_n}}E_{t_i}^\kappa$. 
    Let $B^{\pi_n}:={(B_{t_i})}_{i=1,...,n}$, which is normally distributed with zero mean and covariance $\Sigma^{\pi_n}$ and denote its measure by $\gamma_{\Sigma^{\pi_n}}$, and let $C^{\pi_n}:=\Pi_{i=1}^n(-\overline x_{t_i},\overline x_{t_i})$, a $k$-cell symmetric about zero. 
    Then, $f^{\pi_n}(\kappa):=\mathbb P(E^{\kappa,\pi_n})=\mathbb P(B^{\pi_n}\in C^{\pi_n}-\theta T^{\pi_n}/\kappa)=\int_{\mathbb R^n}\mathbf{1}_{C^{\pi_n}}(x+\theta T^{\pi_n}/\kappa)\varphi_{\Sigma^{\pi_n}}(x) \diff x$, where $\varphi_{\Sigma^{\pi_n}}$ denotes the density of $B^{\pi_n}$ and $\mathbf{1}_A(\cdot):=1\{\cdot \in A\}$ the indicator function.
    As $C^{\pi_n}$ is a convex set, then $\mathbf{1}_{C^{\pi_n}}(x+\theta T^{\pi_n}/\kappa)$ is log-concave in $(x,1/\kappa)$ in the extended sense, over the extended real line $\mathbb R\cup\{-\infty,\infty\}$. 
    As the Gaussian density is strictly log-concave, then, by the Pr\'{e}kopa-Leindler inequality $f^{\pi_n}$ is log-concave in $1/\kappa$.
    An even log-concave function is single-peaked with a maximum at 0, and, therefore, $f^{\pi_n}$ is non-decreasing in $\kappa>0$.
    We continuously extend $f^{\pi_n}$ to $\mathbb R$.
    Since $B^{\pi_n}\overset{d}{=} -B^{\pi_n}$, then $f^{\pi_n}(\kappa)=f^{\pi_n}(-\kappa)$.
    As $f^{\pi_n}$ is non-decreasing in $\kappa>0$, we extend it to $\kappa=0$ by continuity.

    We now pass the result to the continuous-time limit.
    Fix a nested sequence of partitions $\pi_n:||\pi_n||\to 0$ and dense union in $[0,T]$; then, 
    $\{\hat \tau^\kappa >T\}
    =\{X_t^\kappa\in(-\overline x_t,\overline x_t) \forall t\le T\}
    =:
    E^{\kappa,T}
    =
    \bigcap_{n=1}^\infty E^{\kappa,\pi_n}
    $.
    Since the events $E^{\kappa,\pi_n}$ are decreasing in $n$, monotone convergence gives 
    $\lim_{N\to \infty}\mathbb P(\bigcap_{n=1}^N E^{\kappa,\pi_n})=\mathbb P(\bigcap_{n=1}^\infty E^{\kappa,\pi_n})=\mathbb P(E^{\kappa,T})$.
    Letting $f_T(\kappa):=\mathbb P(\hat \tau^\kappa >T)$, it is straightforward to see that, for any $T>0$, $f_T$ extended to $\mathbb R$ is also continuous, log-concave in $1/\kappa>0$, and non-increasing in $|\kappa|\geq 0$, as inequalities are preserved in the limit. 
    Consequently, $\hat \tau^{\kappa} \geq_{FOSD}\hat \tau^{\kappa'}$ for $\kappa>\kappa'>0$.

    Now suppose that $\hat \tau^\kappa\ne 0$ and is finite almost surely and that the FOSD relation is not strict, i.e., $\exists \kappa>\kappa'>0$ such that $\hat \tau^{\kappa} \overset{d}{=} \hat \tau^{\kappa'}$. 
    Then, for any $T>0$, $f_T(\kappa)=f_T(\kappa')$. 
    Take any $T>0$ such that $f_T(\kappa)>0$, which must exist since by assumption $\mathbb P(\hat \tau^\kappa=0)=0$.

    As $f_T$ is non-decreasing, for our fixed $T>0$, the probability $f_T(\kappa'') = \mathbb{P}(\hat \tau^{\kappa''} > T)$ is constant on the interval $[\kappa', \kappa]$. 
    By Girsanov's theorem, we can express the probability of the path remaining within the boundaries under the drifted measure by taking the expectation under the standard Wiener measure $\mathbb{W}$ (where $X_t^0 = B_t$ has zero drift):
    $$f_T(\kappa) = \exp\left(-\frac{1}{2} \kappa^{-2} \theta^2 T\right)\mathbb{E}^{\mathbb{W}}\left[ \mathbf{1}_{E^{0,T}} \exp\left(\kappa^{-1} \theta B_T \right) \right],$$
    where $E^{0,T} = \{B_t \in (-\overline x_t,\overline x_t), \forall t \le T\}$. 
    We have that the second term on the right-hand side, 
    $h(\kappa^{-1}):=\mathbb{E}^{\mathbb{W}}\left[ \mathbf{1}_{E^{0,T}} \exp(\kappa^{-1} \theta B_T) \right]$, is the moment-generating function of the random variable $\theta B_T$ restricted to the event $E^{0,T}$.
    Letting $Y:=\mathbf{1}_{E^{0,T}}\theta B_T$, as the paths on $E^0$ are strictly bounded at time $T$, $|Y|=\mathbf{1}_{E^{0,T}}|\theta B_T|\leq |\theta||\overline x_T|$ and $Y$ is bounded.
    Then, we expand $h(\kappa^{-1})=\mathbb E^{\mathbb W}\left[\sum_{j=0}^\infty\frac{\kappa^{-j}Y^j}{j!}\right]$.
    Since $\frac{|\kappa^{-j}Y^j|}{j!}\leq \frac{|\kappa|^{-j}M^j}{j!}$ and $\sum_{j=0}^\infty\frac{|\kappa|^{-j}M^j}{j!}=\exp(|\kappa|^{-1}M)$, we can swap the expectation and the summation and get $h(\kappa^{-1})=\sum_{j=0}^\infty\frac{\kappa^{-j}\mathbb E^{\mathbb W}[Y^j]}{j!}$. 
    Therefore, $h$ is a real analytic function of $\kappa^{-1}$ and so is $g_T(\kappa^{-1}):=\exp\left(-\frac{1}{2} \kappa^{-2} \theta^2 T\right)h(\kappa^{-1})$, and hence $f_T$ is real analytic on $\mathbb R_{++}$.

    Since $f_T$ is assumed to be constant on $[\kappa',\kappa]$, by the identity theorem for real analytic functions, if two analytic functions agree on an open interval of $\mathbb R_{++}$, they must agree everywhere on $\mathbb R_{++}$.
    Thus, $f_T$ is constant on $\mathbb R_{++}$ and equal to $f_T(\kappa)>0$.
    However, as $f_T$ is continuous on $\mathbb R$ and since 
    $f_T(\tilde\kappa)=\mathbb P(\hat\tau^{\tilde\kappa}>T)\leq \mathbb P(X^{\kappa}_T \in (-\overline x_T,\overline x_T))=\mathbb P(\theta T+\tilde\kappa B_T \in \tilde\kappa(-\overline x_T,\overline x_T)) \to 0$ as $\tilde\kappa\downarrow 0$, $f_T\equiv 0$, contradicting $f_T(\kappa)>0$.
\end{proof}

\begin{corollary}
    \label{corollary:exogenous-stopping:symmetric-thresholds:speed}
    Let $X_t^{\kappa}:=\theta\kappa^{-1} t + B_t$, where $\kappa>0$ and $B_t$ is a standard Brownian motion independent of $\theta$.
    Let $\theta \sim \nu$ where $\nu$ is a measure on $\mathbb R$ such that $\nu(\mathbb R_{++})\in (0,1)$ and define $p_t:=\mathbb P(\theta >0\mid \mathcal F^{X^\kappa})$ and $\pi^\kappa_t(x):=\mathbb P(\theta>0\mid X^\kappa_t=x)$.
    Suppose $\overline x:\mathbb R_+\to \mathbb R_+$ is continuous and define $\overline p_t:=\pi^\kappa_t(\overline x_t)$ and $\underline p_t^:=\pi^\kappa_t(-\overline x_t)$, $C_t^:=(\underline p_t,\overline p_t)$, and $\tau^\kappa:=\inf\{t>0|p_t\notin C_t\}$.
    Then $\tau^\kappa$ increases in $\kappa$ in the first-order stochastic dominance sense; strictly so if $\tau^\kappa\ne 0$ and $<\infty$ almost surely.
\end{corollary}
\begin{proof}
    Let $A(t,x,\kappa):=\int_{\mathbb R_{++}}\exp(\theta \kappa^{-1}x - (1/2)\theta^2\kappa^{-2}t)\diff \nu(\theta)$ and $B(t,x,\kappa):=\int_{\mathbb R_{-}}\exp(\theta \kappa^{-1}x - (1/2)\theta^2\kappa^{-2}t)\diff \nu(\theta)$.
    Then, 
    $\pi^\kappa_t(x)=\frac{A(t,x,\kappa)}{A(t,x,\kappa)+B(t,x,\kappa)}$.
    Since  
    $A'_x(t,x,\kappa)=\int_{\mathbb R_{++}}\theta\kappa^{-1}\exp(\theta \kappa^{-1}x - (1/2)\theta^2\kappa^{-2}t)\diff \nu(\theta)>0$ 
    and 
    $B'_x(t,x,\kappa)=\int_{\mathbb R_{-}}\theta\kappa^{-1}\exp(\theta \kappa^{-1}x - (1/2)\theta^2\kappa^{-2}t)\diff \nu(\theta)<0$, 
    we obtain  
    $\partial_x \pi^\kappa_t(x)=
    \frac{A'_x(t,x,\kappa)B(t,x,\kappa)-B'_x(t,x,\kappa)A(t,x,\kappa)}{(A(t,x,\kappa)+B(t,x,\kappa))^2}>0$, and $\pi^\kappa_t(x)$ admits an inverse.
    Consequently, $\tau^\kappa=\inf\{t>0|p_t\notin C_t\}=\{t>0|\pi^\kappa_t(X^\kappa_t)\notin (\pi_t^\kappa(-\underline x_t),\pi_t^\kappa(\overline x_t))\}=\{t>0|X^\kappa_t\notin (-\underline x_t,\overline x_t)\}$.

    Let $\kappa>\kappa'$. 
    From \hyref{proposition:exogenous-stopping:symmetric-thresholds:speed}[Proposition], we have that, conditional on $\theta$, $\mathbb P(\tau^{\kappa'}\leq T \mid \theta)=\mathbb E[\mathbf 1_{\tau^{\kappa}\leq T}\mid \theta]\leq \mathbb E[\mathbf 1_{\tau^{\kappa'}\leq T}\mid \theta]=\mathbb P(\tau^{\kappa'}\leq T \mid \theta)$, for any $\theta \in \mathbb R$ and $T\geq 0$.
    Then, 
    $\mathbb P(\tau^{\kappa'}\leq T)=\mathbb E[\mathbb E[\mathbf 1_{\tau^{\kappa}\leq T}\mid \theta]]\leq \mathbb E[\mathbb E[\mathbf 1_{\tau^{\kappa'}\leq T}\mid \theta]]=\mathbb P(\tau^{\kappa'}\leq T)$, for any $T\geq 0$.
    If $\tau^\kappa\ne 0$ a.s. and $\tau^\kappa<\infty$, then it must be non-zero for some subset of $\mathbb R$ with positive measure. 
    Strictness follows from \hyref{proposition:exogenous-stopping:symmetric-thresholds:speed}[Proposition].
\end{proof}

The \hyref{theorem:exogenous-stopping:speed-accuracy}[Theorem][(1)] under \hyref{assumption:continuation-region:symmetry}[][(a)] follows immediately by recognising that, in the setting described in the main text, as $\logit(p_0)=0$, $\pi^\kappa_t(x)=\logit^{-1}(2\kappa^{-1} x)$ and then $\logit^{-1}(2\kappa^{-1} \overline x_t)=\overline p_t=1-\underline p_t=1-\logit^{-1}(2\kappa^{-1} \underline x_t) \implies \overline x=-\underline x$. 
As per \hyref{corollary:exogenous-stopping:symmetric-thresholds:speed}[Corollary], the result extends to cases in which $p_0\ne 1/2$ under appropriate adjustments.
Moreover, the case in which $C_t^\kappa:=(\logit^{-1}(\underline x_t),\logit^{-1}(\overline x_t))$ follows as corollary.
To see this, recall $C_t^\kappa:=(\logit^{-1}(2\kappa^{-1}\underline x_t),\logit^{-1}(2\kappa^{-1}\overline x_t))$ and define $\tau^{\kappa,r}:=\inf\{t>0|p_t^\kappa \notin C_t^r\}$.
From the above, for any $s>1$ and $r>0$, 
$\mathbb E[\tau^{\kappa,r}]\leq \mathbb E[\tau^{s \kappa,s r}]$.
Then, 
$\mathbb E[\tau^{\kappa,1}]\leq \mathbb E[\tau^{s\kappa,s}]\leq \mathbb E[\tau^{s\kappa,1}]$, where the last inequality follows from \hyref{proposition:exogenous:speed-accuracy:nested:general}[Proposition] by observing that $C_t^\kappa$ decreases in the subset order in $\kappa$.

We conclude by examining the case when \hyref{assumption:continuation-region:symmetry}[][(b)] holds.
When $C_t^\kappa:=(\logit^{-1}(\underline x_t),\logit^{-1}(\overline x_t))=(\logit^{-1}(\underline x_0),\logit^{-1}(\overline x_0))$, we have that $\overline p'_t=\underline p'_t=0$, and \hyref{assumption:continuation-region:belief-rate} holds.
We focus then on the case in which $C_t^\kappa:=(\logit^{-1}(2\kappa^{-1}\underline x_t),\logit^{-1}(2\kappa^{-1}\overline x_t))=(\logit^{-1}(2\kappa^{-1}\underline x_0),\logit^{-1}(2\kappa^{-1}\overline x_0))$. 
Differently from the cases analysed above, the argument crucially hinges on $p_0=1/2$. 
We have then that $\tau^\kappa=\inf\{t>0|p_t\notin C_t\}=\inf\{t>0|X_t\notin (\underline x,\overline x)\}$ for some $\underline x, \overline x \in \mathbb R$. 
From the moment-generating function in \citet[p. 309, equation 3.0.1]{BorodinSalminen2002Book}, we obtain
$
    \mathbb E[\tau^\kappa]=
    \kappa (\overline x-\underline x)\frac{\sinh(\kappa^{-1}\overline x)\sinh(-\kappa^{-1}\underline x)}{\sinh(\kappa^{-1}(\overline x-\underline x))}
$.
Taking the logarithm and differentiating with respect to $\kappa$, we have
\begin{align*}
    &
    \frac{\partial}{\partial \kappa^{-1}}\ln(\mathbb E[\tau^\kappa])
    =
    -\kappa
    +\underline x \coth(\kappa^{-1}\underline x)
    +\overline x \coth(\kappa^{-1}\overline x)
    -(\overline x-\underline x) \coth(\kappa^{-1}(\overline x-\underline x))
    <0
    \\
    \iff &
    \kappa^{-1}(\overline x-\underline x) \coth(\kappa^{-1}(\overline x-\underline x))+1>
    -\kappa^{-1}\underline x \coth(-\kappa^{-1}\underline x)
    +\kappa^{-1}\overline x \coth(\kappa^{-1}\overline x).
\end{align*}
Letting $g(x):=x \coth(x)$, we note that $g$ is even and strictly convex and $g(0)=1$.
Since for any even and strictly convex function $g$ we have that $g(x-y)+g(0)>g(x)+g(y)=g(x)+g(-y)$ for any $x>0>y$, then, $\mathbb E[\tau^\kappa]$ increases in $\kappa$, strictly so whenever $\mathbb E[\tau^\kappa]>0$.

\subsubsection{Failure of \hyref{theorem:exogenous-stopping:speed-accuracy}[Theorem][(1)] Without \ref{assumption:continuation-region:belief-rate} and \ref{assumption:continuation-region:symmetry}}

We briefly note that shrinking continuation regions generally pose a major challenge due to a selection effect. 
Suppose that 
$(\underline p_t,\overline p_t)=(\underline p,\overline p_H)$ for $t<$ and $(\underline p_t,\overline p_t)=(\underline p,\overline p_L)$ for $t\in [1/10,1/2)$, and $(\underline p_t,\overline p_t)=\emptyset$ for $t>T_2$, with $1>\overline p_H>\overline p_L>p_0>\underline p > 0$, for arbitrarily small $\underline p$.
Recall log-odds posteriors evolve as $\logit(p_t)=2\kappa^{-2}\theta t+\kappa^{-1}B_t$. 
Then, conditional on $\theta=1$, a smaller $\kappa$ pushes the belief process faster toward the upper boundary. 
Consequently, for low enough $\kappa$, almost all paths will stop, conditional on $\theta =1$, whereas for higher $\kappa$ many do survive beyond $T_1$. 
This corresponds to the usual effect: more complex problem, slower stopping.
However, conditional on $\theta = -1$, a smaller $\kappa$ pushes the belief process away from the more binding upper boundary.
This creates a selection effect: by $T_1$, for low enough $\kappa$, almost all paths survive conditional on $\theta =-1$, whereas for higher $\kappa$ many do not. 

Balancing the difference between the tightening of the upper boundary, $\overline p_H-\overline p_L$, the time at which it tightens, $T_1$, and vanishes, $T_2$, as well as the drift rates, $\kappa>\kappa'$, it is possible to construct perverse cases in which $\mathbb E[\tau^\kappa]<\mathbb E[\tau^{\kappa'}]$.
For instance, when 
$\overline p_H:=\logit^{-1}(2)$, 
$\overline p_L:=\logit^{-1}(1/200)$, 
$\underline p$ is arbitrarily close to zero, 
$T_1:=1/10$, $T_2:=1/2$, $\kappa=10/7$, and $\kappa'=10/14$, the expected stopping time in the more complex problem $\kappa$ is smaller than the one in the simpler $\kappa'$.
Moreover, this perverse example can be adjusted for the case in which stopping thresholds are based not on posterior beliefs but rather on the evidence process $X_t^\kappa$.

\subsection{Proof of \hyref{proposition:optimal-stopping}[Proposition]}
\label{appendix:proofs:proposition:optimal-stopping}
For completeness, provide a proof relying on arguments somewhat different from the guess-and-verify approach in \citet[Theorem 21.1]{PeskirShiryaev2006Book}.

First, we show that the continuation region is given by $(\underline p,\overline p)\subset (0,1)$. 
Observe that the value function is convex. 
This is a property that holds generally and which we show in \hyref{lemma:convexity-prior-general}[Lemma].
Moreover, $V$ is also continuous. 
This is because, it is continuous on $(0,1)$ (owing to it being convex) and, any discontinuity at $p = 0 $ requires that $\lim_{\epsilon\downarrow 0}v(\epsilon)\leq \lim_{\epsilon\downarrow 0}V(\epsilon)<C\leq  V(0)=v(0)$, which contradicts continuity of $v$; a similar argument holds for discontinuities at $p=1$.
Then, $\{p\in [0,1] | V(p)>v(p)\}$ is open. 
As $v$ is linear on either side of $\tilde p$, $V-v$ is convex on either side of $\tilde p$, and as $V-v=0$ on $\{0,1\}$, then $V-v$ is quasi-concave.
Immediately, $V-v$ has convex upper contour sets and, therefore, 
$\exists \underline p,\overline p\in (0,1): V(p)>v(p)\iff p \in ( \underline p,\overline p)$, implying that the decision-maker stops as soon as their beliefs exit a constant continuation region: $\tau^*=\inf\{t>0|p_t\notin(\underline p,\overline p)\}$.

Then, by \citet[Proposition 3.1.]{OksendalReikvam1998SSR}, the value function $V$ is the unique viscosity solution of the free-boundary problem $0=-c +2 \kappa^{-2}\left(p(1-p)\right)^2V''(p)$ on $(\underline p, \overline p)$ and $0=V(p)-\max_\alpha u(\alpha,p)$ on $[0,1]\setminus (\underline p, \overline p)$.

Since $V''$ exists a.e. by Alexandrov's theorem, from \citet[p. 15]{CrandallIshiiLions1992BullAMS}, the Hamilton--Jacobi--Bellman equation will hold with equality almost everywhere on $(\underline p,\overline p)$. 
Furthermore, $0\leq V''\leq \kappa^{2} c \max\{(\underline p(1-\underline p))^{-2}, (\overline p(1-\overline p))^{-2}\}$, and so $V''$ is bounded whenever it exists, and, owing to $V$ being convex, $V'$ is bounded wherever it exists.
Since the HJB is given by $F(p,V,V',V'')=-c+2\kappa^{-2}(p(1-p))^2V''$ and $F$ is continuous in $p$ and $V''$ and invariant with respect to $V$ and $V'$, then,  
for any sequence $(p_n)\subset (\underline p,\overline p)$ such that $p_n \to p^*\in (\underline p,\overline p)$, taking further subsequences such that $(V'(p_n),V''(p_n))\to (V'_\infty,V''_\infty)$ converge, $(V_{\infty}',V_{\infty}'')$ belongs to both the super- and subjet of $V$ at $p^*$. 
This implies $V$ is $\mathcal C^2$ everywhere except possibly at $\{\underline p,\overline p\}$ \citep[see][Section 3]{CrandallIshiiLions1992BullAMS}.

We now show that $V$ is $\mathcal C^1$.
$V$ can only fail to be $\mathcal C^1$ at $\{\underline p,\overline p\}$.
For the purpose of contradiction, suppose it is not $\mathcal C^1$ at $\underline p$; the proof that $V$ is $\mathcal C^1$ at $\overline p$ is symmetric.
Since $V$ is convex, $V'(\underline p^+)>V'(\underline p^-)=v'(\underline p^-)$.
Choose $k\in(V'(\underline p^-),V'(\underline p^+))$ and any $M>0$, and set 
$\psi(p):=V(\underline p)+k(p-\underline p)+\frac M 2 (p-\underline p)^2$, which then touches $V$ from below at $\underline p$.
Since $V(\underline p)=v(\underline p)$, the viscosity supersolution property gives 
$0\leq \min\{0,\,-c+2\kappa^{-2}(\underline p(1-\underline p))^2 M\}$.
Choosing 
$0<M<\frac{c\kappa^2}{2(\underline p(1-\underline p))^2}$ 
makes the second term strictly negative, a contradiction.

Finally, we conclude our proof with a characterisation of the optimal stopping thresholds. 

$V$ is $\mathcal C^2$ on $(\underline p,\overline p)$ it must satisfy, for some $\beta_0,\beta_1\in \mathbb R$, 
\begin{align*}
    V(p)&=\frac{c}{2\kappa^{-2}}\frac{\exp(\logit(p))-1}{\exp(\logit(p))+1}\logit(p)+\frac{\exp(\logit(p))}{\exp(\logit(p))+1}\beta_1 + \beta_0.
\end{align*}
As $V$ is $\mathcal C^1$ on $[0,1]$, we obtain the boundary conditions $V=v$ and $V'=v'$ on $\{\underline p,\overline p\}$.
These yield
$V(\underline p)=V(\overline p)-\delta(\overline p-\tilde p)$ and $V'(\underline p)=V'(\overline p)-\delta$,
from whence we obtain the optimality conditions characterising the stopping thresholds: 
\begin{align*}
    \begin{pmatrix*}[c]
        2\kappa^{-2}\delta/c\\
        \tilde p
    \end{pmatrix*}
    =
    \begin{pmatrix*}[c]
        (A(\overline p)+B(\overline p))-(A(\underline p)+B(\underline p))\\
        [C(\overline p)-C(\underline p)]/[(A(\overline p)+B(\overline p))-(A(\underline p)+B(\underline p))]
    \end{pmatrix*},
\end{align*}
with $A(p):=\exp(\logit(p))-\exp(-\logit(p))$, $B(p):=2\logit(p)$, and $C(p):=\logit(p)+\exp(\logit(p))$. 
By uniqueness of the viscosity solution, the system has exactly one solution $(\underline p,\overline p)\subset(0,1)$.

From the first equation, we have that $\underline p<\overline p$.
Let then $d:=(\logit(\overline p)-\logit(\underline p))/2>0$ and $C:=\logit(\tilde p)-\logit(\underline p)$.
It can be shown that, for $C\leq 0$, the left-hand side of the second equation is always strictly smaller than the right-hand side, and for $C\geq 2 d$ it is always strictly larger.
We conclude that $ 0<\underline p<\tilde p<\overline p<1$.

\subsection{Proof of \hyref{theorem:optimal-stopping:speed-accuracy}[Theorem]}
\label{appendix:proofs:theorem:optimal-stopping:speed-accuracy}
We now turn to the proof of \hyref{theorem:optimal-stopping:speed-accuracy}[Theorem].
To avoid confusion with notation defined elsewhere, we first state the problem formally. 

Let $X_t^{\kappa}:=\theta/\kappa t + B_t$, where $\kappa>0$ and $B_t$ is a standard Brownian motion independent of $\theta$, and let $\alpha^\theta$ be such that $\alpha^\theta:=a$ if $\theta >0$ and $\alpha^\theta:=b$ if otherwise. 
Suppose $\mathbb P(\theta = 1)=\mathbb P(\theta =-1) = 1/2$ and define $p_t^\kappa:=\mathbb P(\theta >0 \mid \mathcal F_t^{X^\kappa})$ and the set $\mathbb T$ of stopping times adapted to the natural filtration induced by $X^\kappa$, $\mathcal F^{X^\kappa}$. 
Let $u:A \times\Theta \to \mathbb R$ such that $u(a,1)>u(b,1)$ and $u(a,-1)<u(b,-1)$, and $V(p_0;\kappa):=\sup_{\tau \in \mathbb T}\mathbb E_{p_0}[\max_{\alpha\in A}\mathbb E[u(\alpha,\theta) \mid \mathcal {F_\tau}^{X^\kappa}] - c\tau]$.
Define $\tau^\kappa :=\inf\{t>0|V(p_t^\kappa;\kappa)=\max_{\alpha\in A}\mathbb E[u(\alpha,\theta)\mid \mathcal F_t^{X^\kappa}]\} $ and 
$\alpha_{t}^\kappa:=a$ if $a\in \argmax_{\alpha\in A}\mathbb E[u(\alpha,\theta)\mid \mathcal F_t^{X^\kappa}]$ and $\alpha_{t}^\kappa:=b$ if otherwise.

We prove that (1) $\mathbb E[\tau^\kappa]$ is non-monotone and quasi-concave in $\kappa$; and (2) $\mathbb P(\alpha^\kappa_{\tau^\kappa}=\alpha^\theta)$ is decreasing in $\kappa$. 
We further show that (1) and (2) hold strict if $\tau^\kappa\ne 0$ almost surely.

We first address (2) using a coupling argument. 
While it follows as a corollary of \hyref{proposition:optimal:speed-accuracy:general}[Proposition], since this is a simple special case, we present a self-contained argument. 
Since $p_t^1$ solves $\diff p_t^1:=2 p_t^1(1-p_t^1) \diff \tilde B_t$, from the scaling property of the Brownian motion we have that $p_t^\kappa \overset{d}{=} p_{\kappa^{-2}t}^1$.
For $C\subset (0,1)$, let $\tilde \tau_{C}:=\inf\{t>0|p_t^1\notin C\}$.
Note that $\tilde \tau_{C}=\inf\{t>0|p_t^1\notin C\}=\inf\{\kappa^{-2}t>0|p_{\kappa^{-2}t}^1\notin C\}=\kappa^{-2}\inf\{t>0|p_{\kappa^{-2}t}^1\notin C\}\overset{d}{=}\kappa^{-2}\inf\{t>0|p_{t}^\kappa\notin C\}$.
Also note that, from \hyref{proposition:optimal-stopping}[Proposition], optimal stopping thresholds are constant in $t$, so $C_t^\kappa=C_0^\kappa=(\underline p^\kappa, \overline p^\kappa)$.
Consequently, $\tilde \tau_{C_0^\kappa}\overset{d}{=}\kappa^{-2}\tau^\kappa$.
The optimality conditions characterising $C_0^\kappa$ are   
\begin{align*}
    \begin{pmatrix*}[c]
        2\kappa^{-2}\delta/c\\
        \tilde p
    \end{pmatrix*}
    =
    \begin{pmatrix*}[c]
        (A(\overline p^\kappa)+B(\overline p^\kappa))-(A(\underline p^\kappa)+B(\underline p^\kappa))\\
        [C(\overline p^\kappa)-C(\underline p^\kappa)]/[(A(\overline p^\kappa)+B(\overline p^\kappa))-(A(\underline p^\kappa)+B(\underline p^\kappa))]
    \end{pmatrix*},
\end{align*}
with $A(p):=\exp(\logit(p))-\exp(-\logit(p))$, $B(p):=2\logit(p)$, and $C(p):=\logit(p)+\exp(\logit(p))$. 
Using the implicit function theorem to express $(\logit(\underline p^\kappa),\logit(\overline p^\kappa))$ as a function of $m:=2\kappa^{-2}\delta/c$, we obtain 
\begin{align*}
    \begin{pmatrix*}[c]
        \frac{\diff}{\diff m}\logit(\overline p^\kappa)\\
        \frac{\diff}{\diff m}\logit(\underline p^\kappa)
    \end{pmatrix*}
    =
    \Delta^{-1}
    \begin{pmatrix*}[c]
        \logit'(\underline p^\kappa)(\tilde p-\underline p^\kappa)\\
        \logit'(\overline p^\kappa)(\tilde p-\overline p^\kappa)
    \end{pmatrix*},
\end{align*}
where $\Delta:=\logit'(\underline p^\kappa)\logit'(\overline p^\kappa)(\overline p^\kappa-\underline p^\kappa)$.
Equivalently, 
\begin{align*}
    \begin{pmatrix*}[c]
        \frac{\diff}{\diff m}\overline p^\kappa\\
        \frac{\diff}{\diff m}\underline p^\kappa
    \end{pmatrix*}
    =
    \begin{pmatrix*}[c]
        \frac{\overline p^2(1-\overline p)^2(\tilde p-\underline p)}{\overline p-\underline p}\\
        \frac{\underline p^2(1-\underline p)^2(\tilde p-\overline p)}{\overline p-\underline p}
    \end{pmatrix*}.
\end{align*}
As, from \hyref{proposition:optimal-stopping}[Proposition], $\overline p^\kappa>\tilde p>\underline p^\kappa$, we conclude that $C_0^\kappa$ strictly decreases in $\kappa$ with respect to the subset order.

Fixing $\kappa>\kappa'\geq 0$, we have $\tilde \tau_{C_0^\kappa}\leq \tilde \tau_{C_0^{\kappa'}}$, which in turn implies 
$\tau^\kappa\overset{d}{=}\kappa^{2}\tilde \tau_{C_0^\kappa}\leq \kappa^{2}\tilde \tau_{C_0^{\kappa'}}=({\kappa/\kappa'})^{2}{\kappa'}^{2}\tilde \tau_{C_0^{\kappa'}}\overset{d}{=}({\kappa/\kappa'})^{2}\tau^{\kappa'}$. 
As $p_t^1$ is a bounded martingale, then $|p_{\tau}^1-1/2|$ is a submartingale for any almost surely finite stopping time $\tau$.
Since $\tau^\kappa\overset{d}{=}\kappa^{2}\tilde \tau_{C_0^\kappa}\leq \kappa^{2}\tilde \tau_{C_0^{\kappa'}}=({\kappa/\kappa'})^{2}{\kappa'}^{2}\tilde \tau_{C_0^{\kappa'}}\overset{d}{=}({\kappa/\kappa'})^{2}\tau^{\kappa'}$ and 
$p_t^\kappa \overset{d}{=} p_{\kappa^{-2}t}^1 = p_{{\kappa'}^{-2}{(\kappa/\kappa')}^{-2}t}^1 \overset{d}{=} p_{{\kappa/\kappa'}^{-2}t}^{\kappa'}$, 
we obtain that $p_{\tau^\kappa}^\kappa\overset{d}{=}p_{\tilde \tau_{C_0^\kappa}}^1$. 
Then, 
$
\mathbb E[|p_{\tau^\kappa}^\kappa-1/2|] 
=
\mathbb E[|p_{\tilde \tau_{C_0^\kappa}}^1-1/2|] 
\leq 
\mathbb E[|p_{\tilde \tau_{C_0^{\kappa'}}}^1-1/2|] 
=
\mathbb E[|p_{\tau^{\kappa'}}^{\kappa'}-1/2|]$.

We now turn to proving (1). 
Let $G(p_0):=\mathbb E_{p_0}[\tau^\kappa]$. 
As $p_t^\kappa$ solves $\diff p_t=\sigma^\kappa(p_t)\diff \tilde B_t$ with $\sigma^\kappa(p):=2 \kappa^{-1}p(1-p)$, denote the infinitesimal generator as $L^\kappa G(p):=(1/2)(\sigma^\kappa(p))^2 G''(p)$. 
Then, $G(p)=\mathbb E_p[\tau^\kappa\wedge t]+\mathbb E_p[G(p_{\tau^\kappa\wedge t}^\kappa)]=\mathbb E_p\left[\int_0^{\tau^\kappa\wedge t}1\diff s\right]+\mathbb E_p[G(p_{\tau^\kappa\wedge t}^\kappa)]$. 
From Dynkin's formula, $\mathbb E_p[G(p_{\tau^\kappa\wedge t}^\kappa)]=G(p)+\mathbb E_p\left[\int_0^{\tau^\kappa\wedge t}L^\kappa G(p_s^\kappa)\diff s\right]$.
Consequently, 
$G(p)=G(p)+\mathbb E_p\left[\int_0^{\tau^\kappa\wedge t}(1+L^\kappa G(p_s^\kappa))\diff s\right]$ and, hence, 
$G$ solves $2 \kappa^{-2}(p(1-p))^2 G''(p)=-1$ on $(\underline p^\kappa,\overline p^\kappa)$, with $G(\underline p)=G(\overline p)=0$.

Letting $D(p):=(2 p-1)\logit(p)$, 
we obtain, for $p_0 \in C^\kappa$, 
$$\mathbb E[\tau^\kappa]=\frac{\kappa^2}{2}\left[\frac{\overline p-p_0}{\overline p-\underline p}D(\underline p)+\frac{p_0-\underline p}{\overline p-\underline p}D(\overline p)-D(p_0)\right].$$

Letting $\logit(\overline p)=e+d$ and $\logit(\underline p)=e-d$ and substituting in the above expressions, 
\begin{align*}
    &\frac{p_0-\underline p}{\overline p-\underline p} = \frac{\cosh((e+d)/2)}{\cosh(\logit(p_0)/2)}\frac{\sinh((\logit(p_0)-(e-d))/2)}{\sinh(d)}=:S(d,e), \\
    &\frac{\overline p-p_0}{\overline p-\underline p} = \frac{\cosh((e-d)/2)}{\cosh(\logit(p_0)/2)}\frac{\sinh(((e+d)-\logit(p_0))/2)}{\sinh(d)}=1-S(d,e), \quad \text{ and }
    \\
    &D({\logit}^{-1}(\ell))=\ell \tanh(\ell/2).
\end{align*}

As, at an optimum, $C(\overline p)-\tilde p(A(\overline p)+B(\overline p))=C(\underline p)-\tilde p(A(\underline p)+B(\underline p))$ and $2\kappa^{-2}\delta/c=D'(\overline p)-D'(\underline p)$, substituting and rearranging, we obtain 
\begin{align*}
    0&=C(\overline p)-\tilde p (A(\overline p)+B(\overline p)) - C(\underline p)-\tilde p (A(\underline p)+B(\underline p))
    \\
    &= d (1-2 \tilde p)+ \sinh(d) \cosh(e)(1-2 \tilde p) + \sinh(d) \sinh(e)\\
    \iff & e=f(d;\tilde p):=\ln\left(
       \frac{
            d(2\tilde p-1)+\sqrt{
                d^2(2\tilde p-1)^2
                + 
                4 \tilde p (1-\tilde p) \sinh(d)^2
            }
        }{
            2\sinh(d)(1-\tilde p)
       } 
    \right),
\end{align*}
and 
\begin{align*}
    2\kappa^{-2}\delta/c=D'(\overline p)-D'(\underline p)=4 (d + \sinh(d)\cosh(e))=:m(d,e).
\end{align*}
We observe that $f$ satisfies $f(d;\tilde p)+d>\logit(\tilde p)>f(d;\tilde p)-d$ and that is strictly increasing in $\tilde p$ for any $d$, $\frac{\partial}{\partial d}f(d;\tilde p)\geq 0\iff \tilde p\leq 1/2$, and convex-concave if $\tilde p\leq 1/2$ and concave-convex if $\tilde p\geq 1/2$.
Furthermore, we have that $\lim_{d\downarrow 0}f(d;\tilde p)=\logit(\tilde p)$ and $\lim_{d\to \infty}f(d;\tilde p)=(1/2)\logit(\tilde p)$.

Then, letting $d^\kappa=(\logit(\overline p^\kappa)-\logit(\underline p^\kappa))/2$ and $e^\kappa=(\logit(\overline p^\kappa)+\logit(\underline p^\kappa))/2$, we have that 
\begin{align*}
    (c/\delta)\mathbb E[\tau^\kappa]&={(2\kappa^{-2}\delta/c)}^{-1}\left[\frac{p_0-\underline p^\kappa}{\overline p^\kappa-\underline p^\kappa}D(\overline p^\kappa)+\frac{\overline p^\kappa-p_0}{\overline p^\kappa-\underline p^\kappa}D(\underline p^\kappa)-D(p_0)\right]
    \\
    &=
    \frac{1}{m(d^\kappa,e^\kappa)}\left(
        S(d^\kappa,e^\kappa)D({\logit}^{-1}(e^\kappa+d^\kappa))
        +
        (1-S(d^\kappa,e^\kappa))D({\logit}^{-1}(e^\kappa-d^\kappa))
        -
        D(p_0)
    \right)
    \\
    &=
    \frac{
        d^\kappa(\cosh(d^\kappa)-\cosh(e^\kappa))
        +
        \tanh(\logit(p_0)/2)\left(
            d^\kappa \sinh(e^\kappa)
            +e^\kappa \sinh(d^\kappa)
            -\sinh(d^\kappa)\logit(p_0)
        \right)
    }{
        4 \sinh(d^\kappa)(d^\kappa + \sinh(d^\kappa)\cosh(e^\kappa))
    }
    \\
    &=:\tilde T(d^\kappa,e^\kappa),
\end{align*}
on $e^\kappa+d^\kappa=\logit(\overline p^\kappa)>\logit(p_0)>\logit(\underline p^\kappa)=e^\kappa-d^\kappa$, and zero elsewhere. 
Since $e^\kappa=f(d^\kappa)$, the optimal stopping time can be written in closed form as a function of $d^\kappa$ alone, with $T(d):=\tilde T(d,f(d;\tilde p))$ on $\mathcal D:=\{d>|\logit(p_0)-f(d;\tilde p)|\}$ and $T(d)=0$ outside $\mathcal D$.

We show that $T$ is non-monotone and equals zero at the boundary of $\mathcal D$ and as $d\to \infty$. 
Without loss of generality, take $\tilde p\geq 1/2$ and recall $f(d;\tilde p)$ is bounded, with $\logit(\tilde p)/2\leq f(d;\tilde p)\leq \logit(\tilde p)$. 
Since when $\tilde p\geq 1/2$, $f(d;\tilde p)$ is decreasing, $\logit(p_0)-f(d;\tilde p)$ is increasing in $d$ and bounded, hence $\exists d_*$ such that $d_*=\logit(p_0)-f(d_*)$.
Analogously, $-\logit(p_0)+f(d;\tilde p)$ is decreasing in $d$ and bounded and $\exists d^*$ such that $d^*=-\logit(p_0)+f(d^*;\tilde p)$. 
Let $\underline d:=\max\{-\logit(p_0)+f(d^*;\tilde p),\logit(p_0)-f(d_*;\tilde p)\}$ and note that $\underline d=\inf\mathcal D$.
As $T$ is continuous on $\mathcal D$, we obtain $\lim_{d\downarrow \underline d}T(d)=0<T(d)$ for any $d \in \mathcal D$. 
Finally, as for any $e$ such that $|e|\leq |\logit(\tilde p)|$ we have that $\lim_{d\to \infty}\tilde T(d,e)=0$, then we conclude that $\lim_{d\to \infty}T(d)=0$. 

Using the obtained expressions, one can then show it is possible to express $T$ as the ratio of a concave and a convex function, both strictly positive, $T(d)=\text{Num}(d)/\text{Den}(p)$, where $\text{Num}(d):=d$ and $\text{Den}(d):=d/T(d)$. 
A straightforward but tedious exercise reveals that $d/T(d)$ is strictly convex on $\mathcal D$ by using the closed-form expressions and examining the terms in $\frac{\diff^2}{(\diff d)^2}(d/T(d))$.\footnote{ 
    While we omit the algebra let us highlight that, making use of the closed-form expressions it is possible to show that 
    $(d/\tilde T(d,f(d;\tilde p)))''_{dd}$ is increasing in $\tilde p \geq p_0$ and decreasing for $\tilde p\leq p_0$. 
    Consequently, fixing $d\in \mathcal D$, $(d/\tilde T(d,f(d;\tilde p)))''_{dd}$ is minimised at $\tilde p=p_0$, at which point we can explicitly obtain $(d/\tilde T(d,f(d;\tilde p)))''_{dd}\geq (d/\tilde T(d,f(d;1/2)))''_{dd}>0$. 
}
We then invoke the following lemma:
\begin{lemma}
    \label{lemma:ratio-quasiconcave}
    Let $X$ be a convex set and $\text{Den},\text{Num}:X\to \mathbb R_+$ where $\text{Num}>0$.
    If $\text{Den}$ is (strictly) convex and $n$ is concave, then $g:=\text{Num}/\text{Den}$ is (resp. strictly) quasiconcave.
\end{lemma}
\begin{proof}
    Let $\psi_t(x):=\text{Num}(x)-t \text{Den}(x)$; $\psi_t(x)$ is strictly decreasing in $t$ and concave in $x$.
    Then, $\{x\in X| g(x)\geq t\geq 0\}=\{x\in X| \psi_t(x)\geq 0\}$, which, as $\psi_t(x)$ is concave in $x$ is a convex set.
    If $d$ is strictly convex, then $\psi_t(x)$ is strictly concave in $x$ and, 
    for $t=g(x)$ and $t'=g(x')$ with $t'>t$ and $\lambda \in (0,1)$,
    $\psi_t(\lambda x+(1-\lambda)x')>0 \iff g(\lambda x+(1-\lambda)x')>\min\{t,t'\}=\min\{g(x),g(x')\}=g(x)$.
\end{proof}
Using \hyref{lemma:ratio-quasiconcave}[Lemma] with $d/(d/T(d))$, we have that $T(d)$ is quasiconcave in $d$ and, as $d^\kappa$ is strictly decreasing in $\kappa$, then $T(d^\kappa)=\mathbb E[\tau^\kappa]$ is quasiconcave in $\kappa$.
Given that $T(d)$ is strictly quasiconcave on $\mathcal D$, then $\mathbb E[\tau^\kappa]$ is too on $\overline \kappa:=\sup\{\kappa>0|\mathbb P(\tau^\kappa>0)>0\}$.

\subsection{Proof of \hyref{theorem:ability:speed-accuracy}[Theorem]}
\label{appendix:proofs:theorem:ability:speed-accuracy}
We start by restating the problem. 
We omit the dependence on $\kappa$ and $\lambda$ so as to not overburden the reader.

Let $X_t:= \theta/\kappa \int_0^t \sqrt{e_s}\diff s + B_t$, where $\kappa>0$, $B_t$ is a standard Brownian motion, and $e_t$ be a non-negative process, predictable with respect to the filtration induced by $X$ and $e$, $\mathcal F^{X,e}_t$, and satisfy $\int_0^t \sqrt{e_s} \diff s<\infty$ almost surely for every finite $t$, denoting the class of all such processes $\mathbb E$.
Let $\alpha^\theta$ be such that $\alpha^\theta:=a$ if $\theta >0$ and $\alpha^\theta:=b$ if otherwise. 
Suppose $\mathbb P(\theta = 1)=\mathbb P(\theta =-1) = 1/2$ and define $p_t:=\mathbb P(\theta >0 \mid \mathcal F_t^{X,e})$ and the set $\mathbb T$ of stopping times adapted to $\mathcal F_t^{X,e}$.

Let $u:A \times\Theta \to \mathbb R$ such that $u(a,1)>u(b,1)$ and $u(a,-1)<u(b,-1)$ and $c:\mathbb R_+\to \mathbb R_+$ be twice differentiable and satisfy $c,c',c''>0$ and let $\tau$ and $e$ solve 
$$\sup_{\tilde \tau \in \mathbb T, \tilde e \in \mathcal E}\mathbb E\left[\max_{\alpha\in A}\mathbb E[u(\alpha,\theta)\mid \mathcal F_{\tilde \tau}^{X,\tilde e}]-\int_0^{\tilde \tau} c(\tilde e_t/\lambda)\diff t\right].$$

Define 
$\alpha_{t}:=a$ if $a\in \argmax_{\alpha\in A}\mathbb E[u(\alpha,\theta)\mid \mathcal F_t^{X,e}]$ and $\alpha_{t}^\kappa:=b$ if otherwise.

Then, (1) $\mathbb E[\tau]$ is non-monotone and quasi-concave in $\lambda$ and $\kappa$; (2) $\mathbb P(\alpha_\tau = \alpha^\theta)$ is increasing in $\lambda$ and decreasing in $\kappa$; and (3) $\mathbb E[\tau]$ is strictly quasi-supermodular in $(\lambda, \kappa)$ whenever $\mathbb E[\tau]>0$.

The problem is as in \citet{MoscariniSmith2001Ecta}, but without discounting. 

Fix an admissible effort policy $\tilde e$ and let $V(p_0;\tilde e):=\sup_{\tilde \tau \in \mathbb T}\mathbb E\left[\max_{\alpha\in A}\mathbb E[u(\alpha,\theta)\mid \mathcal F_{\tilde \tau}^{X,\tilde e}]-\int_0^{\tilde \tau} c(\tilde e_t/\lambda)\diff t\right].$ 
$V(\cdot;\tilde e)$ is well-defined, since it bounded above by $||u||_\infty$ and below by by $\max_{\alpha \in A}u(\alpha,p_0)$. 
Since, by \hyref{lemma:convexity-prior-general}[Lemma], $V(\cdot;\tilde e)$ is convex in $p_0$, and $\max_{\alpha\in A}u(\alpha,p_0)$ is convex and piecewise linear with a single kink at $\tilde p$, we have that $\exists \underline p^e,\overline p^e$ such that $(\underline p^{\tilde e},\overline p^{\tilde e})=\{p \in (0,1)\mid V(p;\tilde e)>\max_{\alpha\in A}u(\alpha,p)\}$.

Then, fixing the stopping thresholds implied by an optimal level of effort $e$, 
from \citet[Theorem 11.2.1]{Oksendal2013Book}, there is an optimal level of effort which is Markovian with respect to the belief process and $V$ on $(\underline p^e,\overline p^e)$ satisfies 
\begin{align*}
    0=\sup_{\hat e\geq 0}\left\{ -c(\hat e/\lambda) +\hat e 2\kappa^{-2}(p(1-p))^2V''(p) \right\}.
\end{align*}
The first-order conditions imply that, at an optimum level of effort, we must have 
$c'(e/\lambda)/\lambda=2\kappa^{-2}(p(1-p))^2V''(p)$ and, from the HJB,
$c(e/\lambda)=e 2\kappa^{-2}(p(1-p))^2V''(p)$.
Hence, $c(e/\lambda)-c'(e/\lambda) e/\lambda=0$.
As $c$ is strictly convex, 
$(c(e)-c'(e) e)'_e=-c''(e) e<0$ for any $e>0$.
At $e=0$, $c(0)>0$, whereas, given $c'''\geq 0$, 
$\lim_{e\to\infty} c(e)-c'(e) e=-\infty$. 
Hence, there is a unique $e^*>0$ such that $c(e^*)-c'(e^*) e^*=0$.

Consequently, any optimal level of effort $e_t$ that is Markovian with respect to the belief process is, while strictly positive, independent of the belief thresholds determining the continuation region and given by $e=\lambda e^*$, entailing a cost $c(e/\lambda)=c(e^*)$, which is independent of $\lambda$ and $\kappa$.
Then, there is a unique viscosity solution to the free-boundary problem in the proof of \hyref{proposition:optimal-stopping}[Proposition] in \hyref{appendix:proofs:proposition:optimal-stopping}[Appendix], replacing the constant $c$ with $c(e^*)$ and $\kappa^{-1}$ with $\kappa^{-1}\sqrt{\lambda e^*}$, and, as per \citet[Theorem 11.2.2]{Oksendal2013Book} and the above necessary conditions, it coincides with the value function $V$.

Claims (1) and (2) follow from the fact that, as per \hyref{theorem:optimal-stopping:speed-accuracy}[Theorem], $\mathbb P(\alpha_t=\alpha^\theta)$ is increasing in $\kappa^{-1}\sqrt{\lambda e^*}$ and $\mathbb E[\tau]$ is non-monotone and quasi-concave in $\kappa^{-1}\sqrt{\lambda e^*}$.

For (3), let $h: \kappa^{-1}\sqrt{\lambda e^*}\mapsto h(\kappa^{-1}\sqrt{\lambda e^*})=\mathbb E[\tau]$ and let $\underline z:=\inf\{z\geq 0| h(z)>0\}$.
Then, $h:\mathbb R_+\to \mathbb R_+$ is a quasi-concave function and strictly quasi-concave on $[\overline z, \infty)$.
Define $f(\kappa^{-1},\lambda):=\kappa^{-1}\sqrt{\lambda e^*}$ which is strictly increasing in $(\kappa^{-1},\lambda)$. 
Then, for any $w=(\kappa^{-1},\lambda),\tilde w=(\tilde \kappa^{-1},\tilde \lambda)$ such that $f(w),f(\tilde w)>\underline z$, $f(w\vee w')>\underline z$.

From \hyref{lemma:quasi-submodularity}[Lemma], $H:=h\circ f$ is strictly quasi-submodular in $(\kappa^{-1},\lambda)$ on the relevant domain. 
Since the map $\kappa\mapsto \kappa^{-1}$ reverses the order in the first coordinate, $g(\kappa,\lambda):=H(\kappa^{-1},\lambda)$ is strictly quasi-supermodular in $(\kappa,\lambda)$.

\subsection{Proof of \hyref{theorem:identification}[Theorem]}
\label{appendix:proofs:theorem:identification}
Since 
$\delta \tilde p =u(b,-1)-u(a,-1)+s$, $\partial_s = (1/\delta) \partial_{\tilde p}$, and $\frac{\partial^2}{\partial \kappa \partial s}\mathbb P(\alpha_{\tau^*}=b)=(1/\delta)\frac{\partial^2}{\partial \kappa \partial \tilde p}\mathbb P(\alpha_{\tau^*}=b)$.

Let $\gamma:=c\kappa^2/(2\delta)$, and write the stopping boundaries as $\underline p$ and $\overline p$. 
Since $p_t$ is a bounded martingale, optional stopping gives
$
    p_0
    =\mathbb E[p_{\tau^*}] 
    =q(\gamma,\tilde p)\underline p+(1-q(\gamma,\tilde p))\overline p.
$
Hence, 
$
    \mathbb P(\alpha_{\tau^*}=b)=q(\gamma,\tilde p)=\frac{\overline p-p_0}{\overline p-\underline p}.
$ 
Since $0<\underline p<\tilde p<\overline p<1$ and, since as $q(\gamma,\tilde p)\in (0,1)$, then $\underline p<p_0<\overline p$.

We show that $q_{\kappa\tilde p}''>0$. 
It is enough to prove $q_{\gamma\tilde p}''>0$, since $\gamma=c\kappa^2/(2\delta)$ implies $\partial_\kappa=c\kappa/\delta\partial_\gamma$.

Letting $F:=A+B$, the boundary equations can be written as
\begin{align*}
    \gamma(F(\overline p)-F(\underline p))=1\quad \text{ and }
    \tilde p=\overline p-\gamma\int_{\underline p}^{\overline p}(F(p)-F(\underline p))\diff p,
\end{align*}
where
$
    F'(p)=\frac{1}{p^2(1-p)^2}.
$
Implicit differentiation gives
\begin{align*}
    \underline p_{\tilde p}'
    &=\frac{1}{\gamma(\overline p-\underline p)F'(\underline p)},&
    \overline p_{\tilde p}'
    &=\frac{1}{\gamma(\overline p-\underline p)F'(\overline p)},&
    \underline p_\gamma'
    &=\frac{\overline p-\tilde p}{\gamma}\underline p_{\tilde p}', \quad\text{ and }&
    \overline p_\gamma'
    &=-\frac{\tilde p-\underline p}{\gamma}\overline p_{\tilde p}'.
\end{align*}
Thus $\underline p_{\tilde p}>0$, $\overline p_{\tilde p}>0$, $\underline p_\gamma>0$, and $\overline p_\gamma<0$.

Now set
$
    x:=1-2\underline p$, and $y:=2\overline p-1.
$
Then $x,y\in(0,1)$, and
$
    q(\gamma,\tilde p)=\frac{y}{x+y}.
$
Using the four derivative formulas above, together with $F'(\underline p)=16/(1-x^2)^2$ and $F'(\overline p)=16/(1-y^2)^2$, direct manipulation of the expression gives
\begin{align*}
    q_{\gamma\tilde p}''
    =
    -\frac{P_0(x,y)^2\{P_1(x,y) P_2(x,y)+2 P_3(x,y)\}}
    {4(1-x^2)^2(1-y^2)^2(x+y)^3},
\end{align*}
where
\begin{align*}
    P_1(x,y):=\log\frac{1+x}{1-x}+\log\frac{1+y}{1-y}>0\quad \text{ and }\quad  
    P_0(x,y):=P_1(x,y)(1-x^2)(1-y^2)+2(x+y)(1-xy)>0,
\end{align*}
and
\begin{align*}
P_2(x,y)
:=&x^6y^2+2x^5y^3-4x^4y^4-2x^4y^2+x^4
+2x^3y^5-6x^3y \\
&+x^2y^6-2x^2y^4+10x^2y^2-2x^2
-6xy^3+8xy+y^4-2y^2-2,\\
P_3(x,y)
:=&-x^6y-3x^5y^2+2x^4y^3+2x^4y
+2x^3y^4+2x^3y^2+2x^3 \\
&-3x^2y^5+2x^2y^3-4x^2y
-xy^6+2xy^4-4xy^2+2y^3.
\end{align*}
Since, for $x,y\in(0,1)$, $P_1(x,y)>2(x+y)$, then, if $P_2(x,y)\leq 0$, then 
$
    P_1(x,y) P_2(x,y)+2P_3(x,y)
    <2((x+y)P_2(x,y) +P_3(x,y)).
$
Thus it is enough to prove that 
$P_2(x,y)<0$ and 
$(x+y)P_2(x,y)+P_3(x,y)<0$.

Both inequalities are elementary polynomial inequalities on $(0,1)^2$. Let
\begin{align*}
    B_i^6(z):=\binom{6}{i}z^i(1-z)^{6-i}.
\end{align*}
Expanding $-P_2$ in the bivariate Bernstein basis gives
$
    -P_2(x,y)=\sum_{i,j=0}^6 c_{ij}B_i^6(x)B_j^6(y),
$
with coefficient matrix
\begin{align*}
(c_{ij})=
\begin{pmatrix}
2&2&32/15&12/5&41/15&3&3\\
2&16/9&76/45&107/60&92/45&43/18&8/3\\
32/15&76/45&4/3&7/6&277/225&68/45&28/15\\
12/5&107/60&7/6&7/10&37/75&3/5&4/5\\
41/15&92/45&277/225&37/75&8/225&0&0\\
3&43/18&68/45&3/5&0&0&0\\
3&8/3&28/15&4/5&0&0&0
\end{pmatrix}.
\end{align*}
All entries are nonnegative and at least one is strictly positive. Since $B_i^6(z)>0$ on $(0,1)$ for every $i$, we have $-P_2(x,y)>0$ on $(0,1)^2$, so $P_2(x,y)<0$.

Next write
$
    (x+y)P_2(x,y)+P_3(x,y)=(x+y)P_4(x,y),
$
where
\begin{align*}
P_4(x,y)
=&x^6y^2+2x^5y^3-x^5y-4x^4y^4-4x^4y^2+x^4
+2x^3y^5 +4x^3y^3-4x^3y\\
&+x^2y^6-4x^2y^4+10x^2y^2-xy^5-4xy^3+2xy+y^4-2.
\end{align*}
Expanding $-P_4$ in the same basis gives
$
    -P_4(x,y)=\sum_{i,j=0}^6 d_{ij}B_i^6(x)B_j^6(y),
$
with coefficient matrix
\begin{align*}
(d_{ij})=
\begin{pmatrix}
2&2&2&2&29/15&5/3&1\\
2&35/18&17/9&28/15&83/45&7/4&3/2\\
2&17/9&26/15&8/5&113/75&133/90&23/15\\
2&28/15&8/5&129/100&76/75&9/10&1\\
29/15&83/45&113/75&76/75&116/225&4/15&4/15\\
5/3&7/4&133/90&9/10&4/15&0&0\\
1&3/2&23/15&1&4/15&0&0
\end{pmatrix}.
\end{align*}
Again all entries are nonnegative and at least one is strictly positive, so $-P_4(x,y)>0$ on $(0,1)^2$. 
Since $x+y=2(\overline p-\underline p)>0$, this implies
$
    (x+y)P_2(x,y)+P_3(x,y)<0.
$
Therefore $P_1(x,y) P_2(x,y)+2P_3(x,y)<0$. 
We conclude that $q_{\gamma\tilde p}''>0$ and, therefore, $q_{\kappa\tilde p}''>0$. 
We further note that whilst our proof uses the fact that $p_0=1/2$, it is possible to show that a sufficient condition on the prior is that $(p_0-\tilde p)(p_0-1/2)\leq 0$ in addition to $p_0\in (\underline p,\overline p)$.

\FloatBarrier \newpage

\section{Exogenous and Optimal Stopping with General Diffusions}
\label{appendix:general}
When considering stopping times defined in terms of beliefs exiting a possibly time dependent interval, it is not essential that the underlying process is in fact a Brownian motion with symmetric drift, but only that they are consistent with it, as we indicated in \hyref{section:setup:setup}[Section]. 

In this appendix, we show that our results extend more broadly still to richer uncertainty about the drift that does not fit the bivariate drift case.
We consider a case with rich uncertainty in which beliefs follow a general class of driftless diffusion processes. 
This nests the baseline case of interest in the paper, in which $\diff p_t=\kappa^{-1}p_t(1-p_t)\diff \tilde B_t$.
It also encompasses cases in which the decision-maker faces richer uncertainty about $\theta$, going beyond binary realisations, as happens in \citet{FudenbergStrackStrzalecki2018AER}.
We prove several of the comparative statics used within this more general setting.

We begin by identifying our the setting of the problems we consider.
We then show how \hyref{theorem:exogenous-stopping:speed-accuracy}[Theorems] and \ref{theorem:optimal-stopping:speed-accuracy} extend to more general settings --- see, in particular, \hyref{proposition:exogenous:speed:shrinking:general}[Propositions], \hyref{proposition:exogenous:accuracy:shrinking:general}, and \hyref{proposition:optimal:speed-accuracy:general}. 
In particular, they apply to case in which $\theta$ is normally distributed, with a known mean and variance, instead of the binary support assumption in the main text.

\subsection{Setting and Assumptions}
\label{appendix:general:assumptions}

Let $D :=(\underline Y,\overline Y) \subseteq {\mathbb R}$ and denote its closure by $\overline D$.
Let $\sigma:\mathbb R_+\times \overline D \times \mathbb R_{+}\to \mathbb R_+$. 
We make the following assumptions:
\begin{enumerate}[label=(P\arabic*)]
    \item $\sigma(t, y, m)>0$ for all $t\geq 0$ and $(y, m)\in D\times \mathbb R_{++}$, and $\sigma(t, y, m)=0$ for all $t\geq 0$ and $(y, m)\notin D\times \mathbb R_{++}$;
    \label{assumption:sigma:positive}

    \item $\sigma$ is continuous in $(t, y, m)$;
    \label{assumption:sigma:continuous}
    
    \item $\sigma$ is locally Lipschitz continuous in $y$ on $D$;
    \label{assumption:sigma:Lipschitz-in-y}
    
    \item $\sigma(t, y, m')>\sigma(t, y, m)$ for all $m'>m\geq 0$, $t\geq 0$ and $y \in D$;
    \label{assumption:sigma:increasing-in-m}
    
    \item $\sigma(t', y, m)\leq \sigma(t, y, m)$ for all $t'>t\geq 0$, and all $m,y$.
    \label{assumption:sigma:decreasing-in-t}

    \item $D$ is bounded or $\sigma$ satisfies that, for any $m\geq 0$, $\int_0^\infty\sup_{y\in D}\sigma(t, y, m)^2 \diff t<\infty$ and $\sup_{t\geq 0,\,y\in D}\sigma(t,y,m)<\infty$.
    \label{assumption:sigma:bounded}

    \item For any $m>0$, $\exists K_m<\infty$ such that 
    $|\sigma(t,y,m)-\sigma(t,y',m)|\leq K_m|y-y'|\sqrt{1+|\ln(|y-y'|)|}$, for all $y,y'\in \overline D$ $y\ne y'$.
    \label{assumption:sigma:log-Lipschitz}
    
    \item 
    $\lim_{m\to 0}\sup_{t\geq 0,y\in D}\sigma(t, y, m)=0$ 
    and, for any compact $K\subset D$ and any $t>0$, 
    $\displaystyle\lim_{m\to \infty}\int_0^t\inf_{y \in K}\sigma(s, y, m)^2 \diff s=\infty$.
    \label{assumption:sigma:infinity}

    \item $\sigma(t, y, m)=\sigma(t, -y, m)$ for all $t,m\geq 0$;
    \label{assumption:sigma:even-in-y}
    
    \item $\sigma$ is quasiconcave in $y$, strictly so on $D$, for all $t\geq 0$ and $m>0$;
    \label{assumption:sigma:quasiconcave-in-y}

    \item $\sigma$ is locally Lipschitz continuous in $(t, y, m)$ on $\mathbb R_{++}\times D\times \mathbb R_{++}$;
    \label{assumption:sigma:jointly-Lipschitz}

    
\end{enumerate}

\begin{remark}
    Let $X_t = \theta m t + \tilde B_t$, where $\tilde B_t $ is a Brownian motion independent from $\theta$. 
    We consider the following canonical cases of general interest:
    \begin{enumerate}[label=(\roman*)]
        \item 
        Suppose $\mathbb P(\theta= 1)=1-\mathbb P(\theta= -1)=p_0$, and $Y_t = 2\mathbb P(\theta >0\mid \mathcal F_t^X)-1$. 
        Then $\diff Y_t = \sigma(t, Y_t, m)\diff B_t$ and $\overline Y=-\underline Y=1$, $\sigma(t,y,m)=\mathbf 1_{y \in D} m (1-y^2)^2$, and $\sigma$ satisfies all the above assumptions.
        
        \item
        Suppose that $\theta$ is normally distributed with mean $m_0$ and variance $v_0$, and $Y_t = 2\mathbb P(\theta >0\mid \mathcal F_t^X)-1$.
        Then, denoting with $\phi$ and $\Phi$ the probability and cumulative density functions, we have $\diff Y_t = \sigma(t, Y_t, m)\diff B_t$ and $\overline Y=-\underline Y=1$, and $\sigma(t, y, m)=\mathbf 1_{y \in D} ((v_0)^{-1}m^{-1}+t)^{-1/2}\phi(\Phi^{-1}((1+y)/2))$, where $\sigma$ again satisfies all the above assumptions.

        If instead $Y_t = \mathbb E[\theta \mid \mathcal F_t^X]$, then $D=\mathbb R$, $\sigma(t, y, m)=m/(v_0^{-1}+m^2 t)$, and $\sigma$ satisfies 
        all but \ref{assumption:sigma:quasiconcave-in-y} --- since it is constant in $y$ --- and \ref{assumption:sigma:increasing-in-m} and 
        \ref{assumption:sigma:infinity}, given that it is not monotone in $m$; indeed it non-monotone in $m$ and, while $\lim_{m\to 0}\sigma(t, y, m)=0$, we have that $\lim_{m\to \infty}\int_0^t\sigma(s, y, m)^2\diff s =v_0<\infty$.
        
        \item
        Suppose that $\theta$ is distributed according to a probability measure $\nu$ on the real line that assignes zero probability to $\{\theta=0\}$, and admits a probability density function that is even and log-concave, $\diff \nu(\tilde \theta)=\exp(-\psi(\tilde \theta))\diff \tilde \theta$ with $\psi\in \mathcal C^2_{\mathrm loc}(\mathbb R_{++})$ strictly convex on a set of positive $\nu$-measure, and $Y_t = 2\mathbb P(\theta >0\mid \mathcal F_t^X)-1$.
        Then, $\diff Y_t = \sigma(t, Y_t, m)\diff B_t$ and $\overline Y=-\underline Y=1$, where $\sigma$ immediately satisfies all the above assumptions but for \ref{assumption:sigma:log-Lipschitz} and \ref{assumption:sigma:infinity}; we prove this in \hyref{appendix:general:belief-process}[Appendix].
        The latter two can be ensured with a straightforward condition, e.g., $\psi''(\tilde\theta)>0$, $\forall \tilde\theta$.
    \end{enumerate}

    Many of our results dispense with some of the assumptions, allowing us to nest more structure. 
    For instance, if $\sigma$ is locally Lipschitz in $(y, m)$ and $M_t:=f(Y_t;M_0)$ with $f$ Lipschitz continuous, then one can write $\tilde \sigma(t,Y_t,M_0):=\sigma(t,Y_t,f(Y_t;M_0))$, allowing one to accomodate Markovian controls.
\end{remark}

\subsection{Comparative Statics with Exogenous Stopping}
\label{appendix:general:exogenous}

In this section, we provide comparative statics for exogenous stopping rules.

We first prove comparative statics for the expected stopping time with respect to the parameter governing the speed of diffusion.

\begin{proposition}
    \label{proposition:exogenous:speed:shrinking:general}
    For $m> 0$, let $Y_t^{m}$ solve 
    $\diff Y_t = \sigma(t, Y_t, m) \diff B_t$, $Y_0^m=Y_0\in D$, where 
    $B_t$ is a Brownian motion, and $\sigma$ satisfies 
    \ref{assumption:sigma:positive}, 
    \ref{assumption:sigma:continuous}, 
    \ref{assumption:sigma:Lipschitz-in-y}, 
    \ref{assumption:sigma:increasing-in-m}, and \ref{assumption:sigma:decreasing-in-t}.
    Let $\overline y,\underline y:\mathbb R_+\to \mathbb R$ be such that $(\underline y_t,\overline y_t)\subset D$ for all $t\geq 0$, $\overline y,\underline y\in \mathcal C^1$, $\underline y_t'\geq 0\geq \overline y_t'$, and $Y_0 \in (\underline y_0,\overline y_0)$.

    Define $C_t:=(\underline y_t,\overline y_t)$, $T^*:=\sup\{T>0:C_t\neq\emptyset\text{ for all }t<T\}$, and, for every $T<T^*$,
    $$
    \underline\rho_t^{m,T}
    :=
    \frac{2}{\overline y_t-\underline y_t}
    \int_{\underline y_t}^{\overline y_t}
    \frac{\overline y_t-y}{\sigma(T, y, m)^2}\diff y
    \quad \text{ and }\quad
    \overline\rho_t^{m,T}
    :=
    \frac{2}{\overline y_t-\underline y_t}
    \int_{\underline y_t}^{\overline y_t}
    \frac{y-\underline y_t}{\sigma(T, y, m)^2}\diff y.
    $$

    Let $\tau^{m,C}:=\inf\{t>0|Y_t^m\notin C_t\}$.
    If, for every $T<T^*$ and every $t<T$, $\underline y_t'\underline\rho_t^{m,T}\leq 1$ and $(-\overline y_t')\overline\rho_t^{m,T}\leq 1$, 
    then $\mathbb E[\tau^{m,C}]>\mathbb E[\tau^{m',C}]$ for all $m'>m$.
\end{proposition}
\begin{proof}
    Since $\underline y_t'\geq0$ and $\overline y_t'\leq0$, the continuation region is shrinking: $C_s\subseteq C_t$ whenever $s\geq t$.

    Fix $m'>m$ and fix $T<T^*$.
    For $i\in\{m,m'\}$, define
    \begin{align*}
        u^{i,T}(t,y) := \mathbb E_{t,y}\left[(\tau^{i,C}\wedge T)-t\right]
    \end{align*}
    on $D_T:=\{(t,y):0\le t<T,\, y\in C_t\}$. 
    Given \ref{assumption:sigma:positive}-\ref{assumption:sigma:Lipschitz-in-y}, by the Feynman--Kac representation for killed diffusions, $u^{i,T}$ solves, in the classical sense under smooth approximation and in the viscosity sense in general,
    \begin{align*}
        \partial_tu^{i,T}
        +
        \frac{1}{2}\sigma(t,y,i)^2\partial_{yy}u^{i,T}
        &=
        -1
        \qquad\text{in }D_T,
        \\
        u^{i,T}(t,\underline y_t)
        =
        u^{i,T}(t,\overline y_t)
        &=
        0,
        \qquad 0\le t<T,
        \\
        u^{i,T}(T,y)&=0,
        \qquad y\in C_T.
    \end{align*}
    
    See, for example, the Feynman--Kac formula in \citet[Section 8.2]{Oksendal2013Book}. 
    For simplicity, the proof uses smooth approximations for which the solutions are classical. 
    The general case follows by stability of one-dimensional stochastic differential equations and viscosity solutions under locally uniform convergence; see \citet[Chapter 5, Theorem 2.5]{Oksendal2013Book} and \citet[Section 6]{CrandallIshiiLions1992BullAMS}.

    We first prove that $u^{m',T}$ is concave in $y$. 
    The argument is written for smooth coefficients; the general case follows by the approximation just described. 
    Write $u:=u^{m',T}$ and let $a(t,y):=\frac{1}{2}\sigma(t,y,m')^2$.
    Thus $u$ solves $u_t+a(t,y)u_{yy}=-1$ in $D_T$.

    Fix $t<T$ and let $v^{t,T}$ solve the problem with a frozen time $t$:
    \begin{align*}
        \frac{1}{2}\sigma(T,y,m')^2\,(v^{t,T})''(y)&=-1,
        && y\in(\underline y_t,\overline y_t),
        \\
        v^{t,T}(y)&=0, && y\in \{\underline y_t,\overline y_t\}.
    \end{align*}
    Since $C_s\subseteq C_t$ for $s\geq t$, and since, from \ref{assumption:sigma:decreasing-in-t}, $\sigma(s,y,m')^2\geq \sigma(T,y,m')^2$ for every $s\in[t,T]$, 
    \begin{align*}
        \partial_sv^{t,T}(y)+\frac{1}{2}\sigma(s,y,m')^2\,(v^{t,T})''(y)
        =
        -\frac{\sigma(s,y,m')^2}{\sigma(T,y,m')^2}
        \leq -1.
    \end{align*}
    Moreover, since $C_s\subseteq C_t$ for $s\geq t$, $v^{t,T}$ is defined on the future moving domain. 
    The preceding inequality shows that $v^{t,T}$ is a supersolution for the parabolic problem solved by $u$ on this domain. 
    Since $v^{t,T}\geq 0$ on $(\underline y_t,\overline y_t)$, while $u$ vanishes on the lateral boundary and at the terminal time, we have $u-v^{t,T}\leq 0$ on the parabolic boundary. 
    By the parabolic maximum principle \citep[Section 7.1.4]{Evans2010Book},
    \begin{equation}
        \label{equation:shrinking-1}
        u(t,y)\leq v^{t,T}(y),
        \qquad y\in(\underline y_t,\overline y_t).
    \end{equation}

    Solving the frozen problem gives
    \begin{align*}
        (v^{t,T})'(\underline y_t)
        =
        \underline \rho^{m',T}_t,
        \qquad \text{and} \qquad 
        -(v^{t,T})'(\overline y_t)
        =\overline \rho^{m',T}_t.
    \end{align*}
    Since $\sigma(T,y,m')>\sigma(T,y,m)$ for $y\in D$, we have
    \begin{align*}
        \underline \rho_t^{m',T}\leq \underline \rho_t^{m,T},
        \qquad
        \overline \rho_t^{m',T}\leq \overline \rho_t^{m,T}.
    \end{align*}
    Hence the assumptions $\underline y_t'\underline\rho_t^{m,T}\leq 1$ and
    $(-\overline y_t')\overline\rho_t^{m,T}\leq 1$ imply
    \begin{align*}
        \underline y_t'\underline\rho_t^{m',T}\leq 1,
        \qquad
        (-\overline y_t')\overline\rho_t^{m',T}\leq 1.
    \end{align*}

    Since $u(t,\cdot)\leq v^{t,T}(\cdot)$ and both functions vanish at the endpoints, (\ref{equation:shrinking-1}) implies
    \begin{equation}
        u_y(t,\underline y_t)\leq (v^{t,T})'(\underline y_t) \quad\text{ and }
        -u_y(t,\overline y_t)\leq -(v^{t,T})'(\overline y_t).
        \label{equation:shrinking-2}
    \end{equation}

    Differentiating the boundary identities $u(t,\underline y_t)=u(t,\overline y_t)=0$ gives
    \begin{align*}
        u_t(t,\underline y_t)
        =
        -\underline y_t'u_y(t,\underline y_t),
        \qquad \text{and} \qquad
        u_t(t,\overline y_t)
        =
        -\overline y_t'u_y(t,\overline y_t).
    \end{align*}
    Using (\ref{equation:shrinking-2}), $\underline y_t'\geq 0$, $\overline y_t'\leq 0$, 
    and the inequalities $\underline y_t'\underline\rho_t^{m',T}\leq 1$ and
    $(-\overline y_t')\overline\rho_t^{m',T}\leq 1$, we obtain
    \begin{align*}
        u_t(t,\underline y_t)&=-\underline y_t'u_y(t,\underline y_t)
        \geq-\underline y_t'(v^{t,T})'(\underline y_t)
        =-\underline y_t'\underline \rho^{m',T}_t\geq -1,\\
        u_t(t,\overline y_t)&=-\overline y_t'u_y(t,\overline y_t)
        \geq-\overline y_t'(v^{t,T})'(\overline y_t)
        =\overline y_t' \overline \rho^{m',T}_t\geq -1.
    \end{align*}

    Set $z:=u_t+1$. 
    Since $u_t+a u_{yy}=-1$, we have $z=-a u_{yy}$. 
    Hence $z\geq 0$ is equivalent to $u_{yy}\leq 0$. 
    Differentiating $u_t+a u_{yy}=-1$ in $t$ gives
    \begin{equation}
        z_t+a z_{yy}-\frac{a_t}{a}z=0.
        \label{equation:shrinking-3}
    \end{equation}
    Since $\sigma$ is non-increasing in $t$, we have $a_t\leq 0$. 
    The zero-order coefficient in (\ref{equation:shrinking-3}) is therefore non-negative. 
    
    We now work toward applying the parabolic maximum principle in \citet[Theorem 12, Ch. 7.1.4]{Evans2010Book}.
    First, let $q:=-a_t/a\geq 0$, so that 
    $z_t+a z_{yy}+qz=0$.
    This equation is written in backward parabolic form. 
    To apply the forward maximum principle, write 
    $$\widetilde z(s,y):=z(T-s,y), \quad
    \widetilde a(s,y):=a(T-s,y), \quad\text{ and }\quad
    \widetilde q(s,y):=q(T-s,y).$$
    Then
    $
        \widetilde z_s-\widetilde a\,\widetilde z_{yy}
        -\widetilde q\,\widetilde z=0.
    $
    Let $\widetilde D_T:=\{(s,y) | 0<s\leq T, y \in C_{T-s}\}$.
    Choose $\lambda\geq \sup_{(s,y)\in \widetilde D_T} \widetilde q$.\footnote{
        In the smooth approximation, $q$ is continuous and, therefore, bounded on compact subcylinder of $D_T$. 
        We apply the maximum principle on such compact subcylinders and then pass to the limit.
    }
    Then, define
    $r(s,y):=e^{-\lambda s}\widetilde z(s,y)$. 
    Then
    $r_s-\widetilde a r_{yy}+(\lambda-\widetilde q)r=0$, 
    where $\lambda-\widetilde q\geq 0$. 
    If $r$ is strictly negative at some point in $(s_0,y_0) \in \widetilde D_T$, then by the strong maximum principle in \citet[Theorem 12, Ch. 7.1.4]{Evans2010Book}, $r(s_0,y)$ is constant in $y$, and therefore strictly negative, which contradicts the fact that $z=1+u_t\geq 0 \implies r\geq 0$ on the lateral boundary.
    At $T$, $z(T,y)=0$ in the sense that $u_t(T^-,y)=-1$ for $y\in C_T$, because $u(T-r,y)=r\mathbb P_{T-r,y}(\tau^{m',C}>T)+o(r)=r+o(r)$ as $r\downarrow 0$ whenever $y\in C_T$. 
    Hence, $r\geq 0$ on $\widetilde D_T$.
    Since the exponential change of variables preserves signs, $z\geq 0$ on 
    $D_T$.
    Thus
    $\partial_{yy}u^{m',T}\leq 0$ in $D_T$.
    
    We now compare $m$ and $m'$. 
    Recall that, by \ref{assumption:sigma:increasing-in-m}, $\sigma(t,y,m)<\sigma(t,y,m')$ for $y\in D$. 
    Using the concavity of $u^{m',T}$ in $y$, we obtain, in the viscosity sense, 
    \begin{align*}
        \partial_t u^{m',T}
        +
        \frac{1}{2}\sigma(t,y,m)^2 \partial_{yy}u^{m',T}
        =
        -1
        &+
        \frac{1}{2}\left(\sigma(t,y,m)^2-\sigma(t,y,m')^2\right)
        \partial_{yy}u^{m',T}
        \\
        &\geq -1=  \partial_t u^{m,T}
        +
        \frac{1}{2}\sigma(t,y,m)^2 \partial_{yy}u^{m,T}.
    \end{align*}
    Thus $u^{m',T}$ is a subsolution of the parabolic problem solved by $u^{m,T}$, 
    As $u^{m',T}$ and $u^{m,T}$ satisfies the same conditions on the lateral and terminal boundaries, by the comparison principle, 
    $
        u^{m',T}\leq u^{m,T}
    $
    on $D_T.$
    In particular, this means that, for every $T<T^*$,
    \begin{align*}
        \mathbb E[\tau^{m',C}\wedge T]
        =
        u^{m',T}(0,Y_0)
        \leq
        u^{m,T}(0,Y_0)
        =
        \mathbb E[\tau^{m,C}\wedge T].
    \end{align*}

    As $\tau^{i,C}\wedge T$ increase pointwise in $T$ as $T\to T^*$,
    by monotone convergence, 
    $u^{i,T}(t,y)\to u^i(t,y) = \mathbb E_{t,y}[\tau^{i,C}]$.
    Moreover, as each $u^{i,T}$ is concave in $y$, the pointwise limit $u^i$ is too.
    We therefore have that $u^{m'}(t,y)\leq u^{m}(t,y)$.

    It remains to prove strictness. 
    From the dynamic programming principle, $u^i$ on the interior of $D_{T^*}$ satisfies, in the viscosity sense, 
    \begin{align*}
        F^i(t,y,u^i):=
        \partial_t u^{i}(t,y)
        +
        \frac{1}{2}\sigma(t,y,i)^2 \partial_{yy}u^{i}(t,y)
        &=
        -1.
    \end{align*}
    From the above, $u^{m}(t,y)=\mathbb E_{t,y}\left[\tau^{m,C}-t\right]$ is  pointwise weakly increases in $m$.
    If $u^{m}=u^{m'}=u$, then they would be the same function solving both $-1=F^m(t,y,u)=F^{m'}(t,y,u)$, which is possible only if $u_{yy}=0$ in the viscosity sense.
    Because $u(t,\cdot)$ is concave, this means $u(t,\cdot)$ is linear on each
    continuation interval. 
    But $u$ vanishes at the two lateral boundaries and is strictly positive in the interior, a contradiction. 
    Hence there exists $(t_0,y_0)\in D_{T^*}$ such that
    $
        u^m(t_0,y_0)>u^{m'}(t_0,y_0).
    $
    By continuity, there is an open neighbourhood $O \subset D_{T^*}$ on
    which $u^m>u^{m'}$.

    Now fix any starting point $(t,y)\in D_{T^*}$. 
    Since the diffusion is locally nondegenerate in the interior and $D_{T^*}$ is connected, the process starting from $(t,y)$ has positive probability of reaching $O$ before exiting the continuation domain. 
    We conclude that $u^m(t,y)>u^{m'}(t,y)$.
    And, in particular, as $(0,Y_0)\in D_{T^*}$, then $\mathbb E[\tau^m]>\mathbb E[\tau^{m'}]$.
\end{proof}

Note that the expectation of the absolute value of the stopped process is the analogous counterpart to the notion of accuracy used in the main text. 
We now provide comparative statics for the expectation of the absolute value of the stopped process with respect to speed of diffusion.

\begin{proposition}
    \label{proposition:exogenous:accuracy:shrinking:general}
    For $m> 0$, let $Y_t^{m}$ solve 
    $\diff Y_t = \sigma(t, Y_t, m) \diff B_t$, $Y_0^m=Y_0\in D$, where 
    $B_t$ is a Brownian motion, and $\sigma$ satisfies \ref{assumption:sigma:positive}, \ref{assumption:sigma:continuous}, \ref{assumption:sigma:Lipschitz-in-y}, \ref{assumption:sigma:increasing-in-m}, and \ref{assumption:sigma:decreasing-in-t}.
    Let $\overline y,\underline y:\mathbb R_+\to \mathbb R$ be such that $\overline y_t \geq \underline y_t$ for all $t\geq 0$, and $Y_0 \in (\underline y_0,\overline y_0)$.
    
    Define $C_t:=(\underline y_t,\overline y_t)$ and $T^*:=\sup\{T>0:C_t\neq\emptyset\text{ for all }t<T\}$. 
    Let $\tau^{m,C}:=\inf\{t>0|Y_t^m\notin C_t\}$.

    If $(\underline y_{t'},\overline y_{t'})\subseteq (\underline y_t,\overline y_t)\subset D$ for all $t'> t\geq 0$,
    then $\mathbb E[|Y_{\tau^{m,C}}^{m}|]\leq \mathbb E[|Y_{\tau^{m',C}}^{m'}|]$ for all $m'>m\geq 0$.
    If, in addition, $\underline y,\overline y$ are continuous and there exist $0<t<t'<T^*$ such that $0\in C_s$ for all $s\leq t'$ and $C_{t'}\subset C_t$, then $
    \mathbb E\left[|Y_{\tau^{m,C}}^m|\right]
    <
    \mathbb E\left[|Y_{\tau^{m',C}}^{m'}|\right]$ for all $m'>m\geq 0$.
\end{proposition}
\begin{proof}
    Let $q$ be a Brownian motion in natural scale started from $Y_0$ and absorbed at $\{\underline Y,\overline Y\}$, and let $\tau_{\mathrm{abs}}:=\inf\{u>0:q_u\in\{\underline Y,\overline Y\}\}$. 
    For each $m>0$, define $f^m(t,y):=1/\sigma(t,y,m)^2$. 
    This is well-defined on $D$ by \ref{assumption:sigma:positive}.
    Let $\rho^m$ solve the random ordinary differential equation
    \begin{align*}
        \frac{\diff}{\diff u}\rho^m(u)
        =
        f^m(\rho^m(u),q_u),
        \qquad
        \rho^m(0)=0,
    \end{align*}
    on $[0,\tau_{\mathrm{abs}})$, and set $\rho^m(u)=\infty$ for $u\geq\tau_{\mathrm{abs}}$. 
    Let $A^m$ be the generalised inverse of $\rho^m$, and define $\widetilde Y_t^m:=q_{A_t^m}$. 
    Then $\widetilde Y^m$ has quadratic variation $A^m$ and satisfies $A_t^m=\int_0^t\sigma(s,\widetilde Y_s^m,m)^2\diff s$. 
    Therefore $\widetilde Y^m$ solves
    \begin{align*}
        \diff \widetilde Y_t^m
        =
        \sigma(t,\widetilde Y_t^m,m)\diff \widetilde B_t
    \end{align*}
    for some Brownian motion $\widetilde B$. 
    By \ref{assumption:sigma:continuous} and \ref{assumption:sigma:Lipschitz-in-y}, weak uniqueness holds, and hence $\widetilde Y^m\overset{d}{=}Y^m$.

    Define $\widetilde\tau^{m,C}:=\inf\{t>0:\widetilde Y_t^m\notin C_t\}$, and define the corresponding intrinsic stopping time $U^m:=A_{\widetilde\tau^{m,C}}^m$. 
    Then
    $
        U^m
        =
        \inf\{u>0:q_u\notin C_{\rho^m(u)}\},
    $
    and $\widetilde\tau^{m,C}=\rho^m(U^m)$, while $\widetilde Y_{\widetilde\tau^{m,C}}^m=q_{U^m}$.

    Now fix $m'>m$. 
    By \ref{assumption:sigma:increasing-in-m}, $\sigma(t,y,m')>\sigma(t,y,m)$ for all $y\in D$, and therefore $f^m(t,y)> f^{m'}(t,y)$. 
    By \ref{assumption:sigma:decreasing-in-t}, $f^{m'}(t,y)$ is non-decreasing in $t$. 
    Hence the comparison principle for the random ordinary differential equation gives
    $
        \rho^m(u) >\rho^{m'}(u)$, $\forall u\geq 0.
    $
    Since $C_t$ is weakly decreasing in $t$ with respect to the subset order, this implies
    $
        C_{\rho^m(u)}\subseteq C_{\rho^{m'}(u)}
        $, $\forall u\geq 0.
    $
    Therefore
    $
        \{q_u\notin C_{\rho^{m'}(u)}\}
        \subseteq
        \{q_u\notin C_{\rho^m(u)}\},
    $
    and consequently $U^m\leq U^{m'}$ almost surely.

    Since $q$ is a martingale and $|\cdot|$ is convex, $|q_u|$ is a submartingale. 
    Moreover, because $C_t\subseteq C_0$ for all $t\geq 0$, the stopped variables $q_{U^m}$ and $q_{U^{m'}}$ are bounded by $\max\{|\underline y_0|,|\overline y_0|\}$. 
    Hence, by optional sampling,
    $
        \mathbb E[|q_{U^m}|]
        \leq
        \mathbb E[|q_{U^{m'}}|].
    $
    Using $\widetilde Y_{\widetilde\tau^{m,C}}^m=q_{U^m}$, $\widetilde Y_{\widetilde\tau^{m',C}}^{m'}=q_{U^{m'}}$, and $\widetilde Y^i\overset{d}{=}Y^i$ for $i\in\{m,m'\}$, we obtain
    $
        \mathbb E\left[|Y_{\tau^{m,C}}^m|\right]
        \leq
        \mathbb E\left[|Y_{\tau^{m',C}}^{m'}|\right].
    $

    We now prove strictness. 
    Assume that $\underline y,\overline y$ are continuous and that there exist $0<t<t'<T^*$ such that $0\in C_s$ for all $s\leq t'$ and $C_{t'}\subset C_t$. 
    We first show that
    $
        \mathbb P(U^m<U^{m'})>0.
    $
    Since $f^m>f^{m'}$ on $D$ and $f^{m'}$ is non-decreasing in its first argument, the comparison for the random ordinary differential equations gives $\rho^m(u)>\rho^{m'}(u)$ for every $u\in(0,\tau_{\mathrm{abs}})$. 
    Fix a continuous deterministic path $\gamma$ with $\gamma(0)=Y_0$ and values in $D$. 
    Along this path, the corresponding solutions of the two ordinary differential equations still satisfy $\rho_\gamma^m(u)>\rho_\gamma^{m'}(u)$ for every $u>0$ for which both are finite.

    Since $C_{t'}\subset C_t$ and $\underline y,\overline y$ are continuous, there exist times $r<s$ with $0<r<s<t'$ and $C_s\subset C_r$, with $s-r$ arbitrarily small, and there exists an open interval $J\subset C_r\setminus C_s$. 
    Choose $u_0>0$ and a continuous path $\gamma$ such that $\rho_\gamma^{m'}(u_0)<r<s<\rho_\gamma^m(u_0)$, $\gamma_{u_0}\in J$, and $\gamma_u\in C_{\rho_\gamma^m(u)}$ for all $u<u_0$. 
    This is possible because the clock gap $\rho_\gamma^m(u)-\rho_\gamma^{m'}(u)$ is strictly positive for every $u>0$, while $s-r$ can be chosen arbitrarily small.

    By continuity of $\rho^m$, $\rho^{m'}$, and of the Brownian paths, the same strict inequalities hold for all Brownian paths sufficiently close to $\gamma$ on $[0,u_0]$. 
    Brownian motion has full support on continuous paths, so this event has strictly positive probability. 
    On this event,
    $
        q_u\in C_{\rho^m(u)}
    $ for all $u<u_0$,
    while
    $
        q_{u_0}\in C_{\rho^{m'}(u_0)}
        \setminus C_{\rho^m(u_0)}.
    $
    Therefore $U^m\leq u_0<U^{m'}$ on this event. 
    Hence
    $
        \mathbb P(U^m<U^{m'})>0.
    $

    Since $0\in C_s$ for all $s\leq t'$ and, on the event above, the later stopping rule remains active up to a calendar time no larger than $t'$, the later stopping rule does not stop the Brownian motion at $0$. 
    Thus, conditional on $\mathcal F_{U^m}$ and on a subset of $\{U^m<U^{m'}\}$ with strictly positive probability, the stopped Brownian motion $q_{U^{m'}}$ has positive probability of lying on either side of $0$.
    Let us denote such an event as $E^*$.

    Since $q$ is a bounded martingale and $U^m\leq U^{m'}$, optional sampling gives
    $
        q_{U^m}
        =
        \mathbb E\left[q_{U^{m'}}\mid\mathcal F_{U^m}\right].
    $
    Therefore, by Jensen's inequality,
    \begin{align*}
        |q_{U^m}|
        =
        \left|
        \mathbb E\left[q_{U^{m'}}\mid\mathcal F_{U^m}\right]
        \right|
        \leq
        \mathbb E\left[|q_{U^{m'}}|\mid\mathcal F_{U^m}\right],
    \end{align*}
    and, as on $E^*$, $\left|
    \mathbb E\left[q_{U^{m'}}\mid\mathcal F_{U^m}\right]
    \right|
    <
    \mathbb E\left[|q_{U^{m'}}|\mid\mathcal F_{U^m}\right]$,
    taking expectations gives
    $
        \mathbb E[|q_{U^m}|]
        <
        \mathbb E[|q_{U^{m'}}|].
    $
    Using $\widetilde Y_{\widetilde\tau^{m,C}}^m=q_{U^m}$, $\widetilde Y_{\widetilde\tau^{m',C}}^{m'}=q_{U^{m'}}$, and $\widetilde Y^i\overset{d}{=}Y^i$ for $i\in\{m,m'\}$, we obtain
    $
        \mathbb E\left[|Y_{\tau^{m,C}}^m|\right]
        <
        \mathbb E\left[|Y_{\tau^{m',C}}^{m'}|\right].
    $
\end{proof}

We consider how expected stopping time and the expected absolute value of the stopped process varies when the continuation region expands.

\begin{proposition}
    \label{proposition:exogenous:speed-accuracy:nested:general}
    For $m> 0$, let $Y_t^{m}$ solve 
    $\diff Y_t = \sigma(t, Y_t, m) \diff B_t$, $Y_0^m=Y_0\in D$, where 
    $B_t$ is a Brownian motion, and $\sigma$ satisfies \ref{assumption:sigma:positive}, \ref{assumption:sigma:continuous}, and \ref{assumption:sigma:Lipschitz-in-y}.
    For a convex- and open-valued correspondence $C:\mathbb R_+\rightrightarrows \mathbb R$, let $\tau^{m,C}:=\inf\{t>0|Y_t^m\notin C_t\}$.

    If $C,C':\mathbb R_+\rightrightarrows \mathbb R$ are convex- and open-valued correspondences such that $C_t\subseteq C_t'$ and $C'$ is uniformly bounded, then  
    $\tau^{m,C}\leq\tau^{m,C'}$ almost surely 
    and 
    $
    \mathbb E\left[|Y_{\tau^{m,C}}^m|\right]
    \leq
    \mathbb E\left[|Y_{\tau^{m,C'}}^{m}|\right]
    $. 
    If, in addition, $C$ and $C'$ are continuous and there is $0<T$ such that $0\in C_T\subset C_T'$, then $\mathbb P(\tau^{m,C}< \tau^{m,C'})>0$ and 
    $
    \mathbb E\left[|Y_{\tau^{m,C}}^m|\right]
    <
    \mathbb E\left[|Y_{\tau^{m,C'}}^{m}|\right]
    $.
\end{proposition}
\begin{proof}
    That $\tau^{m,C}\leq\tau^{m,C'}$ almost surely follows immediately from the definition of $\tau^{m,C}$.
    Then, $
    \mathbb E\left[|Y_{\tau^{m,C}}^m|\right]
    \leq
    \mathbb E\left[|Y_{\tau^{m,C'}}^{m}|\right]
    $ is immediate from Doob's optional stopping theorem, noting that $|Y_t^m|$ is a submartingale and $|Y_{\tau^{m,i}}^m|$ is bounded by $\sup_t\sup C_t,|\inf_t\inf C_t|<\infty$.

    Write $C_t=(\underline y_t,\overline y_t)$ and $C_t'=(\underline z_t,\overline z_t)$. 
    Since $C_t\subseteq C_t'$ for every $t$, we have $\underline z_t\leq \underline y_t<\overline y_t\leq \overline z_t$.

    Since $C_T\subset C_T'$, either $\underline z_T<\underline y_T$ or $\overline y_T<\overline z_T$. 
    Suppose, without loss of generality, that $\overline y_T<\overline z_T$. 
    Choose $z$ such that $\overline y_T<z<\overline z_T$. 
    By continuity of $C$ and $C'$, after choosing $z$ slightly away from the endpoints if necessary, there exists $\varepsilon>0$ such that
    $
        (z-\varepsilon,z+\varepsilon)\subset C_T',
    $ $
        (z-\varepsilon,z+\varepsilon)\cap C_T=\emptyset.
    $

    We now construct a continuous path $\gamma:[0,T]\to D$ such that $\gamma_0=Y_0$, $\gamma_s\in C_s'$ for all $s\in[0,T]$, and $\gamma_T=z$. 
    Since $C'$ is continuous and open interval-valued, define
    \begin{align*}
        \lambda_0:=
        \frac{Y_0-\underline z_0}{\overline z_0-\underline z_0},
        \qquad
        \lambda_T:=
        \frac{z-\underline z_T}{\overline z_T-\underline z_T}.
    \end{align*}
    Then $\lambda_0,\lambda_T\in(0,1)$. 
    Let $\lambda:[0,T]\to(0,1)$ be continuous with $\lambda(0)=\lambda_0$ and $\lambda(T)=\lambda_T$, and set
    $
        \gamma_s
        :=
        \underline z_s
        +
        \lambda(s)(\overline z_s-\underline z_s).
    $
    Then $\gamma_0=Y_0$, $\gamma_T=z$, and $\gamma_s\in C_s'$ for every $s\in[0,T]$. 
    Since $[0,T]$ is compact and $\gamma_s\in C_s'$ for every $s$, there exists $\eta>0$ such that every continuous path $\omega$ with $\omega_0=Y_0$ and $\sup_{0\leq  s\leq  T}|\omega_s-\gamma_s|<\eta$ satisfies $\omega_s\in C_s'$ for every $s\in[0,T]$. 
    Taking $\eta$ smaller if necessary, such a path also satisfies $\omega_T\notin C_T$.

    By \ref{assumption:sigma:positive}, \ref{assumption:sigma:continuous}, and \ref{assumption:sigma:Lipschitz-in-y}, the diffusion is non-degenerate on compact subsets of $D$ and has full support on continuous paths that remain in $D$. 
    Hence
    $
        \mathbb P\left(
        \sup_{0\leq  s\leq  T}|Y_s^m-\gamma_s|<\eta
        \right)>0.
    $
    On this event, $Y_s^m\in C_s'$ for all $s\leq  T$, while $Y_T^m\notin C_T$. 
    Therefore
    $
        \tau^{m,C}\leq  T<\tau^{m,C'}.
    $
    Hence
    $
        \mathbb P(\tau^{m,C}<\tau^{m,C'})>0.
    $
    The case $\underline z_T<\underline y_T$ is identical, choosing $z\in(\underline z_T,\underline y_T)$ instead.

    Since $0\in C_T\subset C_T'$, and $C'$ is continuous and open-valued, there exist $\delta>0$ and $\eta>0$ such that
    $
        (-\eta,\eta)\subset C_s'
    $ for all $s\in[T,T+\delta].
    $
    From the previous argument, there is an event $E$ with $\mathbb P(E)>0$ on which
    $
        \tau^{m,C}\leq T<\tau^{m,C'}.
    $
    Shrinking $E$ if necessary, we may assume that $Y_T^m$ lies in a compact subinterval of $C_T'\setminus C_T$. 
    On $E$, conditional on $\mathcal F_{\tau^{m,C}}$, the process remains inside $C'$ at least until time $T$ with positive probability. 
    Give that $0\in C_s'$ for $s\in[T,T+\delta]$ and $\sigma(s,y,m)>0$ in the interior by \ref{assumption:sigma:positive}, the diffusion has positive conditional probability, before leaving $C'$, of moving to the right of $0$ and also positive conditional probability of moving to the left of $0$. 
    Hence, on a subset $E^*\subset E$ with $\mathbb P(E^*)>0$,
    $
    \mathbb P\left(Y_{\tau^{m,C'}}^m>0\mid\mathcal F_{\tau^{m,C}}\right)>0
    $ and $
    \mathbb P\left(Y_{\tau^{m,C'}}^m<0\mid\mathcal F_{\tau^{m,C}}\right)>0.
    $ 
    Moreover, as $Y^m$ is a bounded martingale up to $\tau^{m,C'}$ and $\tau^{m,C}\leq \tau^{m,C'}$, optional sampling gives
    $
        Y_{\tau^{m,C}}^m
        =
        \mathbb E\left[Y_{\tau^{m,C'}}^m\mid\mathcal F_{\tau^{m,C}}\right].
    $
    Therefore, by Jensen's inequality,
    \begin{align*}
        |Y_{\tau^{m,C}}^m|
        =
        \left|
        \mathbb E\left[Y_{\tau^{m,C'}}^m\mid\mathcal F_{\tau^{m,C}}\right]
        \right|
        \leq
        \mathbb E\left[|Y_{\tau^{m,C'}}^m|\mid\mathcal F_{\tau^{m,C}}\right],
    \end{align*}
    and, as on $E^*$, $\left|
    \mathbb E\left[Y_{\tau^{m,C'}}^m\mid\mathcal F_{\tau^{m,C}}\right]
    \right|
    <
    \mathbb E\left[|Y_{\tau^{m,C'}}^m|\mid\mathcal F_{\tau^{m,C}}\right]$,
    taking expectations gives
    $
    \mathbb E[|Y_{\tau^{m,C}}^m|]
    <
    \mathbb E[|Y_{\tau^{m,C'}}^m|].
    $
\end{proof}

Finally, we conclude the section by showing how expected stopping time and the expected absolute value of the stopped process varies when the continuation region shifts upward or downward.

\begin{proposition}
    \label{proposition:exogenous:choice:shift:general}
    For $m> 0$, let $Y_t^{m}$ solve 
    $\diff Y_t = \sigma(t, Y_t, m) \diff B_t$, $Y_0^m=Y_0\in D$, where 
    $B_t$ is a Brownian motion, and $\sigma$ satisfies \ref{assumption:sigma:positive}, \ref{assumption:sigma:continuous}, and \ref{assumption:sigma:Lipschitz-in-y}. 
    Let $\overline y,\overline z,\underline y,\underline z:\mathbb R_+\to D$ be continuous and such that, for all $t\geq 0$, $\overline y_t\geq \underline y_t$ and $\overline z_t \geq \underline z_t$, define $C_t:=(\underline y_t,\overline y_t)$ and $C_t'=(\underline z_t, \overline z_t)$, and assume $Y_0\in C_0\cap C_0'$. 
    Let $\tau^{m,C}:=\inf\{t>0|Y_t^m\notin C_t\}$ and $\tau^{m,C'}:=\inf\{t>0|Y_t^m\notin C_t'\}$.

    If $\overline y_t\geq \overline z_t$ and $\underline y_t\geq \underline z_t$, 
    then 
    $\mathbb P(Y_{\tau^{m,C}}^m=\overline y_{\tau^{m,C}} \text{ and }\tau^{m,C}\leq T)\leq \mathbb P(Y_{\tau^{m,C'}}^m=\overline z_{\tau^{m,C'}} \text{ and }\tau^{m,C'}\leq T)$ 
    and 
    $\mathbb P(Y_{\tau^{m,C}}^m=\underline y_{\tau^{m,C}} \text{ and }\tau^{m,C}\leq T)\geq \mathbb P(Y_{\tau^{m,C'}}^m=\underline z_{\tau^{m,C'}} \text{ and }\tau^{m,C'}\leq T)$
    for all $T\geq 0$.
    If, in addition, for some $T^*>0$ we have $C_{T^*}\ne C_{T^*}'$ and $C_t,C_t'\ne \emptyset$ $\forall t\leq T^*$, then the above inequalities are strict for all $T>T^*$.
\end{proposition}
\begin{proof}
    Since $Y^m$ has continuous paths and $\underline y,\overline y,\underline z,\overline z$ are continuous, whenever $Y^m$ exits $C$ before time $T$, it exits through one of the two boundaries, and similarly for $C'$.

    First consider the event 
    $
    E^+_T:=\left\{
        Y_{\tau^{m,C}}^m=\overline y_{\tau^{m,C}}
        \text{ and }
        \tau^{m,C} \leq  T
    \right\}.
    $
    On $E^+_T$, let $\tau^{m,C}=t\leq T$. 
    Then $Y_s^m>\underline y_s\geq \underline z_s$ for all $s<t$, and $Y_t^m=\overline y_t\geq \overline z_t$. 
    Since $Y_0\in C_0'$, we have $Y_0^m<\overline z_0$. 
    Hence, by continuity, there exists $s\leq t$ such that $Y_s^m=\overline z_s$ and, consequently, on $E^+_T$, $\tau^{m,C'}\leq t\leq T$.
    We conclude that 
    \begin{align*}
        &
        \{Y_{\tau^{m,C}}^m=\overline y_{\tau^{m,C}}\text{ and }\tau^{m,C} \leq T\}
        \subseteq 
        \{Y_{\tau^{m,C'}}^m=\overline z_{\tau^{m,C'}}\text{ and }\tau^{m,C'} \leq T\}
        \\
        \implies & 
        \mathbb P(Y_{\tau^{m,C}}^m=\overline y_{\tau^{m,C}} \text{ and }\tau^{m,C}\leq T)
        \leq 
        \mathbb P(Y_{\tau^{m,C'}}^m=\overline z_{\tau^{m,C'}} \text{ and }\tau^{m,C'}\leq T).
    \end{align*}
    Via a symmetric argument,
    $
        \mathbb P(Y_{\tau^{m,C}}^m=\underline y_{\tau^{m,C}} \text{ and }\tau^{m,C}\leq T)
        \geq 
        \mathbb P(Y_{\tau^{m,C'}}^m=\underline z_{\tau^{m,C'}} \text{ and }\tau^{m,C'}\leq T).
    $

    We now prove strictness. 
    Suppose that $C_{T^*}\ne C_{T^*}'$ and that $C_t,C_t'\ne\emptyset$ for all $t\leq T^*$. 
    Since $C_t$ strong-set dominates $C_t'$, either $\overline y_{T^*}>\overline z_{T^*}$ or $\underline y_{T^*}>\underline z_{T^*}$.

    Assume first that $\overline y_{T^*}>\overline z_{T^*}$. 
    By continuity, there exist $0<r_1<r_2<T^*$ and an open interval $J$ such that, for every $s$ in a neighbourhood of $r_1$,
    $
        J\subset C_s\setminus C_s'
    $
    and every point of $J$ lies above $\overline z_s$. 
    Since $\sigma(t,y,m)>0$ in the interior by \ref{assumption:sigma:positive}, and since $\sigma$ is continuous and locally Lipschitz in $y$ by \ref{assumption:sigma:continuous} and \ref{assumption:sigma:Lipschitz-in-y}, the diffusion has full support on continuous paths that remain in $D$. 
    Hence there is positive probability that $Y^m$ stays in $C'$ up to time $r_1$, hits $J$ at time $r_1$, and then remains below $\overline y_s$ up to time $T^*$. 
    On this event, $Y^m$ exits $C'$ through the upper boundary before $T^*$, but does not exit $C$ through the upper boundary before $T^*$. 
    Therefore, for every $T>T^*$,
    $
        \mathbb P(Y_{\tau^{m,C}}^m=\overline y_{\tau^{m,C}} \text{ and }\tau^{m,C}\leq T)
        <
        \mathbb P(Y_{\tau^{m,C'}}^m=\overline z_{\tau^{m,C'}} \text{ and }\tau^{m,C'}\leq T).
    $

    To get strictness for the lower-boundary inequality in the same case, choose a continuous path which stays in $C$ up to time $r_1$, hits the open upper shell $C_{r_1}\setminus C_{r_1}'$ at time $r_1$, and then hits the lower boundary $\underline y_s$ of $C$ before time $T^*$. 
    By the same full-support argument, this event has strictly positive probability. 
    On this event, $Y^m$ exits $C'$ through the upper boundary before it can exit $C'$ through the lower boundary, while $Y^m$ exits $C$ through the lower boundary before $T^*$. 
    Hence, for every $T>T^*$,
    $
        \mathbb P(Y_{\tau^{m,C}}^m=\underline y_{\tau^{m,C}} \text{ and }\tau^{m,C}\leq T)
        >
        \mathbb P(Y_{\tau^{m,C'}}^m=\underline z_{\tau^{m,C'}} \text{ and }\tau^{m,C'}\leq T).
    $
    The case $\underline y_{T^*}>\underline z_{T^*}$ is symmetric. 
\end{proof}

\subsection{Comparative Statics with Optimal Stopping}
\label{appendix:general:optimal}

We now consider optimal stopping.

For $m> 0$, let $Y_t^{m}$ solve 
$\diff Y_t = \sigma(t, Y_t, m) \diff B_t$, $Y_0^m=Y_0\in D$, where 
$B_t$ is a Brownian motion, and $\sigma$ satisfies \ref{assumption:sigma:positive}, \ref{assumption:sigma:continuous}, \ref{assumption:sigma:Lipschitz-in-y}, \ref{assumption:sigma:increasing-in-m}, \ref{assumption:sigma:decreasing-in-t}, \ref{assumption:sigma:bounded}, and \ref{assumption:sigma:log-Lipschitz}. 
Denote $Y_s^{t,y,m}$ as the solution to $\diff Y_s = \sigma(t+s, Y_s, m) \diff B_s$, $Y_0^{t,y,m}=Y_0=y\in D$.

Let $v:\bar D \to \mathbb R_+$ be a convex function which is $\mathcal C^2$ everywhere but at $\tilde y \in D$, at which point it is continuous but with a convex kink.
Specifically, we consider the case $v(y):=\max\{a_0 + a_1 y, b_0 + b_1 y\}$, where $a_1>0>b_1$ and $\tilde y := (b_0-a_0)/(a_1-b_1) \in D$.\footnote{
    For $v\geq 0$, it suffices that $\min_y v(y)=v(\tilde y)\geq 0\iff -a_0b_1+a_1 b_0\geq 0$. 
} 

Let $\mathbb T^{t,y,m}$ denote the set of stopping times taking values in $\mathbb R_+$ and adapted to $\mathcal F_s^{Y^{t,y,m}}$ and let 
$V^m(t,y):=\sup_{\tau \in \mathbb T^{t,y,m}}
    \mathbb E\left[
        \exp(-r (t+\tau)) v(Y_\tau^{t,y,m}) - \int_t^{t+\tau}\exp(-r s) c \diff s
    \right].
$
For notational convenience, define $\widehat V^m(t,y):=\exp(r t)V^m(t,y)$. 
Throught, we assume that $c,r\geq 0$ and that $c>0$, or $r>0$ and $v>0$.
Define $\tau^{t,y,m}:=\inf\{s>0| V^m(s+t,Y_{s}^{t,y,m})=\exp(-rt)v(Y_{s}^{t,y,m})\}$ and $\tau^{m}:=\tau^{0,y,m}$.
It is well-known that $\tau^{t,y,m}$ corresponds to the earliest optimal stopping time for the stopping problem above.

We begin by proving a number of properties satisfied by the value function.
\begin{proposition}
    \label{proposition:optimal:properties-value-function:general}
    The value function $V^m$ satisfies the following properties:
    \begin{enumerate}[label=(\arabic*)]
        \item $V^m$ weakly increases in $m$, strictly so on $\{(t,y)|V^m(t,y)>\exp(-rt)v(y)\}$;
        \label{proposition:optimal:properties-value-function:general:increasing-in-m}
        \item $V^m$ weakly decreases in $t$, strictly so on $\{(t,y)|V^m(t,y)>\exp(-rt)v(y)\}$ if $\sigma$ is strictly decreasing in $t$ for any $y\in D$ and $m>0$;
        \label{proposition:optimal:properties-value-function:general:decreasing-in-t}
        \item $V^m$ is continuous in $(t,y)$, $\mathcal C^1$ in $y$, and convex in $y$;
        \label{proposition:optimal:properties-value-function:general:continuity-convexity}
        \item if $\sigma$ is constant in $t$, then $V^m(t,y)=\exp(-rt)V^m(0,y)$ for any $(t, y, m)$; 
        \label{proposition:optimal:properties-value-function:general:constant-in-t}
        and
        \item if \ref{assumption:sigma:even-in-y} holds and $v$ is even, then $V^m$ is even in $y$.
        \label{proposition:optimal:properties-value-function:general:even}
    \end{enumerate}
\end{proposition}

\begin{proof}[Proof of Proposition \ref*{proposition:optimal:properties-value-function:general}\ref*{proposition:optimal:properties-value-function:general:increasing-in-m}]
    We use a coupling similar to that in the proof of \hyref{proposition:exogenous:speed-accuracy:nested:general}[Proposition].
    Fix $(t,y)\in \mathbb R_+\times D$ and fix $m'>m$. 
    Let $q$ be a Brownian motion in natural scale started from $y$ and absorbed at the boundary of $D$. 
    For $i\in\{m,m'\}$, define
    $
        f^i(s,z):=1/\sigma(s,z,i)^2.
    $
    This is well-defined on $D$ by \ref{assumption:sigma:positive}. 
    Let $\rho^i$ solve the random ordinary differential equation
    $
        \frac{\diff}{\diff u}\rho^i(u)
        =
        f^i(t+\rho^i(u),q_u),
        $ $
        \rho^i(0)=0,
    $
    up to absorption, and set $\rho^i(u)=\infty$ after absorption. 
    Let $A^i$ be the generalised inverse of $\rho^i$, and define
    $
        \widetilde Y_s^i:=q_{A_s^i}.
    $
    Then $\widetilde Y^i$ has quadratic variation $A^i$ and satisfies
    $
        A_s^i
        =
        \int_0^s\sigma(t+r,\widetilde Y_r^i,i)^2\diff r.
    $
    Therefore $\widetilde Y^i$ solves
    $
        \diff \widetilde Y_s^i
        =
        \sigma(t+s,\widetilde Y_s^i,i)\diff \widetilde B_s
    $
    for some Brownian motion $\widetilde B$. 
    By \ref{assumption:sigma:continuous} and \ref{assumption:sigma:Lipschitz-in-y}, weak uniqueness holds, and hence $\widetilde Y^i\overset{d}{=}Y^{t,y,i}$.

    By \ref{assumption:sigma:increasing-in-m}, $\sigma(s,z,m')>\sigma(s,z,m)$ for $z\in D$, and therefore
    $
        f^m(s,z)>f^{m'}(s,z).
    $
    By \ref{assumption:sigma:decreasing-in-t}, $s\mapsto f^{m'}(s,z)$ is non-decreasing. 
    Hence, the comparison principle for the random ordinary differential equations gives
    $
        \rho^m(u)\geq \rho^{m'}(u)
    $
    for all $u\geq 0$, with strict inequality for $u>0$ before absorption.

    Now let $\tau$ be any stopping time for $\widetilde Y^m$, and set
    $
        U:=A_\tau^m.
    $
    By the standard time-change filtration argument, $U$ is a stopping time for the filtration of $q$. 
    Define
    $
        \tau':=\rho^{m'}(U).
    $
    Then $\tau'$ is a stopping time for $\widetilde Y^{m'}$, and we obtain
    $
        \tau'
        =
        \rho^{m'}(U)
        \leq 
        \rho^m(U)
        =
        \tau,
        $ and $
        \widetilde Y_{\tau'}^{m'}=q_U=\widetilde Y_\tau^m.
    $
    Since $v\geq 0$, $r\geq 0$, and $c\geq 0$, 
    $
        \exp(-r(t+s))v(q_U)
        -
        \int_t^{t+s} \exp(-r \ell)c \diff \ell
    $ 
    is weakly decreasing in $s$. 
    Therefore, pathwise,
    \begin{align*}
        \exp(-r(t+\tau'))v(\widetilde Y_{\tau'}^{m'})
        -
        \int_t^{t+\tau'}\exp(-r s)c\diff s
        \geq 
        \exp(-r(t+\tau))v(\widetilde Y_\tau^m)
        -
        \int_t^{t+\tau}\exp(-r s)c\diff s.
    \end{align*}
    Taking expectations gives
    \begin{align*}
        &\mathbb E\left[
        \exp(-r(t+\tau'))v(\widetilde Y_{\tau'}^{m'})
        -
        \int_t^{t+\tau'}\exp(-rs)c\diff s
        \right]
        \\
        &\geq 
        \mathbb E\left[
        \exp(-r(t+\tau))v(\widetilde Y_\tau^m)
        -
        \int_t^{t+\tau}\exp(-rs)c\diff s
        \right].
    \end{align*}
    Taking the supremum over $\tau$ and using $\widetilde Y^i\overset{d}{=}Y^{t,y,i}$ gives
    $
        V^{m'}(t,y)\geq V^m(t,y).
    $

    We now prove strictness on the continuation region for $m$. 
    Suppose $V^m(t,y)>\exp(-rt)v(y)$, and let $\tau^m:=\tau^{t,y,m}$ be the earliest optimal stopping time for the problem with $m$. 
    Since $(t,y)$ is in the continuation region, $\tau^m>0$ with positive probability. 
    Construct $U:=A_{\tau^m}^m$ and $\tau':=\rho^{m'}(U)$ as above. 
    Then $\tau'$ is admissible for the problem with $m'$, $\widetilde Y_{\tau'}^{m'}=\widetilde Y_{\tau^m}^m$, and, on $\{\tau^m>0\}$,
    $
        \tau'
        =
        \rho^{m'}(U)
        <
        \rho^m(U)
        =
        \tau^m.
    $

    If $c>0$, then the flow cost term is strictly smaller at $\tau'$ than at $\tau^m$ on $\{\tau^m>0\}$. 
    Hence
    $
        \exp(-r(t+\tau'))v(\widetilde Y_{\tau'}^{m'})
        -
        \int_t^{t+\tau'}\exp(-rs)c\diff s
        >
        \exp(-r(t+\tau^m))v(\widetilde Y_{\tau^m}^m)
        -
        \int_t^{t+\tau^m}\exp(-rs)c\diff s
    $
    on an event of strictly positive probability, and weakly everywhere. 
    Taking expectations gives
    \begin{align*}
        V^{m'}(t,y)
        \geq 
        \mathbb E\left[
        \exp(-r(t+\tau'))v(\widetilde Y_{\tau'}^{m'})
        -
        \int_t^{t+\tau'}\exp(-rs)c\diff s
        \right]
        >
        V^m(t,y).
    \end{align*}

    It remains to consider the case $c=0$. 
    Then $r>0$. 
    Since $V^m(t,y)>\exp(-rt)v(y)$ and $\tau^m$ is optimal, we have
    $
        \mathbb E\left[\exp(-r(t+\tau^m))v(\widetilde Y_{\tau^m}^m)\right]>\exp(-rt)v(y)\geq 0.
    $
    Hence
    $
        \mathbb P(v(\widetilde Y_{\tau^m}^m)>0)>0.
    $
    On the event $\{\tau^m>0, v(\widetilde Y_{\tau^m}^m)>0\}$, which has strictly positive probability, we have $\tau'<\tau^m$ and $\widetilde Y_{\tau'}^{m'}=\widetilde Y_{\tau^m}^m$. 
    Therefore
    $
        \exp(-r(t+\tau'))v(\widetilde Y_{\tau'}^{m'})
        >
        \exp(-r(t+\tau^m))v(\widetilde Y_{\tau^m}^m)
    $
    on an event of strictly positive probability, and weakly everywhere. 
    Taking expectations gives
    $
        V^{m'}(t,y)>V^m(t,y).
    $
    Thus $V^m$ is weakly increasing in $m$, and strictly increasing in $m$ on $\{(t,y):V^m(t,y)>\exp(-rt)v(y)\}$.
\end{proof}

\begin{proof}[Proof of Proposition \ref*{proposition:optimal:properties-value-function:general}\ref*{proposition:optimal:properties-value-function:general:decreasing-in-t}]
    Fix $0\leq t<t'$ and $y\in D$. 
    Consider the two processes started from $y$ at $s=0$,
    \begin{align*}
        \diff Y_s^{t,y,m}&=\sigma(t+s,Y_s^{t,y,m},m)\diff B_s,\\
        \diff Y_s^{t',y,m}&=\sigma(t'+s,Y_s^{t',y,m},m)\diff B_s.
    \end{align*}
    By \ref{assumption:sigma:decreasing-in-t},
    $
        \sigma(t+s,z,m)\geq \sigma(t'+s,z,m)
    $
    for all $s\geq 0$ and $z\in D$. 
    Thus moving the initial calendar time from $t$ to $t'$ weakly decreases the diffusion coefficient pointwise along the entire future problem.

    The time-change argument in the proof that $V^m$ is increasing in $m$ therefore applies verbatim, with $\sigma(t+s,z,m)$ playing the role of the larger volatility coefficient and $\sigma(t'+s,z,m)$ playing the role of the smaller one. 
    Equivalently, after normalising by $\widehat V^m(t,y):=\exp(rt)V^m(t,y)$, this is the same volatility-comparison argument. 
    In particular, for every stopping time $\tau'$ admissible for $Y^{t',y,m}$, the same intrinsic Brownian path construction gives a stopping time $\tau$ admissible for $Y^{t,y,m}$ such that
    $
        \tau\leq \tau'
    $
    and
    $
        Y_{\tau}^{t,y,m}=Y_{\tau'}^{t',y,m}
    $
    pathwise.
\end{proof}

\begin{proof}[Proof of Proposition \ref*{proposition:optimal:properties-value-function:general}\ref*{proposition:optimal:properties-value-function:general:continuity-convexity}]
    Fix $m>0$ and recall that 
    $
        \widehat V^m(t,y):=\exp(rt)V^m(t,y).
    $
    Then
    \begin{align*}
        \widehat V^m(t,y)
        =
        \sup_{\tau\in\mathbb T^{t,y,m}}
        \mathbb E\left[
            \exp(-r\tau)v(Y_\tau^{t,y,m})
            -
            \int_0^\tau\exp(-rs)c\diff s
        \right].
    \end{align*}
    It is enough to prove that $\widehat V^m$ is continuous in $(t,y)$, $\mathcal C^1$ in $y$, and convex in $y$, because $V^m(t,y)=\exp(-rt)\widehat V^m(t,y)$.

    We first prove continuity. 
    For $T<\infty$, define the truncated value
    \begin{align*}
        \widehat V_T^m(t,y)
        :=
        \sup_{\tau\in\mathbb T^{t,y,m},\,\tau\leq T}
        \mathbb E\left[
            \exp(-r\tau)v(Y_\tau^{t,y,m})
            -
            \int_0^\tau\exp(-rs)c\diff s
        \right].
    \end{align*}
    The time-augmented process $(t+s,Y_s^{t,y,m})$ is a Feller diffusion on compact subsets of $\mathbb R_+\times D$ because $\sigma$ is continuous and locally Lipschitz in $y$ on $D$. 
    Moreover, $v$ is continuous and locally Lipschitz on $D$. 
    Hence, by the finite-horizon continuity theorem for optimal stopping of Feller diffusions, $\widehat V_T^m$ is continuous in $(t,y)$ locally on $\mathbb R_+\times D$; see \citet[Proposition 3.3]{Pham1998}.\footnote{
        To be precise, \citet[Proposition 3.3]{Pham1998} applies locally on compact subsets $[0,T]\times \bar I$ where $\bar I\subset D$, on which $\sigma$ is Lipschitz and uniformly bounded away from 0 and $\infty$. 
        This gives local continuity of the finite-horizon value in the interior.
    } 

    We now let $T\to\infty$. 
    We show that $\widehat V_T^m\to\widehat V^m$ locally uniformly. 
    Let $K\subset\mathbb R_+\times D$ be compact.

    Suppose first that $D$ is bounded. 
    Then $v$ is bounded on $D$. 
    Let $M:=\sup_{y\in D}v(y)<\infty$. 
    If $r>0$, then for every stopping time $\tau$,
    \begin{align*}
        \mathbf 1_{\{\tau>T\}}\exp(-r\tau)v(Y_\tau^{t,y,m})
        \leq
        M\exp(-rT).
    \end{align*}
    Since the running cost is nonnegative, truncating $\tau$ at $T$ can lose at most this amount. 
    Therefore
    \begin{align*}
        0
        \leq
        \widehat V^m(t,y)-\widehat V_T^m(t,y)
        \leq
        M\exp(-rT),
    \end{align*}
    uniformly in $(t,y)$.

    If $r=0$, then $c>0$. 
    Let $\tau$ be $\varepsilon$-optimal for $\widehat V^m(t,y)$. 
    Since stopping immediately gives payoff $v(y)\geq 0$, we have
    \begin{align*}
        \mathbb E\left[
            v(Y_\tau^{t,y,m})
            -
            c\tau
        \right]
        \geq
        -\varepsilon.
    \end{align*}
    Since $v\leq M$, it follows that
    $
        c\mathbb E[\tau]\leq M+\varepsilon.
    $
    Let $\tau_T:=\tau\wedge T$. 
    Then
    \begin{align*}
        &\left(
            v(Y_\tau^{t,y,m})
            -
            c\tau
        \right)
        -
        \left(
            v(Y_{\tau_T}^{t,y,m})
            -
            c\tau_T
        \right) \leq
        \mathbf 1_{\{\tau>T\}}v(Y_\tau^{t,y,m})
        \leq
        M\mathbf 1_{\{\tau>T\}}.
    \end{align*}
    Therefore
    \begin{align*}
        \widehat V^m(t,y)-\widehat V_T^m(t,y)
        \leq
        \varepsilon+
        M\mathbb P(\tau>T)
        \leq
        \varepsilon+
        \frac{M(M+\varepsilon)}{cT}.
    \end{align*}
    Letting $\varepsilon\to 0$ gives local uniform convergence also in this case.

    Suppose now that $D\subseteq \mathbb R$ is unbounded.
    Then, from \ref{assumption:sigma:decreasing-in-t} and \ref{assumption:sigma:bounded}, for every compact interval $K_t\subset\mathbb R_+$, we have that 
    \begin{align*}
        \sup_{t\in K_t}
        \int_0^\infty
        \sup_{z\in\mathbb R}\sigma(t+s,z,m)^2\diff s
        \leq
        \int_0^\infty
        \sup_{z\in\mathbb R}\sigma(s,z,m)^2\diff s
        <
        \infty.
    \end{align*}
    Since $v$ is the maximum of two affine functions, there exist constants $A,B<\infty$ such that
    $
        0\leq v(y)\leq A+B|y|
    $
    for every $y\in\mathbb R$. 
    Write
    $
        Y_s^{t,y,m}=y+Z_s^{t,y,m},
        $ and $
        Z_s^{t,y,m}:=\int_0^s\sigma(t+\ell,Y_\ell^{t,y,m},m)\diff B_\ell.
    $
    By the Burkholder--Davis--Gundy inequality,
    \begin{align*}
        \sup_{(t,y)\in K}
        \mathbb E\left[
            \sup_{s\geq0}|Z_s^{t,y,m}|^2
        \right]
        &\leq
        C
        \sup_{(t,y)\in K}
        \mathbb E\left[
            \int_0^\infty\sigma(t+s,Y_s^{t,y,m},m)^2\diff s
        \right]
        \leq
        C
        \sup_{t\in K_t}
        \int_0^\infty
        \sup_{z\in\mathbb R}\sigma(t+s,z,m)^2\diff s
        <
        \infty.
    \end{align*}
    Hence
    \begin{align*}
        \sup_{(t,y)\in K}
        \mathbb E\left[
            \sup_{s\geq0}v(Y_s^{t,y,m})^2
        \right]
        <
        \infty.
    \end{align*}
    In particular, the family
    $
        \{\sup_{s\geq0}v(Y_s^{t,y,m}):(t,y)\in K\}
    $
    is uniformly integrable and a constant $\tilde M<\infty$ such that $\sup_{(t,y)\in K} \mathbb E\left[ \sup_{s\geq0}v(Y_s^{t,y,m}) \right]\leq \tilde M$.

    If $r>0$, then, for every stopping time $\tau$,
    \begin{align*}
        \mathbf 1_{\{\tau>T\}}\exp(-r\tau)v(Y_\tau^{t,y,m})
        \leq
        \exp(-rT)\sup_{s\geq0}v(Y_s^{t,y,m}).
    \end{align*}
    Taking expectations and then suprema gives
    \begin{align*}
        &
        \mathbb E[\mathbf 1_{\{\tau>T\}}\exp(-r\tau)v(Y_\tau^{t,y,m})]
        \leq
        \sup_{(t,y)\in K}\sup_{\tau\in\mathbb T^{t,y,m}}
        \mathbb E\left[
            \mathbf 1_{\{\tau>T\}}\exp(-r\tau)v(Y_\tau^{t,y,m})
        \right]
        \\
        &\leq
        \exp(-rT)
        \sup_{(t,y)\in K}
        \mathbb E\left[
            \sup_{s\geq0}v(Y_s^{t,y,m})
        \right]\leq \exp(-rT)\tilde M \to 0.
    \end{align*}
    Let $\tau_T:=\tau\wedge T$; then
    \begin{align*}
        &\mathbb E\left[\left(
            \exp(-r\tau)v(Y_\tau^{t,y,m})
            -
            \int_0^\tau\exp(-rs)c\diff s
        \right)\right]
        -
        \mathbb E\left[\left(
            \exp(-r\tau_T)v(Y_{\tau_T}^{t,y,m})
            -
            \int_0^{\tau_T}\exp(-rs)c\diff s
        \right)\right]
        \\
        &\qquad\leq
        \mathbb E\left[\mathbf 1_{\{\tau>T\}}\exp(-r\tau)v(Y_\tau^{t,y,m})\right],
    \end{align*}
    which converges to 0 as $T\to \infty$.

    If $r=0$, then $c>0$. 
    Let $\tau$ be $\varepsilon$-optimal for $\widehat V^m(t,y)$, with $(t,y)\in K$. 
    Since stopping immediately gives payoff $v(y)\geq0$, we have
    $
        \mathbb E\left[
            v(Y_\tau^{t,y,m})
            -
            c\tau
        \right]
        \geq
        -\varepsilon.
    $
    Therefore
    $
        c\mathbb E[\tau]
        \leq
        \mathbb E\left[v(Y_\tau^{t,y,m})\right]+\varepsilon
        \leq
        \mathbb E\left[\sup_{s\geq0}v(Y_s^{t,y,m})\right]+\varepsilon.
    $
    The right-hand side is bounded uniformly over $(t,y)\in K$. 
    Thus $\varepsilon$-optimal stopping times have uniformly bounded first moments on $K$. 
    With $\tau_T:=\tau\wedge T$,
    \begin{align*}
        &
        \mathbb E\left[\left(
            v(Y_\tau^{t,y,m})
            -
            c\tau
        \right)\right]
        -
        \mathbb E\left[\left(
            v(Y_{\tau_T}^{t,y,m})
            -
            c\tau_T
        \right)\right]
        \leq
        \mathbb E\left[\mathbf 1_{\{\tau>T\}}v(Y_\tau^{t,y,m})\right]
        \leq
        \mathbb E\left[\mathbf 1_{\{\tau>T\}}\sup_{s\geq0}v(Y_s^{t,y,m})\right].
    \end{align*}
    By the uniform integrability just shown and the uniform first-moment bound on $\tau$, the right-hand side converges to $0$ uniformly over $(t,y)\in K$ along $\varepsilon$-optimal stopping times.
    Hence
    \begin{align*}
        \lim_{T\to\infty}
        \sup_{(t,y)\in K}
        \left(
            \widehat V^m(t,y)-\widehat V_T^m(t,y)
        \right)
        =
        0.
    \end{align*}
    To see the last convergence, fix $L>0$. 
    Then, uniformly over $(t,y)\in K$ and over the $\varepsilon$-optimal stopping times considered above,
    \begin{align*}
        \mathbb E\left[
            \mathbf 1_{\{\tau>T\}}\sup_{s\geq0}v(Y_s^{t,y,m})
        \right]
        &\leq
        \mathbb E\left[
            \mathbf 1_{\{\sup_{s\geq0}v(Y_s^{t,y,m})>L\}}
            \sup_{s\geq0}v(Y_s^{t,y,m})
        \right]
        +
        L\mathbb P(\tau>T).
    \end{align*}
    The first term is uniformly small for large $L$ by uniform integrability, while the second term is bounded by
    $
        L\sup \mathbb E[\tau]/T
    $
    and therefore vanishes as $T\to\infty$ for fixed $L$. 

    Thus, in all cases, $\widehat V_T^m\to\widehat V^m$ locally uniformly. 
    Since each $\widehat V_T^m$ is continuous, $\widehat V^m$ is continuous in $(t,y)$. 
    Therefore $V^m$ is continuous in $(t,y)$.

    We next prove that $\widehat V^m$ is $\mathcal C^1$ in $y$. 
    Fix a compact cylinder $Q:=[t_0,t_1]\times I$, where $0\leq t_0<t_1<\infty$ and $\bar I\subset D$ where $\bar I$ is a compact interval. 
    On $Q$, $\sigma(t,y,m)^2$ is continuous and uniformly bounded away from $0$ and $\infty$. 
    Moreover,
    $
        v(y)=\max\{a_0+a_1y,b_0+b_1y\}
    $
    is the maximum of two affine functions, hence is convex with a single kink. 
    By the dynamic programming principle, $\widehat V^m$ satisfies the dynamic programming variational inequality
    \begin{align*}
        \max\left\{
            v(y)-\widehat V^m(t,y),
            \partial_t\widehat V^m(t,y)
            +
            \frac{1}{2}\sigma(t,y,m)^2\partial_{yy}\widehat V^m(t,y)
            -
            r\widehat V^m(t,y)
            -
            c
        \right\}
        =
        0
    \end{align*}
    in the viscosity sense on $Q$; see \citet[Section 8.2]{Oksendal2013Book} and \citet[Chapter 3]{Pham2009Book}.
    On the continuation region, $\{\widehat V^m>v\}$, $\widehat V^m$ solves the linear parabolic equation
    \begin{align*}
        \partial_t\widehat V^m(t,y)
        +
        \frac{1}{2}\sigma(t,y,m)^2\partial_{yy}\widehat V^m(t,y)
        -
        r\widehat V^m(t,y)
        -
        c
        =
        0.
    \end{align*}
    This variational inequality is the parabolic problem to which \citet[Corollary 1]{DurandardStrulovici2026WP} applies locally.
    Therefore $\widehat V^m$ is continuously differentiable in the space variable on $Q$. 
    Since $Q\subset\mathbb R_+\times D$ was arbitrary, $\widehat V^m$ is $\mathcal C^1$ in $y$ on $\mathbb R_+\times D$. 
    Hence $V^m$ is also $\mathcal C^1$ in $y$.

    It remains to prove convexity in $y$. 
    On the continuation region, $\widehat V^m$ solves
    \begin{align*}
        r\widehat V^m(t,y)
        &
        =
        -
        c
        +
        \partial_t\widehat V^m(t,y)
        +
        \frac{1}{2}\sigma(t,y,m)^2\partial_{yy}\widehat V^m(t,y)
        \\
        \iff 
        \partial_{yy}\widehat V^m(t,y)
        &
        =
        \frac{2}{\sigma(t,y,m)^2}
        \left(
            r\widehat V^m(t,y)+c-\partial_t\widehat V^m(t,y)
        \right).
    \end{align*}
    By the monotonicity in $t$ proved above, $\widehat V^m$ is weakly decreasing in $t$, which implies $\partial_t\widehat V^m\leq 0$ in a viscosity sense. 
    Since, on the continuation region, $\widehat V^m> v\geq0$, and  $r,c\geq0$, it follows that, there, $\widehat V^m$ satisfies 
    $
        \partial_{yy}\widehat V^m(t,y)> 0
    $ in a viscosity sense.

    On the stopping region, $\widehat V^m(t,y)=v(y)$, and $v$ is convex. 
    Moreover, since $\widehat V^m$ is $\mathcal C^1$ in $y$, smooth pasting holds at every interior boundary point of the continuation region. 
    Hence $\partial_y\widehat V^m(t,\cdot)$ cannot jump downward when passing from a stopping interval into a continuation interval or from a continuation interval back into a stopping interval. 
    Since $\partial_y\widehat V^m(t,\cdot)$ is strictly increasing inside every continuation interval and agrees with the weakly increasing derivative of the convex function $v$ on stopping intervals, $\partial_y\widehat V^m(t,\cdot)$ is weakly increasing on $D$. 
    Therefore $\widehat V^m(t,\cdot)$ is convex on $D$. 
    Since $V^m(t,\cdot)=\exp(-rt)\widehat V^m(t,\cdot)$ and $\exp(-rt)>0$, $V^m(t,\cdot)$ is convex on $D$.

    Thus $V^m$ is continuous in $(t,y)$, $\mathcal C^1$ in $y$, and convex in $y$.
\end{proof}

\begin{proof}[Proof of Proposition \ref*{proposition:optimal:properties-value-function:general}\ref*{proposition:optimal:properties-value-function:general:constant-in-t}]
    This follows immediately from the fact that $Y^{t,y,m}\overset{d}{=}Y^{0,y,m}$, hence $V^m(t,y)=\exp(-r t)V^m(0,y)$.
\end{proof}

\begin{proof}[Proof of Proposition \ref*{proposition:optimal:properties-value-function:general}\ref*{proposition:optimal:properties-value-function:general:even}]
    This is immediately from the fact that 
    $V^m(t,y)=\sup_{\tau\in \mathbb T^{t,y,m}}\mathbb E[\exp(-r (t+\tau))v(Y_\tau)-\int_t^{t+\tau}c\diff s]=\sup_{\tau\in \mathbb T^{t,y,m}}\mathbb E[\exp(-r (t+\tau))v(-Y_\tau^{t,y,m})-\int_t^{t+\tau}c\diff s]$ and 
    $Y^{t,y,m}\overset{d}{=}-Y^{t,-y,m}$.
\end{proof}

We now show how the properties shown in \hyref{proposition:optimal:properties-value-function:general}[Proposition] deliver properties for the stopping thresholds defining the continuation region.

\begin{proposition}
    \label{proposition:optimal:stopping-thresholds:general}
    There are $\overline y^m,\underline y^m:\mathbb R_+\to D$ satisfying $\underline y^m\leq \overline y^m$ such that all of the following properties hold:
    \begin{enumerate}[label=(\arabic*)]
        \item $\tau^{m}=\inf\{t>0|Y_{t}^{m}\notin (\underline y_t^m,\overline y_t^m)\}$;
        \item $\overline y^m$ and $\underline y^m$ are continuous; 
        \item $\overline y^m$ is weakly decreasing and $\underline y^m$ is weakly increasing, strictly whenever $(\underline y^m_t,\overline y^m_t)\ne \emptyset$ if $\sigma$ is strictly decreasing in $t$ for any $y\in D$ and $m>0$;
        \item for any $m'>m$, $(\underline y^m_t, \overline y^m_t)\subseteq (\underline y^{m'}_t, \overline y^{m'}_t)$ for any $s\geq 0$, with strict inclusion whenever the right-hand side is non-empty;
        \item if $\sigma$ is constant in $t$, then so are $\overline y^m$ and $\underline y^m$; 
        and
        \item if $\sigma$ satisfies \ref{assumption:sigma:even-in-y} and $v$ is even, then $\overline y^m=-\underline y^m$.
    \end{enumerate}
\end{proposition}

\begin{proof}
    Let
    $
        h(t,y):=\widehat V^m(t,y)-v(y),
    $
    and $D_t:=\{y\in \bar D | h(t,y)>0\}$, which denotes the continuation region at $t$.

    Observe that since $\widehat V^m$ is convex and $\mathcal C^1$ in $y$ and equals $v$ outside $D_t$, then $\tilde y\in D_t$ while $D_t\ne \emptyset$.
    Since $\widehat V^m-v$ is convex on either side of $\tilde y$,  $h\geq 0$ is quasiconcave.
    Moreover, $h$ is $\mathcal C^1$ in $y$ except at $\tilde y$ and $\sign(y-\tilde y)\partial_y h(t,y)\leq 0$ for $y \ne \tilde y$, with equality outside of $D_t$.
    Hence, $D_t$ must be an interval in $D$ and we can write $\tau^{t,y,m}:=\inf\{s>0|Y_s^{t,y,m}\notin D_t\}$.
    
    \textbf{Step 1: $\sup D_t<\overline Y$ and $\inf D_t>\underline Y$.}

    We prove that $\sup D_t<\overline Y$.
    We consider first the case in which $\overline Y<\infty$.
    Fix $t\geq0$ and $m>0$, and suppress $m$ from the notation. 
    Choose $y_0\in(\widetilde y,\overline Y)$, where $\widetilde y$ is the kink of $v$. 
    On $[y_0,\overline Y]$, $v$ is affine, so $v_{yy}=0$. 
    Since $v\geq0$ and one of $r,c$ is strictly positive, we may choose $y_0$ so that there exists $\lambda>0$ with
    $
        rv(y)+c\geq\lambda
    $
    for all $y\in[y_0,\overline Y]$.

    On the continuation region, $\widehat V^m$ solves
    \begin{align*}
        \partial_t\widehat V^m(t,y)
        +
        \frac{1}{2}\sigma(t,y,m)^2\partial_{yy}\widehat V^m(t,y)
        -
        r\widehat V^m(t,y)
        -
        c
        =
        0.
    \end{align*}
    Therefore, on $D_t\cap (y_0,\overline Y)$,
    \begin{align*}
        \partial_t h(t,y)
        +
        \frac{1}{2}\sigma(t,y,m)^2\partial_{yy}h(t,y)
        -
        rh(t,y)
        =
        rv(y)+c.
    \end{align*}
    By monotonicity in $t$, $\partial_t h(t,y)=\partial_t\widehat V^m(t,y)\leq 0$. 
    Also $h\geq 0$. 
    By \ref{assumption:sigma:positive}, $\sigma(t,\overline Y,m)=0$, and then, by \ref{assumption:sigma:log-Lipschitz}, 
    $
        \sigma(t,y,m)
        \leq
        K_m(\overline Y-y)
        \sqrt{1+\left|\ln(\overline Y-y)\right|}
    $.  
    Hence, on $D_t\cap(y_0,\overline Y)$,
    \begin{align*}
        \frac{1}{2}
        K_m^2
        (\overline Y-y)^2
        \left(1+\left|\ln(\overline Y-y)\right|\right)
        \partial_{yy}h(t,y)
        &\geq
        \frac{1}{2}\sigma(t,y,m)^2\partial_{yy}h(t,y)
        \\
        &=
        rv(y)+c
        -
        \partial_t h(t,y)
        +
        rh(t,y)
        \geq
        \lambda.
    \end{align*}
    Therefore
    \begin{align*}
        \partial_{yy}h(t,y)
        \geq
        \frac{2\lambda}
        {
            K_m^2
            (\overline Y-y)^2
            \left(1+\left|\ln(\overline Y-y)\right|\right)
        }.
    \end{align*}
    Suppose, toward a contradiction, that $\sup D_t = \overline Y$. 
    Then $(y_0,\overline Y)\subset D_t$.
    On $(y_0,\overline Y)$,
    \begin{align*}
        \partial_{yy}h(t,y)
        \geq
        \frac{2\lambda}
        {
            K_m^2
            (\overline Y-y)^2
            \left(1+\left|\ln(\overline Y-y)\right|\right)
        }.
    \end{align*}
    Fix $y\in(y_0,\overline Y)$. 
    Since $\widehat V^m(t,\cdot)$ is $\mathcal C^1$ in $y$, $\partial_yh(t,y)$ is finite. 
    Integrating, for $y<z<\overline Y$,
    \begin{align*}
        \partial_yh(t,z)
        \geq
        \partial_yh(t,y)
        +
        \frac{2\lambda}{K_m^2}
        \int_y^z
        \frac{1}
        {
            (\overline Y-q)^2
            \left(1+\left|\ln(\overline Y-q)\right|\right)
        }
        \diff q,
    \end{align*}
    and 
    $
        \partial_yh(t,z)\to+\infty
    $
    as $z\to\overline Y$.

    Integrating once more, for $y<z<\overline Y$,
    \begin{align*}
        h(t,z)-h(t,y)
        &\geq
        \partial_yh(t,y)(z-y)
        +
        \frac{2\lambda}{K_m^2}
        \int_y^z
        \frac{z-q}
        {
            (\overline Y-q)^2
            \left(1+\left|\ln(\overline Y-q)\right|\right)
        }
        \diff q.
    \end{align*}
    The last integral diverges as $z\to\overline Y$. 
    To see this, let $u=\overline Y-q<1$ and $\epsilon=\overline Y-z<1$,
    \begin{align*}
        \int_y^z
        \frac{z-q}
        {
            (\overline Y-q)^2
            \left(1+\left|\ln(\overline Y-q)\right|\right)
        }
        \diff q
        &=
        \int_\epsilon^{\overline Y-y}
        \frac{u-\epsilon}
        {
            u^2(1+\ln (1/u))
        }
        \diff u
        \\
        &\geq
        \int_{2\epsilon}^{\overline Y-y}
        \frac{u-\epsilon}
        {
            u^2\left(1+\ln(1/u)\right)
        }
        \diff u
        \geq
        \frac{1}{2}
        \int_{2\epsilon}^{\overline Y-y}
        \frac{1}
        {
            u\left(1+\ln(1/u)\right)
        }
        \diff u.
    \end{align*}
    With a change of variables, $w:=1+\ln(1/u)$, implying $\diff w=-\diff u/u$
    \begin{align*}
        \int_{2\epsilon}^{\overline Y-y}
        \frac{1}
        {
            u\left(1+\ln(1/u)\right)
        }
        \diff u
        &=
        \int_{1+\ln(1/{(\overline Y-y)})}^{1+\ln(1/(2\epsilon))}
        \frac{1}{w}\diff w
        \\
        &=
        \ln\left(1+\ln\left(\frac{1}{2\epsilon}\right)\right)
        -
        \ln\left(1+\ln\left(\frac{1}{\overline Y-y}\right)\right)\to \infty
    \end{align*}
    as $\epsilon\to 0$. 
    Thus $h(t,z)\to\infty$ as $z\to\overline Y$.

    This contradicts the boundedness of $h(t,\cdot)$ near the finite boundary $\overline Y$, given that $h$ is continuous and satisfies $h(t,\overline Y)=0$. 
    We conclude that if $\overline Y<\infty$, then $\sup D_t<\overline Y$.

    Suppose now $\overline Y=\infty$.
    By \ref{assumption:sigma:bounded}, $\sup_{s\geq t,\,y\in D}\sigma(s,y,m)\leq \overline\sigma<\infty$.
    Let $\overline W$ be the value of the constant-volatility problem
    $
        \overline W(y)
        :=
        \sup_{\tau}
        \mathbb E\left[
            e^{-r\tau}v(y+\overline\sigma B_\tau)
            -
            \int_0^\tau e^{-rs}c\diff s
        \right].
    $
    Since $\sigma(s,y,m)\leq \overline\sigma$ for all $s\geq t$ and $y\in D$, the volatility-comparison argument used above gives
    $
        \widehat V^m(t,y)\leq \overline W(y)
    $
    for all $y\in D$. 
    Since immediate stopping is feasible,
    $
        \widehat V^m(t,y)\geq v(y)
    $.
    Hence it is enough to prove that $\overline W(y)=v(y)$ for all sufficiently large $y$.
    
    First, observe that $v(y)=\max\{a_0+a_1 y,b_0+b_1y\}\leq \max\{|a_0|+|a_1||y|,|b_0|+|b_1||y|\}\leq C_0(1+|y|)$ for $C_0=\max_{i\in\{0,1\}}\{|a_i|,|b_i|\}$.
    Assume first that $c>0$. 
    Note that if $c>0$, $\sup_{\tau}\mathbb E[k |B_\tau|-c\tau]=k^2/(4c)$, whereas for $r>0$, $\sup_{\tau}\mathbb E[\exp(-r\tau)|B_\tau|]=(2 r)^{-1/2}(\sinh(z^*))$, for $z^*:z^*\sinh(z^*)=\cosh(z^*)$.
    Then, for any optimal stopping time $\tau$, when $c>0$,
    \begin{align*}
        \overline W(y)
        &=
        \sup_{\tau}\mathbb E\left[
            e^{-r\tau}v(y+\overline\sigma B_\tau)
            -
            \int_0^\tau e^{-rs}c\diff s
        \right]
        \leq 
        \sup_{\tau}\mathbb E\left[
            v(y+\overline\sigma B_\tau)
            -
            \tau c
        \right]
        \\
        &\leq 
        \sup_{\tau}\mathbb E\left[
            C_0(1+|y|+\overline\sigma |B_\tau|)
            -
            \tau c
        \right]
        =
        C_0(1+|y|)+
        \sup_\tau\mathbb E\left[
            C_0\overline\sigma |B_\tau|
            -
            \tau c
        \right]
        \\
        &=
        C_0(1+|y|)+
        (C_0\overline \sigma)^2(4 c )^{-1}.
    \end{align*}
    When instead $r>0=c$, 
    \begin{align*}
        \overline W(y)
        &=
        \sup_{\tau}\mathbb E\left[
            e^{-r\tau}v(y+\overline\sigma B_\tau)
        \right]
        \leq 
        \sup_{\tau}\mathbb E\left[
            e^{-r\tau}v(y+\overline\sigma B_\tau)
        \right]
        \\
        &\leq 
        \sup_{\tau}\mathbb E\left[
            e^{-r\tau}C_0(1+|y|+\overline\sigma |B_\tau|)
        \right]
        \leq
        C_0(1+|y|)+
        C_0\overline\sigma \sup_\tau\mathbb E\left[
            e^{-r\tau} |B_\tau|
        \right]
        \\
        &=
        C_0(1+|y|)+
        (C_0\overline \sigma)(2 r)^{-1/2}(\sinh(z^*)).
    \end{align*}
    Hence, in any case, $\exists \tilde C_1>0$ and $\tilde C_0\in \mathbb R$ such that 
    $\overline W(y)\leq\tilde C_1 |y|+\tilde C_0$.

    Now fix $y_0>\widetilde y^+$ and note that for $y\geq y_0$, $v(y)=a_0+a_1y$. 
    Suppose, for the purpose of a contradiction, that the continuation region of $\overline W$ is unbounded above. 
    On $y>y_0$, $\overline W(y)\geq \widehat V^m(t,y)>v(y)$.
    Set 
    $
        h(y):=\overline W(y)-v(y).
    $
    Then $h>0$ on $(y_0,\infty)$, and, on this interval, $\overline W$ solves
    \begin{align*}
        \frac{1}{2}\overline\sigma^2 \overline W''(y)
        -
        r\overline W(y)
        -
        c
        =
        0.
    \end{align*}
    Since $v''=0$ on $(y_0,\overline Y)$, this gives
    \begin{align*}
        h''(y)
        =
        2\overline\sigma^{-2}(r h(y)+rv(y)+c)\geq 2\overline\sigma^{-2}(r v(y_0) + c)=:\lambda>0.
    \end{align*}
    Integrating twice we have that, for some $B_2>0$ and $B_1,B_0\in \mathbb R$, 
    $\overline W(y)\geq B_2 y^2+B_1 y + B_0 \iff 0\geq \overline W(y) - \tilde C_1 |y|-\tilde C_0 \geq B_2 y^2+(B_1-\tilde C_1) y + (B_0-\tilde C_0) \to \infty$, a contradiction.
    We conclude that $\sup\{y\geq y_0|\overline W(y)>v(y)\}<\infty$.
    The argument is symmetric for the case in which $\underline Y=-\infty$.

    For all the cases above, we find that $\sup D_t<\overline Y$; via symmetric arguments, we have that $\inf D_t >\underline Y$.
    
    \textbf{Proof of Property (1).}

    Define $\overline y^m_t:=\sup D_t$ and $\underline y^m_t:=\inf D_t$.
    By step 1 above, these are well-defined.
    As $D_t$ is an open interval, by definition, $D_t=(\underline y^m_t,\overline y^m_t)$.
    Then, $y\notin (\underline y^m_t,\overline y^m_t) \iff y\notin D_t \iff \hat V^m(t,y)>v(y)$.
    Property (1) follows immediately.

    \textbf{Proof of Properties (2) and (3).}

    Since $\hat V^m(t,y)$ is continuous in $t$ (\hyref{proposition:optimal:properties-value-function:general}[Proposition]\ref{proposition:optimal:properties-value-function:general:continuity-convexity}), then so are $\overline y^m$ and $\underline y^m$. 

    As $\hat V^m(t,y)$ decreases in $t$, then $D_t$ is decreasing in $t$ with respect to the subset order.
    If $\hat V^m(t,y)$ strictly decreases in $t$, as is the case when $\sigma$ is strictly decreasing in $t$ (\hyref{proposition:optimal:properties-value-function:general}[Proposition]\ref{proposition:optimal:properties-value-function:general:decreasing-in-t}), then, whenever non-empty, so is $D_t$.

    \textbf{Proof of Property (4).}

    As $\hat V^m(t,y)$ strictly increases in $m$ on $y\in D_t$ (\hyref{proposition:optimal:properties-value-function:general}[Proposition]\ref{proposition:optimal:properties-value-function:general:increasing-in-m}), then $D_t$ is strictly increasing in $t$ with respect to the subset order whenever non-empty.

    \textbf{Proof of Property (5).}

    As $\hat V^m(t,y)=\hat V^m(0,y)$ whenever $\sigma$ is constant with respect to $t$ (\hyref{proposition:optimal:properties-value-function:general}[Proposition]\ref{proposition:optimal:properties-value-function:general:constant-in-t}), then $D_t=D_0$.

    \textbf{Proof of Property (6).}

    As $\hat V^m(t,y)=\hat V^m(t,-y)$ whenever $\sigma$ and $v$ are even with respect to $y$, (\hyref{proposition:optimal:properties-value-function:general}[Proposition]\ref{proposition:optimal:properties-value-function:general:even}), then $D_t$ is symmetric about 0.
\end{proof}

We proceed to proving comparative statics. 
Our first result is that the expected absolute value of the stopped process is monotone in the speed of diffusion, whereas the expected optimal stopping time is non-monotone.

\begin{proposition}
    \label{proposition:optimal:speed-accuracy:general}
    $\mathbb E\left[|Y_{\tau^{m}}^m|\right]$
    is weakly increasing in $m$ and strictly so if $\exists T>0$ such that $V^{m}(T,0)>v(0)$ and $\tau^m\ne 0$ almost surely. 
    If \ref{assumption:sigma:infinity} holds and $D$ is bounded, then $\mathbb E[\tau^{m}]$ is non-monotone in $m$. 
\end{proposition}
\begin{proof}
    For a fixed continuous correspondence $C:\mathbb R_+\rightrightarrows D$ such that $C_t$ is interval-valued, let $\tau^{m,C}:=\inf\{t>0|Y_t^m\notin C_t\}$. 
    From \hyref{proposition:optimal:stopping-thresholds:general}[Proposition], there is a continuous interval-valued correspondence $C^m:\mathbb R_+\rightrightarrows D$ such that $\tau^m=\inf\{t>0|Y_t^m\notin C_t^m\}$, where $C_t^m$ strictly increases in the subset order respect to $m$ whenever non-empty and weakly decreases in the subset order with respect to $t$.
    Then, for $m'>m>0$, from \hyref{proposition:exogenous:accuracy:shrinking:general}[Propositions] and \ref{proposition:exogenous:speed-accuracy:nested:general}, 
    $\mathbb E\left[|Y_{\tau^{m}}^m|\right]
    =\mathbb E\left[|Y_{\tau^{m,C^m}}^m|\right]
    \leq 
    \mathbb E\left[|Y_{\tau^{m',C^m}}^{m'}|\right]
    \leq 
    \mathbb E\left[|Y_{\tau^{m',C^{m'}}}^{m'}|\right]
    \leq \mathbb E\left[|Y_{\tau^{m'}}^{m'}|\right]
    $, 
    with strict inequality if $0 , Y_0 \in C^{m'}_T$ for some $T>0$ --- which, owing to the fact that $C_t^{m'}$ is decreasing in $t$, implies $0,Y_0\in C_t^{m'}$ $\forall t\in [0, T]$.
    Additionally, note that $\exists T>0:Y_0\in C_T^{m'} \iff \tau^{m'}\ne 0$ a.s.

    We now prove that if \ref{assumption:sigma:infinity} holds and $D$ is bounded, then $\tau^{m}$ is non-monotone in $m$.
    Let
    $
    \bar\sigma(m):=\sup_{t\geq 0,y\in D}\sigma(t,y,m).
    $
    By \ref{assumption:sigma:increasing-in-m}, $\bar\sigma(m)$ is monotone in $m$, and by \ref{assumption:sigma:infinity}, $\lim_{m\to 0}\bar \sigma(m)=0$. 
    For any a.s. finite stopping time $\tau$,
    $\mathbb E[|Y_\tau^m-Y_0|^2]=\mathbb E[|\int_0^\tau \sigma(t, Y_t^m, m)\diff B_t|^2]=\mathbb E[\int_0^\tau \sigma(t, Y_t^m, m)^2\diff t]\leq \mathbb E[\tau]\bar\sigma(m)^2$.
    On the other hand, since $v$ is Lipschitz continuous with a bound $L:=a_1\vee|b_1|$, then 
    $\mathbb E[|v(Y_\tau^m)-v(Y_0)|]\leq L\mathbb E[|Y_\tau^m-Y_0|]$.
    By the Cauchy-Schwarz inequality, 
    $\mathbb E[|Y_\tau^m-Y_0|]\leq \mathbb E[|Y_\tau^m-Y_0|^2]^{1/2}\leq \mathbb E[\tau]^{1/2}\bar\sigma(m)$.
    
    Assume $c>0$. 
    From optimality, 
    \begin{align*}
        v(Y_0)
        &\leq 
        V^m(0,Y_0)
        =\mathbb E[\exp(-r\tau^m)v(Y_{\tau^m}^m)-c\int_0^{\tau^m}\exp(-rs)\diff s]
        \leq 
        \mathbb E[v(Y_{\tau^m}^m)]-c\mathbb E[{\tau^m}] 
        \\
        \implies c\mathbb E[\tau]&\leq \mathbb E[v(Y_{\tau^m}^m)-v(Y_0)].
    \end{align*}
    Since $v$ is Lipschitz, $c\mathbb E[{\tau^m}]\leq L\mathbb E[|Y_{\tau^m}^m-Y_0|] \leq L \mathbb E[{\tau^m}]^{1/2}\bar\sigma(m)\iff \mathbb E[{\tau^m}]\leq (L/c)^{2}\bar\sigma(m)^2 \to 0$ as $m\to 0$.

    Now assume instead $c=0<r$ and $v>0$ and set
    $
        v_*:=\inf_{y\in D}v(y)>0.
    $
    We work directly with the normalized value
    $
        \widehat V^m(t,y)
        :=
        \sup_{\tau}
        \mathbb E[
            e^{-r\tau}v(Y_\tau^{t,y,m})
        ].
    $
    Since immediate stopping is feasible,
    $
        v(y)
        \leq
        \mathbb E[
            e^{-r\tau^{t,y,m}}
            v(Y_{\tau^{t,y,m}}^{t,y,m})
        ].
    $
    Therefore
    \begin{align*}
        v_*
        \mathbb E\left[
            1-e^{-r\tau^{t,y,m}}
        \right]
        &\leq
        \mathbb E\left[
            e^{-r\tau^{t,y,m}}
            \left(
                v(Y_{\tau^{t,y,m}}^{t,y,m})-v(y)
            \right)
        \right]
        \leq
        L
        \mathbb E\left[
            e^{-r\tau^{t,y,m}}
            |Y_{\tau^{t,y,m}}^{t,y,m}-y|
        \right].
    \end{align*}
    Now write
    \begin{align*}
        Y_s^{t,y,m}-y
        =
        \int_0^s\sigma(t+u,Y_u^{t,y,m},m)\diff B_u.
    \end{align*}
    Then, by Ito's formula,
    \begin{align*}
        \mathbb E\left[
            e^{-r\tau^{t,y,m}}
            |Y_{\tau^{t,y,m}}^{t,y,m}-y|^2
        \right]
        &=
        \mathbb E\left[
            \int_0^{\tau^{t,y,m}}e^{-rs}\sigma(t+s,Y_s^{t,y,m},m)^2\diff s
            -
            r\int_0^{\tau^{t,y,m}}\exp(-rs)B_t^2\diff s
        \right]
        \\
        &\leq
        \bar\sigma(m)^2
        \mathbb E\left[
            \int_0^{\tau^{t,y,m}}e^{-rs}\diff s
        \right]
        =
        \bar\sigma(m)^2
        \frac{1-\mathbb E[e^{-r\tau^{t,y,m}}]}{r}
        \leq
        \frac{\bar\sigma(m)^2}{r}.
    \end{align*}
    Hence, by Cauchy--Schwarz,
    $
        \mathbb E[
            e^{-r\tau^{t,y,m}}
            |Y_{\tau^{t,y,m}}^{t,y,m}-y|
        ]
        \leq
        \frac{\bar\sigma(m)}{\sqrt r}.
    $
    Consequently, uniformly over $t\geq 0$ and $y\in D$,
    \begin{align}
        \mathbb E[
            1-e^{-r\tau^{t,y,m}}
        ]
        \leq
        \frac{L}{v_*\sqrt r}\bar\sigma(m).
    \end{align}

    Fix $T>0$. 
    Since for any stopping time $\tau$ we have 
    $
        1-e^{-r\tau}\geq (1-e^{-r T})\mathbf 1_{\{\tau>T\}},
    $
    then, taking expectations and rearranging, we obtain 
    $\mathbb P(\tau^{t,y,m}>T)\leq \mathbb E[
        1-e^{-r\tau^{t,y,m}}
    ](1-e^{-r T})^{-1}$.
    As $\sup_{t\geq 0, y \in D}\mathbb E[
            1-e^{-r\tau^{t,y,m}}
        ]\leq \frac{L}{v_*\sqrt r}\bar\sigma(m)$, 
    we have 
    \begin{align*}
        \sup_{t\geq 0, y \in D}\mathbb P(\tau^{t,y,m}>T)
        \leq 
        \frac{L}{v_*\sqrt r}(1-e^{-r T})^{-1}\bar\sigma(m)
        =:p_m(T),
    \end{align*}
    where $\lim_{m\to 0}p_m(T)=0$ for any finite $T\geq 0$.

    Now we use the strong Markov property and the fact that $\tau^{t,y,m}$ is the earliest optimal stopping time. 
    For the initial point $Y_0$, we write $\tau^m:=\tau^{0,Y_0,m}$. 
    On $\{\tau^m>n T \}$, the shifted remaining stopping time is the earliest optimal stopping time for the problem starting from $(n T,Y_{n T}^m)$. 
    Therefore, 
    $
        \mathbb P(\tau^m>(n+1) T \mid\mathcal F_{n T })
        \leq
        p_m( T )
        $ on $\{\tau^m>n T \}.
    $
    Iterating,
    $
        \mathbb P(\tau^m>n T )
        \leq
        p_m( T )^n.
    $
    Hence, for all small enough $m$ such that $p_m( T )<1$,
    \begin{align*}
        \mathbb E[\tau^m]
        &=
        \int_0^\infty\mathbb P(\tau^m>s)\diff s
        \leq
        T 
        \sum_{n=0}^\infty
        \mathbb P(\tau^m>n T )
        \leq
        \frac{ T }{1-p_m( T )}.
    \end{align*}
    Taking $m\to 0$ gives
    $
        \limsup_{m\to 0}\mathbb E[\tau^m]
        \leq
        T.
    $
    Since $ T >0$ was arbitrary,
    $
        \lim_{m\to 0}\mathbb E[\tau^m]=0.
    $

    For bounded $D=(\underline Y,\overline Y)$, let $G(y):=\frac{\overline Y-y}{\overline Y-\underline Y}v(\underline Y) + \frac{y - \underline Y}{\overline Y-\underline Y}v(\overline Y)$.
    Since $G$ is linearly interpolating between the values of $v$ at the boundary of $D$, we have that, for any $m>0$, $v\leq V^m< G$ on $D$.
    Fix $\delta>0$ and set $K_\delta:=[\underline Y+\delta,\overline Y-\delta]$ and let $\tau_\delta^m:=\inf\{t>0|Y_t^m\notin K_\delta\}$.
    Given that, by \ref{assumption:sigma:infinity}, its quadratic variation term diverges on compact subsets of $D$, $Y_t^m$ exits $K_\delta$ arbitrarily fast for any $\delta>0$ and, for any $T>0$, $\lim_{m\to \infty}\mathbb P(\tau_\delta^m>T)=0$.
    
    By optimality, $V^m(0, Y_0)\geq \mathbb E[\exp(-\tau_\delta^m)v(Y_{\tau_\delta^m}^m)-\int_0^{\tau_\delta^m}\exp(-r s)c\diff s]$.
    Taking limits, $\lim_{m\to \infty}V^m(0,Y_0)\geq G_\delta(Y_0)$, where $G_\delta(y):=\frac{\overline Y-y-\delta}{\overline Y-\underline Y-2\delta}v(\underline Y+\delta) + \frac{y - \underline Y-\delta}{\overline Y-\underline Y-2\delta}v(\overline Y-\delta)$ is a linear interpolation between the boundaries of $K_\delta$.
    Letting $\delta>0$, $G(Y_0)\leq \lim_{m\to \infty}V^m(0,Y_0)\leq G(Y_0)$.

    Assume first that $c>0$.
    Again, by optimality, 
    \begin{align*}
        V^m(0,Y_0)
        &=
        \mathbb E[\exp(-r\tau^m)v(Y_{\tau^m}^m)-\int_0^{\tau^m}\exp(-rs)c\diff s]
        \leq 
        \mathbb E[v(Y_{\tau^m}^m)-c\tau^m]
        \leq 
        G(Y_0)-c\mathbb E[\tau^m]
        \\
        \iff 
        c\mathbb E[\tau^m]&\leq G(Y_0)-V^m(0,Y_0).
    \end{align*}
    Hence, $\lim_{m\to \infty}c\mathbb E[\tau^m]\leq \lim_{m\to \infty}G(Y_0)-V^m(0,Y_0)=0$.
    
    If instead $r>c=0$ and $v>0$.
    Note that $\widehat V^m(t,y)\geq v(y)$ and $v^*:=\sup_y v(y)\geq v\geq \inf_y v(y)=v(\tilde y)=:v_*$.
    As $\exp(-r\tau^m)v(Y_{\tau^m}^m)=v(Y_{\tau^m}^m)-(1-\exp(-r\tau^m))v(Y_{\tau^m}^m)\leq G(Y_{\tau^m}^m)-(1-\exp(-r\tau^m))v_*$,
    taking expectations, 
    $V^m(0,Y_0)=\mathbb E[\exp(-r\tau^m)v(Y_{\tau^m}^m)]\leq G(Y_0)-\mathbb E[(1-\exp(-r\tau^m))]v_*$.
    Therefore, $\mathbb E[(1-\exp(-r\tau^m))]\leq (G(Y_0)-V^m(0,Y_0))/v_*$ and, taking limits $\lim_{m\to \infty}\mathbb E[(1-\exp(-r\tau^m))]=0$, which implies that $\tau^m$ converges in probability to zero as $m\to \infty$.
    
    If $\exists T: D_T=\emptyset$, then $\tau^m\leq T$ and then $\mathbb E[\tau^m]\to 0$.
    If otherwise, we need to bound the tail distribution of $\tau^m$ and proceed as follows: 
    as $\tilde y \in D_t\ne \emptyset,\forall t\geq 0$, then $v_*< \widehat V^m(t,y)=\mathbb E[\exp(-r \tau^{t,y,m})v(Y^{t,y,m}_{\tau^{t,y,m}})]\leq \mathbb E[\exp(-r \tau^{t,y,m})]v^*$.
    Consequently, for $T>0$,
    \begin{align*}
        &v_*/v^* 
        \leq 
        \mathbb E[\exp(-r \tau^{t,y,m})]
        \leq 
        \mathbb E[\mathbf 1_{\tau^{t,y,m}\leq T} + \mathbf 1_{\tau^{t,y,m}> T}\exp(-r T)]
        =
        1-(1-\exp(-rT))\mathbb P(\tau^{t,y,m}> T)
        \\
        \implies &
        \sup_{t\geq 0, y \in D, m> 0}\mathbb P(\tau^{t,y,m}> T)
        \leq \frac{1-v_*/v^*}{1-\exp(-rT)}=:q.
    \end{align*}
    Then, using the strong Markov property as above, we obtain $\mathbb P(\tau^{t,y,m}> nT)\leq q^n$, 
    and therefore 
    $\int_0^\infty \mathbb P(\tau^{m}> s)\diff s = \sum_{n=0}^\infty \int_{nT}^{(n+1)T} \mathbb P(\tau^{m}> s)\diff s\leq T \sum_{n=0}^\infty \mathbb P(\tau^{m}> nT) =T/(1-q)$, hence $(\tau^m)$ is uniformly integrable for any $m>0$.
    Then, as $\tau^m$ converges in probability to zero as $m\to \infty$ and is uniformly integrable, $\lim_{m\to \infty}\mathbb E[\tau^m]=0$.
    This concludes the proof.
\end{proof}

Our second result of this section is that a shift in payoffs akin to a subsidy leads to a monotone shift in the joint distribution of choices and stopping time.

\begin{proposition}
    \label{proposition:optimal:choice:shift:general}
    Let $\overline y^m,\underline y^m:\mathbb R_+\to D$ be such that 
    $\overline y^m_t:=\sup D_t^m$ and $\underline y^m_t:=\inf D_t^m$, where 
    $D_t^m:=\{y\in D | \exp(rt)V^m(t,y)>v(y)\}$.
    Then, 
    (1) $\overline y^m$ and $\underline y$ pointwise strictly decrease (increase) in $a_0$ ($b_0$); and 
    (2) $\mathbb P(Y_{\tau^{m}}^m=\overline y_{\tau^{m}} \text{ and }\tau^{m}\leq T)$ increases (decreases) in $a_0$ ($b_0$) whereas
    $\mathbb P(Y_{\tau^{m}}^m=\underline y_{\tau^{m}} \text{ and }\tau^{m}\leq T)$ decreases (increases) in $a_0$ ($b_0$), $\forall T> 0$, strictly so when $\tau^{m}\ne 0$ almost surely.
\end{proposition}
\begin{proof}
    Let $v(y):=\max\{a_0+a_1y,b_0+b_1y\}$, with $a_1>0>b_1$, and let
    $
        \tilde v(y):=\max\{\tilde a_0+a_1y,b_0+b_1y\}
    $
    with $\tilde a_0>a_0$. 
    Set $\Delta:=\tilde a_0-a_0>0$. 
    Let $\widehat V^m$ and $\widehat{\tilde V}^m$ denote the normalized value functions associated with $v$ and $\tilde v$, respectively, and 
    let $D_t^m=(\underline y_t^m,\overline y_t^m)$ and 
    $\tilde D_t^m=(\tilde{\underline y}_t^m,\tilde{\overline y}_t^m)$ denote the corresponding continuation regions.

    Given that we have that $0\leq \tilde v(y)- v(y)\leq \Delta$ for all $y \in D$, then, for every $(t,y)$, $
        \widehat V^m(t,y)
        \leq
        \widehat{\tilde V}^m(t,y)
        \leq
        \widehat V^m(t,y)+\Delta.
    $
    Therefore, it immediately follows that 
    (1) $\widehat V^m(t,y)-\tilde v(y)= -\Delta  \implies \widehat{\tilde V}^m(t,y)=\tilde v(y)$ and (2) 
    $\widehat{\tilde V}^m(t,y)=\tilde v(y) = v(y) \implies \widehat V^m(t,y) =v(y)$.
    (1) and (2) then imply, respectively, that  $\overline y_t\geq \tilde{\overline y}_t$ 
    and 
    $\underline y_t\geq \tilde{\underline y}_t$.

    We now prove that the inequalities regarding the stopping thresholds are strict.

    Suppose, toward a contradiction, that, at $t_1$, 
    $
        \tilde{\overline y}_{t_1}^m=\overline y_{t_1}^m
    $
    and $\overline y_{t_1}^m>\underline y_{t_1}^m$. 
    Since the continuation regions are open intervals and the upper boundaries coincide at $t_1$ and $\underline y_{t_1}^m\geq \underline {\tilde y}_{t_1}^m$, 
    $
        (\underline y_{t_1}^m,\overline y_{t_1}^m)\subset D_{t_1}^m\cap \tilde D_{t_1}^m.
    $
    For $t_0<t_1$, we have that
    $
        Q:=(t_0,t_1]\times(\underline y_{t_0}^m\,\overline y_{t_1}^m)
    $
    is contained in the common continuation region, with .

    Define
    $
        w(t,y):=\widehat{\tilde V}^m(t,y)-\widehat V^m(t,y),
    $ $
        z(t,y):=\Delta-w(t,y).
    $
    On $Q$, 
    \begin{align*}
        z_t(t,y)
        +
        \frac{1}{2}\sigma(t,y,m)^2 z_{yy}(t,y)
        -
        rz(t,y)
        =
        -r\Delta
        \leq 0.
    \end{align*}
    At the common upper boundary point $\overline y_{t_1}$,
    \begin{align*}
        z(t_1,\overline y_{t_1})
        &=
        \Delta
        -
        \left(
            \widehat{\tilde V}^m(t_1,\overline y_{t_1})-\widehat V^m(t_1,\overline y_{t_1})
        \right)
        =
        \Delta
        -
        \left(
            \tilde v(\overline y_{t_1})-v(\overline y_{t_1})
        \right)
        =
        0.
    \end{align*}
    In contrast, $z>0$ in $Q$. 
    Indeed, if $z(t,y)=0$ at some interior point of $Q$, then $w(t,y)=\Delta$. Since $0\leq w\leq\Delta$ and $w$ solves
    \begin{align*}
        w_t
        +
        \frac{1}{2}\sigma(t,y,m)^2 w_{yy}
        -
        rw
        =
        0
    \end{align*}
    on $Q$, the strong maximum principle \citep[Section 7.1.4]{Evans2010Book} implies that $w\equiv\Delta$ on $Q$. 
    But this is impossible because, at points sufficiently close to $(t_0,\underline y_{t_0})$, the payoff difference $\tilde v-v<\Delta$ since $\underline y_{t_0}<\tilde y$. 
    Thus $z>0$ in $Q$.

    As $z=\Delta -(\widehat{\tilde V}^m-\widehat V^m)$ is $\mathcal C^1$ in $y$ by \hyref{proposition:optimal:properties-value-function:general}[Proposition], $z>0$ on $Q$ and $z(t_1,\overline y_{t_1})=0$, then $\partial_y z(t_1,\overline y_{t_1})<0\iff \partial_y w(t_1,\overline y_{t_1})>0$.
    However, smooth pasting for both value functions at the common upper boundary $\overline y_{t_1}$ gives
    $\partial_y w(t_1,\overline y_{t_1})=\partial_y (\widehat{\tilde V}^m-\widehat V^m)(t_1,\overline y_{t_1})=\partial_y ({\tilde v}-v)(\overline y_{t_1})=0$, a contradiction.
    We conclude that $\tilde{\overline y}_{t_1}^m<\overline y_{t_1}^m$.
    The proof for $\tilde{\underline y}_{t_1}^m<\underline y_{t_1}^m$ is analogous.
    This proves (1), that is, that increasing $a_0$ shifts both stopping thresholds strictly downward.
    
    We now prove (2).
    Now define
    $
        C_t:=(\underline y_t^m,\overline y_t^m),
    $ $
        \tilde C_t:=(\tilde{\underline y}_t^m,\tilde{\overline y}_t^m).
    $
    Since
    $
        \tilde{\underline y}_t^m<\underline y_t^m
    $
    and
    $
        \tilde{\overline y}_t^m<\overline y_t^m
    $,
    the correspondence $\tilde C$ is a downward shift of $C$.
    The result then follows immediately from \hyref{proposition:exogenous:choice:shift:general}[Proposition].
\end{proof}

\subsection{Properties of the Belief Process under Rich Uncertainty}
\label{appendix:general:belief-process}

We show how rich uncertainty about $\theta$ as discussed in \hyref{appendix:general:assumptions}[Appendix] delivers the assumptions made about the volatility term in the belief process.
More specifically, we consider the case in which the decision-maker observes $X_t^m:=\theta m t + B_t$, where $m>0$ and $B_t$ is a standard Brownian motion independent of $\theta$, and $\theta$ and $m$ are fixed at $t=0$.

We focus on the case in which $\theta$ is uncertain; this is but a matter of normalisation.
Our analysis extends immediately to settings in which both $\theta$ and $m$ are uncertain.
For instance, if $\tilde m\sim \diff \nu_m$ and $\tilde \theta\sim \diff \nu_\theta$ are independently distributed, then $\theta :=\tilde m \tilde\theta \sim  \diff \nu_m\diff \nu_\theta $ and $m \theta \sim \frac{1}{m}\diff \nu_m\diff \nu_\theta$, where $m$ can be interpreted as a shift in the distribution of $\tilde m$.

Our main result in this section consists of proving that, for a general class of priors on $\theta$, we obtain the assumptions stated in \hyref{appendix:general:assumptions}[Appendix] and used in obtaining the results in \hyref{appendix:general:exogenous}[Appendices] and \hyref{appendix:general:optimal}.

\begin{proposition}
    \label{proposition:uncertainty-belief-properties}    
    Let $X_t^m:=\theta m t + B_t$, where $m>0$ and $B_t$ is a standard Brownian motion independent of $\theta$. 
    Suppose $\theta$ is fixed at $t=0$, distributed according to a probability measure $\nu$ on the real line that assigns zero probability to $\{\theta =0\}$, admits a probability density function that is even and log-concave, $\diff \nu(\theta)=\exp(-\psi(\theta))\diff \theta$, with $\psi\in \mathcal C^2_{\text{loc}}(\mathbb R_{++})$, and is independent of $B_t$ and $m$. 
    Define $p_t^m:=\mathbb P(\theta>0\mid \mathcal F_t^{X^m})=\pi^m_t(X_t^m)$, $Y_t^m:=2p_t^m-1$ and, for $y \in D:=(-1,1)$, let 
    $\sigma(t, y, m):=2\partial_x\pi_t^m\left((\pi_t^m)^{-1}((1+y)/2)\right)$.
    Extend $\sigma$ to $y\in \{-1,1\}$ and $m=0$ as $0$.

    Then $\sigma:\mathbb R_+\times\overline D\times\mathbb R_+\to\mathbb R_+$ satisfies:
    \begin{enumerate}[label=(\arabic*)]
        \item $\sigma(t,y,m)>0$ for every $(t,y,m)\in\mathbb R_+\times D\times\mathbb R_{++}$, and $\sigma(t,y,m)=0$ for $(y,m)\notin D\times\mathbb R_{++}$;
        \item $\sigma$ is continuous on $\mathbb R_+\times D\times\mathbb R_{++}$;
        \item $\sigma$ is locally Lipschitz continuous on $\mathbb R_{++}\times D\times\mathbb R_{++}$;
        \item $\sigma(t,y,m)=\sigma(t,-y,m)$ for every $(t,y,m)\in\mathbb R_+\times\overline D\times\mathbb R_+$;
        \item $\sigma$ is increasing in $m$, strictly so for every $t>0$ and $y\in D$ if $\psi'>0$ on a set of positive $\nu$-measure;
        \item $\sigma$ is strictly decreasing in $t$;
        \item $\sigma$ is continuous on $\mathbb R_+\times \overline D\times\mathbb R_+$, if $\lim_{z\to\infty}\frac{A'(z)A(-z)-A'(-z)A(z)}{(A(z)+A(-z))^2}=0$, where $A(z):=\int_0^\infty \exp(z \theta-\psi(\theta))\diff \theta$;\footnote{
            It can be shown that a simple sufficient condition for this is that 
            $\lim_{z\to \infty}\psi(z)/z=\infty$.
        }
        \item $\lim_{m\to 0}\sup_{t\geq0,\,y\in D}\sigma(t,y,m)=0$;
        \item for every compact $K\subset D$ and every $T>0$,
        \begin{align*}
            \lim_{m\to\infty}
            \int_0^T\inf_{y\in K}\sigma(t,y,m)^2\diff t
            =\infty.
        \end{align*}
    \end{enumerate}
\end{proposition}
\begin{proof}

    \textbf{Proof of (1).}
    Fix $t>0$; conditional on $\theta$, $X_t^m|\theta \sim N(\theta m t, t)$. 
    Let $A(t, x, m):=\int_0^\infty \exp(\theta m x -(1/2)\theta^2 m^2 t -\psi(\theta))\diff \theta$.
    Since $\diff \nu(\theta)=\exp(-\psi(\theta))\diff \theta$ and $\diff \nu(\theta)=\diff \nu(-\theta)$ for any $\theta\in \mathbb R$, then we have  
    \begin{align*}
        \pi_t^m(x)
        =\frac{\int_{\mathbb R_{++}}\exp(\theta m x-(1/2)\theta^2 m^2 t)\diff \nu(\theta)}{\int_{\mathbb R}\exp(\theta m x-(1/2)\theta^2 m^2 t)\diff \nu(\theta)}
        &=\frac{A(t,x,m)}{A(t,x,m)+A(t,-x,m)}
        \\
        &=1-\frac{A(t,-x,m)}{A(t,x,m)+A(t,-x,m)}
        =1-\pi_t^m(-x).
    \end{align*}

    Since for $m>0$ we have $A'_x>0$, then 
    \begin{align*}
        \partial_x\pi_t^m(x)
        &=\frac{A(t, -x, m)A'_x(t, x, m)+A(t, x, m)A'_x(t, -x, m)}{(A(t, x, m)+A(t, -x, m))^2}>0, 
    \end{align*}
    and $\pi_t^m(x)$ is strictly increasing in $x$ for any $t\geq 0$ and $m>0$. 
    An analogous argument shows that $\pi_t^m(x)$ is strictly decreasing (increasing) in $t$ for $x>0$ ($x<0$). 
    Since $\partial_x \pi_t^m(x)>0$ for any $t\geq 0$, $x\in \mathbb R$, and $m>0$, ${(\pi_t^m)}^{-1}$ exists, and $\sigma$ is well-defined and strictly positive on $\mathbb R_+\times D\times \mathbb R_{++}$. 
    Setting $\sigma$ to zero outside of $\mathbb R_+\times D\times \mathbb R_{++}$ delivers (1).

    \textbf{Proof of (2).}
    
    For (2), we observe that $\pi_t^m(x)$ and $\partial_x\pi_t^m(x)$ are continuous in $(t, x, m)\in \mathbb R_{+}\times \mathbb R \times \mathbb R_{++}$; consequently, $\sigma$ is continuous for $(t,y,m)\in \mathbb R_+\times D \times \mathbb R_{++}$.

    \textbf{Proof of (3).}
    
    Fix a compact set 
    $
        Q\subset \mathbb R_{++}\times D\times \mathbb R_{++}.
    $
    It is enough to prove that $\sigma$ is Lipschitz on $Q$.

    Since $Q$ is compact and $Q\subset \mathbb R_{++}\times D\times \mathbb R_{++}$, there exist constants 
    $
        0<t_0<t_1<\infty
    $, $
        0<m_0<m_1<\infty
    $, and $
        -1<y_0<y_1<1
    $
    such that
    $
        t\in[t_0,t_1]
    $, $
        m\in[m_0,m_1]
    $, and $
        y\in[y_0,y_1]
    $
    on $Q$.

    Recall that
    \begin{align*}
        A(t,x,m)
        :=
        \int_0^\infty 
        \exp\left(\theta m x-\frac{1}{2}\theta^2m^2t-\psi(\theta)\right)\diff\theta.
    \end{align*}
    It is immediate that $A(t,x,m)$ is $\mathcal C^1$ in $(t,x,m)$, and the same is true of $\pi_t^m(x)$ and $\partial_x\pi_t^m(x)$.

    Moreover, $\partial_x\pi_t^m(x)>0$ on $\mathbb R_+\times D \times\mathbb R_{++}$. 
    Therefore, by the implicit function theorem,  
    $(\pi_t^m)^{-1}((1+y)/2)$
    is $\mathcal C^1$ locally on $\mathbb R_+\times D \times\mathbb R_{++}$. Consequently,
    $
        2\partial_x\pi_t^m\left((\pi_t^m)^{-1}((1+y)/2)\right)
    $
    is $\mathcal C^1$ locally on $\mathbb R_{++}\times D\times\mathbb R_{++}$. 
    Since every $\mathcal C^1$ function is Lipschitz on compact subsets of its domain, $\sigma$ is locally Lipschitz continuous on $\mathbb R_{++}\times D\times\mathbb R_{++}$.

    \textbf{Proof of (4).}
    
    As $\diff \nu(\theta)=\diff \nu(-\theta) \implies \psi(\theta)=\psi(-\theta)\implies \pi_t^m(x)= 1-\pi_t^m(-x)$ and $\partial_x \pi_t^m(x)=\partial_{x} \pi_t^m(-x)$, 
    then 
    ${(\pi_t^m)}^{-1}((1+y)/2)=-{(\pi_t^m)}^{-1}((1-y)/2)$ and, 
    $\sigma(t,y,m)=2\partial_x \pi_t^m({(\pi_t^m)}^{-1}((1+y)/2))
    =
    2\partial_x \pi_t^m(-{(\pi_t^m)}^{-1}((1-y)/2))=\sigma(t,-y,m)$.

    \textbf{Proof of (5).}

    We proceed to showing that $\sigma$ is weakly increasing in $m$.
    By symmetry, it is enough to prove the claim for $y\geq 0$, as the case for $y<0$ follows by (4).
     
    For fixed $y\geq 0$, let $x_t^m(y):={(\pi_t^m)}^{-1}((1+y)/2)$.
    Equivalently $\logit(\pi_t^m(x_t^m(y)))=\logit((1+y)/2)$.

    Fix $t$ and define $L(t,x,m):=\logit(\pi_t^m(x))=\ln(A(t,x,m))-\ln(A(t,-x,m))$ and $L_x(t,x,m):=\partial_x L(t,x,m)$.
    Since $\partial_x \pi_t^m(x)=\pi_t^m(x)(1-\pi_t^m(x))L_x(t,x,m)$, we have that $\sigma(t,y,m)=(1/2)(1-y^2)L_x(t,x_t^m(y),m)$.
    It suffices to show $L_x(t,x_t^m(y),m)$ is strictly increasing in $m$ for $y\geq 0$.

    For $x\geq 0$, define probability measures $\mu_+$ and $\mu_-$ on $\mathbb R_{+}$ as 
    \begin{align*}
        \diff \mu_+(z)
        :=
        \frac{
            \exp\left(x z - \frac{1}{2}tz^2-\psi(z/m)\right)
        }{
            \int_0^\infty 
            \exp\left(x \tilde z - \frac{1}{2}t\tilde z^2-\psi(\tilde z/m)\right)\diff \tilde z
        }\diff z,
        \quad\text{ and }\quad
        \diff \mu_-(z)
        :=
        \frac{
            \exp\left(-x z - \frac{1}{2}tz^2-\psi(z/m)\right)
        }{
            \int_0^\infty 
            \exp\left(-x \tilde z - \frac{1}{2}t\tilde z^2-\psi(\tilde z/m)\right)\diff \tilde z
        }\diff z,
    \end{align*}
    with expectations, variances, and covariances under $\mu_\pm$ denoted by $\mathbb E_\pm$, $\operatorname{Var}_\pm$, and $\operatorname{Cov}_\pm$.
    Note that, since  
    $A(t, x, m)=(1/m)\int_0^\infty \exp(x z - \frac{1}{2}tz^2-\psi(z/m))\diff z$,  
    $L(t,x,m)=\ln A(t, x, m) - \ln A(t, -x, m)$, 
    and 
    $\frac{\partial}{\partial m}\left[-\psi(z/m)\right] = z/m^2\psi'(z/m)$, 
    direct differentiation gives 
    \begin{align*}
        L_x&:=\partial_x L(t,x,m) = \mathbb E_+[Z]+\mathbb E_-[Z]\\
        L_{xx}&:=\partial_{xx} L(t,x,m) =\operatorname{Var}_+(Z)-\operatorname{Var}_-(Z)\\
        L_{m}&:=\partial_{m} L(t,x,m) = \mathbb E_+[Z/m^2 \psi'(Z/m)]-\mathbb E_-[Z/m^2\psi'(Z/m)] \\
        L_{xm}&:=\partial_{xm} L(t,x,m) = \operatorname{Cov}_+(Z,Z/m^2 \psi'(Z/m))+\operatorname{Cov}_-(Z,Z/m^2 \psi'(Z/m)),
    \end{align*}
    whereas implicit differentiation of $L(x_t^m(y),m)$ yields 
    \begin{align*}
        \frac{\diff}{\diff m}x_t^m(y)
        =
        -\frac{L_m}{L_x}
        =
        -\frac{\mathbb E_+[Z/m^2 \psi'(Z/m)]-\mathbb E_-[Z/m^2\psi'(Z/m)]}{\mathbb E_+[Z]+\mathbb E_-[Z]}.
    \end{align*}
    Consequently,
    $$
        \frac{\diff}{\diff m}L_x(x_t^m(y),m)=L_{xm}(x_t^m(y),m)+L_{xx}(x_t^m(y),m)\frac{\diff}{\diff m}x_t^m(y)
    $$
    which expands to
    \begin{align*}
        \frac{\diff}{\diff m}L_x(x_t^m(y),m)
        &=
        \operatorname{Cov}_+(Z,Z/m^2 \psi'(Z/m))+\operatorname{Cov}_-(Z,Z/m^2 \psi'(Z/m))
        \\
        &\qquad
        -
        (\operatorname{Var}_+(Z)-\operatorname{Var}_-(Z))\frac{\mathbb E_+[Z/m^2 \psi'(Z/m)]-\mathbb E_-[Z/m^2 \psi'(Z/m)]}{\mathbb E_+[Z]+\mathbb E_-[Z]}.
    \end{align*}
    Hence, it is sufficient to prove that 
    \begin{align*}
        &(\mathbb E_+[Z]+\mathbb E_-[Z])(\operatorname{Cov}_+(Z,Z/m^2 \psi'(Z/m))+\operatorname{Cov}_-(Z,Z/m^2 \psi'(Z/m)))
        \\
        &\qquad 
        -
        (\operatorname{Var}_+(Z)-\operatorname{Var}_-(Z))(\mathbb E_+[Z/m^2 \psi'(Z/m)]-\mathbb E_-[Z/m^2 \psi'(Z/m)])\geq 0.
    \end{align*}
    Writing $w(z):=1/m^2 \psi'(z/m)$ and expanding the left-hand side gives
    \begin{align*}
        &
        \mathbb E_+[Z]\operatorname{Cov}_+(Z,Z w(Z))
        +
        \mathbb E_+[Z]\operatorname{Cov}_-(Z,Z w(Z))
        +
        \mathbb E_-[Z]\operatorname{Cov}_+(Z,Z w(Z))
        +
        \mathbb E_-[Z]\operatorname{Cov}_-(Z,Z w(Z))
        \\
        &\qquad
        -
        \operatorname{Var}_+(Z)\mathbb E_+[Z w(Z)]
        +
        \operatorname{Var}_+(Z)\mathbb E_-[Z w(Z)]
        +
        \operatorname{Var}_-(Z)\mathbb E_+[Z w(Z)]
        -
        \operatorname{Var}_-(Z)\mathbb E_-[Z w(Z)]
        \\
        &
        =
        \left(
            \mathbb E_+[Z]\operatorname{Cov}_+(Z,Z w(Z))
            -
            \operatorname{Var}_+(Z)\mathbb E_+[Z w(Z)]
        \right)
        +
        \left(
            \mathbb E_-[Z]\operatorname{Cov}_-(Z,Z w(Z))
            -
            \operatorname{Var}_-(Z)\mathbb E_-[Z w(Z)]
        \right)
        \\
        &\qquad
        +
        \mathbb E_+[Z]\operatorname{Cov}_-(Z,Z w(Z))
        +
        \mathbb E_-[Z]\operatorname{Cov}_+(Z,Z w(Z))
        +
        \operatorname{Var}_+(Z)\mathbb E_-[Z w(Z)]
        +
        \operatorname{Var}_-(Z)\mathbb E_+[Z w(Z)].
    \end{align*}
    We note that the last four terms are non-negative.
    Indeed, it is immediate that $\mathbb E_\pm[Z]> 0$, $\operatorname{Var}_\pm(Z)> 0$, and, $\mathbb E_\pm[Z/m^2 \psi'(Z/m)]\geq 0$.
    Given that $\psi$ is convex and even, $\psi'(z)\geq 0$ for $z\geq 0$. 
    Then, $w(z)=1/m^2 \psi'(z/m)$ is non-decreasing in $z$, so, for $z\geq 0$, $z w(z)$ is increasing in $z$ and $\operatorname{Cov}_\pm(Z,Z w(Z))\geq 0$.
    It only remains to show that 
    $$\mathbb E_\pm[Z]\operatorname{Cov}_\pm(Z,Z w(Z))-\operatorname{Var}_\pm(Z)\mathbb E_\pm[Z w(Z)]\geq 0.$$
    We prove this for a generic probability distribution on $\mathbb R_{++}$, dropping the subscripts.

    \begin{lemma}
        \label{lemma:cov12exp1-var1exp2}
        Let $\mu$ be a probability measure on $\mathbb R_{++}$, $Z\sim \mu$, $w:\mathbb R_{++}\to \mathbb R_{+}$ be a non-decreasing function, and $H(z):=z w(z)$.
        Define $\mu_1:=\mathbb E[Z]$, $v:=\operatorname{Var}(Z)$, $h:=\mathbb E[H(Z)]$, and $c:=\operatorname{Cov}(Z,H(Z))$.
        Then, $c  \mu_1-v h\geq 0$.
    \end{lemma}
    \begin{proof}
        Note that
        \begin{align*}
            c\mu_1-vh
            &=
            \left(
                \mathbb E[Z H(Z)]-\mathbb E[Z]\mathbb E[H(Z)]
            \right)\mathbb E[Z]
            -
            \left(
                \mathbb E[Z^2]-(\mathbb E[Z])^2
            \right)\mathbb E[H(Z)]
            \\
            &=
            \mathbb E[Z]\mathbb E[ZH(Z)]
            -
            \mathbb E[Z^2]\mathbb E[H(Z)]
            \\
            &=
            \mathbb E[Z]\mathbb E[Z^2w(Z)]
            -
            \mathbb E[Z^2]\mathbb E[Zw(Z)].
        \end{align*}
        Now define probability measure $\widetilde\mu$ as $\diff\widetilde\mu(z) := \frac{z}{\mathbb E[Z]}\,\diff\mu(z)$. 
        Then, $\widetilde{\mathbb E}[f(Z)] = \frac{\mathbb E[Zf(Z)]}{\mathbb E[Z]}$. 
        Therefore, we have 
        \begin{align*}
            c\mu_1-vh
            &=
            (\mathbb E[Z])^2
            \left(
                \widetilde{\mathbb E}[Zw(Z)]
                -
                \widetilde{\mathbb E}[Z]\,
                \widetilde{\mathbb E}[w(Z)]
            \right)
            =
            (\mathbb E[Z])^2
            \operatorname{Cov}_{\widetilde\mu}(Z,w(Z)).
        \end{align*}
        As $w$ is non-decreasing, $\operatorname{Cov}_{\widetilde\mu}(Z,w(Z))\geq 0$, and we conclude that $c\mu_1-vh\geq 0$.
    \end{proof}
    Applying \hyref{lemma:cov12exp1-var1exp2}[Lemma] to $\mu_+$ and $\mu_-$ gives $\mathbb E_\pm[Z] \operatorname{Cov}_\pm(Z,Z w(Z)) - \operatorname{Var}_\pm(Z) \mathbb E_\pm[Z w(Z)]\geq 0$.
    We conclude that the inequality holds, strictly so given if $\psi'\ne 0$ on a set of positive $\nu$-measure. 
    This proves (5).

    \textbf{Proof of (6).}

    We proceed to show that $\sigma(t,y,m)$ is strictly decreasing in $t$. 
    The proof strategy is the same as for the proof of (5); for the same reasons as before, it is enough to prove the claim for $y\geq 0$.
    Fix $m>0$ and write $L(t,x,m):=\logit(\pi_t^m(x))$. 
    With $\mu_+$ and $\mu_-$ defined as above, direct differentiation gives
    \begin{align*}
        L_t&=-\frac{1}{2}\left(\mathbb E_+[Z^2]-\mathbb E_-[Z^2]\right),\\
        L_{xt}&=-\frac{1}{2}\left(\operatorname{Cov}_+(Z,Z^2)+\operatorname{Cov}_-(Z,Z^2)\right).
    \end{align*}
    Since $L(t,x_t^m(y),m)=\logit((1+y)/2)$, implicit differentiation gives 
    \[
        \frac{\diff}{\diff t}x_t^m(y)
        =
        -\frac{L_t}{L_x}.
    \]
    Therefore, using the expressions for $L_x$ and $L_{xx}$ already obtained above,
    \begin{align*}
        \frac{\diff}{\diff t}L_x(t,x_t^m(y),m)
        &=
        L_{xt}+L_{xx}\frac{\diff}{\diff t}x_t^m(y)
        \\
        &=
        -\frac{1}{2}\left(\operatorname{Cov}_+(Z,Z^2)+\operatorname{Cov}_-(Z,Z^2)\right)
        \\
        &\quad
        +
        \frac{1}{2}
        \left(\operatorname{Var}_+(Z)-\operatorname{Var}_-(Z)\right)
        \frac{\mathbb E_+[Z^2]-\mathbb E_-[Z^2]}{\mathbb E_+[Z]+\mathbb E_-[Z]}.
    \end{align*}
    Hence, by exactly the same argument used to prove (5), now applying \hyref{lemma:cov12exp1-var1exp2}[Lemma] with $H(z)=z^2$ and $w(z)=z$, we obtain
    $\frac{\diff}{\diff t}L_x(t,x_t^m(y),m)< 0$.
    Since $\sigma(t,y,m)=1/2 (1-y^2)L_x(t,x_t^m(y),m)$, $\sigma(t,y,m)$ is strictly decreasing in $t$. 
    The case $y<0$ follows by symmetry.

    \textbf{Proof of (7) and (8).}
    Recall that 
    \begin{align*}
        \partial_x\pi_t^m(x)
        &=\frac{A(t, -x, m)A'_x(t, x, m)+A(t, x, m)A'_x(t, -x, m)}{(A(t, x, m)+A(t, -x, m))^2}>0. 
    \end{align*}
    Since, $\partial_x \pi_0^m(x)= m \partial_x \pi_0^1(m x)$, and, by assumption, 
    \begin{align*}
        \lim_{x \to \infty}\frac{A(0, -x, 1)A'_x(0, x, 1)-A(0, x, 1)A'_x(0, -x, 1)}{(A(0, x, 1)+A(0, -x, 1))^2}=0,
    \end{align*}
    it remains to be shown that 
    \begin{align*}
        \lim_{x \to \infty}\frac{A(0, x, 1)A'_x(0, -x, 1)}{(A(0, x, 1)+A(0, -x, 1))^2}=0,
    \end{align*}
    so that $\lim_{x\to \infty}\partial_x \pi_0^1(x)=0$.
    As $\lim_{x\to \infty}A'_x(0, -x, 1)=0$ and $\lim_{x\to \infty}A(0, x, 1)=\infty$, then, 
    \begin{align*}
        \lim_{x \to \infty}\frac{A(0, x, 1)A'_x(0, -x, 1)}{(A(0, x, 1)+A(0, -x, 1))^2}
        \leq 
        \lim_{x \to \infty}\frac{A'_x(0, -x, 1)}{A(0, x, 1)}
        =
        0.
    \end{align*}
    Moreover, $\lim_{x\to \infty}\pi_0^m(x)=1$ and $\lim_{x\to -\infty}\pi_0^m(x)=0$.
    Then, $\lim_{y\to 1}{(\pi_0^m)}^{-1}((1+y)/2)=\infty$ and 
    $\lim_{y\to 1}\sigma(0,y,m)=2\partial_x \pi_0^m({(\pi_0^m)}^{-1}((1+y)/2))=0$.
    Since we have shown that $\sigma$ is decreasing in $t$, for any $t\geq 0$ and any $m>0$, $\lim_{y\to 1}\sigma(t,y,m)\leq \lim_{y\to 1}\sigma(0,y,m)=0$.
    A symmetric argument shows that, for any $t\geq 0$ and $m>0$, 
    $\lim_{y\to -1}\sigma(t,y,m)=0$.
    
    Finally, we check continuity at $m=0$.
    As $\partial_x\pi_0^1(x)$ is continuous in $x\in \mathbb R$, positive, and $\lim_{x\to \infty}\partial_x\pi_0^1(x)=\lim_{x\to-\infty}\partial_x\pi_0^1(x)=0$, then $\exists C<\infty:$ $\sup_{x\in \mathbb R}\partial_x\pi_0^1(x)\leq C$.
    Since $\partial_x\pi_0^m(x)=m \partial_x\pi_0^1(mx)$, then $\partial_x\pi_0^m(x)\leq m C$.
    This implies that 
    \begin{align*}
        &0\leq \sup_{t\geq 0,y\in D}\sigma(t,y,m)\leq \sup_{y\in D}\sigma(0,y,m)= 2m C
        \implies
        \lim_{m\to 0}\sup_{t\geq 0,y\in D}\sigma(t,y,m)=0.
    \end{align*}
    This proves (8). 
    Under the additional condition in (7), the preceding boundary argument also proves continuity at $y=\pm1$; together with interior continuity and continuity at $m=0$, this proves (7).

    \textbf{Proof of (9).}

    Fix a compact $K\subset D$ and choose $\varepsilon\in(0,1/2)$ such that 
    $
        K\subset[-1+2\varepsilon,1-2\varepsilon].
    $

    For $n>0$ and $\xi\in\mathbb R$, define
    \begin{align*}
        B(\xi,n)
        &:=
        \int_0^\infty 
        \exp\left(\xi \theta-\frac{1}{2}\theta^2-\psi(\theta/n)\right)\diff \theta,
        \\
        \Pi^n(\xi)
        &:=
        \frac{B(\xi,n)}{B(\xi,n)+B(-\xi,n)}.
    \end{align*}
    Then, $\pi_t^m(x)=\Pi^{m\sqrt{t}}(x/\sqrt{t})$ and $\partial_x\pi_t^m(x)=(1/\sqrt{t})\partial_\xi \Pi^{m\sqrt{t}}(x/\sqrt{t})$.

    We claim that there is $c_\varepsilon>0$ such that 
    $\inf_{n \geq 1}\inf_{y\in[-1+2\varepsilon,1-2\varepsilon]}\partial_\xi \Pi^{n}({(\Pi^n)}^{-1}((1+y)/2)) \geq c_\varepsilon$. 
    Let $K_\xi :=[-\bar\xi,\bar\xi]$, where $\bar \xi := {(\Pi^\infty)}^{-1}(1-\varepsilon)$.

    First, note that, since $\psi$ is convex and even, it is minimised at $0$, so, for any $\xi\in [-\bar\xi,\bar \xi]$, 
    $B'_\xi(\xi,n)=\int_0^\infty \theta\exp\left(\xi \theta-\frac{1}{2}\theta^2-\psi(\theta/n)\right)\diff \theta\leq \int_0^\infty \theta\exp\left(\bar\xi \theta-\frac{1}{2}\theta^2-\psi(0)\right)\diff \theta=B'_\xi(\bar\xi,\infty)$ 
    and, similarly, 
    $B''_{\xi\xi}(\xi,n)\leq B''_{\xi\xi}(\bar\xi,\infty)= \int_0^\infty \theta^2\exp\left(\bar\xi \theta-\frac{1}{2}\theta^2-\psi(0)\right)\diff \theta$.
    Consequently, for $\xi\in [-\bar\xi,\bar \xi]$, 
    $|B(\xi,n)-B(\xi',n)|\leq |\xi-\xi'||B'_\xi(\bar\xi,\infty)|$, and 
    $|B'_{\xi}(\xi,n)-B'_{\xi}(\xi',n)|\leq |\xi-\xi'||B''_{\xi\xi}(\bar\xi,\infty)|$.
    For $n\in[1,N_\varepsilon]$, where $N_\varepsilon<\infty$ is to be determined, continuity and strict positivity give
    $
        \inf_{1\leq n\leq N_\varepsilon}
        \inf_{p\in[\varepsilon,1-\varepsilon]}
        \partial_\xi \Pi^{n}({(\Pi^n)}^{-1}(p))
        >0.
    $ 
    By dominated convergence and equicontinuity, $B(\cdot,n)$ and
    $\partial_\xi B(\cdot,n)$ converge uniformly to $B(\cdot,\infty)$
    and $\partial_\xi B(\cdot,\infty)$ as $n\to \infty$ for $\xi \in [-\bar \xi,\bar \xi]$.
    Then, 
    as $n\to \infty$, $\Pi^n$ converges uniformly to $\Pi^\infty$ and $\partial_\xi \Pi^n$ to $\partial_\xi \Pi^\infty$ for $\xi \in [-\bar \xi,\bar \xi]$, 
    where 
    $$\Pi^\infty(\xi):=\frac{\int_0^\infty \exp(\xi \theta - 1/2\theta^2)\diff \theta}{\int_{-\infty}^\infty \exp(\xi \theta - 1/2\theta^2)\diff \theta}=\Phi(\xi)\text{ and }\partial_\xi \Pi^\infty=\phi,$$
    where $\Phi$ and $\phi$ denote the cumulative and probability density functions of the standard normal distribution. 
    Since $\Pi^n\to\Phi$ locally uniformly and $\Phi(-\bar\xi)<\varepsilon<1-\varepsilon<\Phi(\bar\xi)$,
    there exists $N_\varepsilon<\infty$ such that, for all $n\geq N_\varepsilon$,$\Pi^n(-\bar\xi)<\varepsilon$ and $\Pi^n(\bar\xi)>1-\varepsilon$.
    Since $\Pi^n$ is strictly increasing, it follows that
    ${(\Pi^n)}^{-1}(p)\in[-\bar\xi,\bar\xi]$
    for all $p\in[\varepsilon,1-\varepsilon]$ and all $n\geq N_\varepsilon$.
    It follows that $\lim_{n\to \infty}\partial_\xi \Pi^{n}({(\Pi^n)}^{-1}((1+y)/2))=\phi(\Phi^{-1}((1+y)/2))$, uniformly for $y\in [-1+2\varepsilon, 1-2\varepsilon]$, i.e., $\exists N_\varepsilon<\infty$, such that $\forall y\in [-1+2\varepsilon, 1-2\varepsilon]$ and $\forall n>N_\varepsilon$, $\partial_\xi \Pi^{n}({(\Pi^n)}^{-1}((1+y)/2))\geq \phi(\Phi^{-1}(1-\varepsilon))/2$.
    We conclude that there is 
    $c_\varepsilon>0$ such that 
    $\inf_{n \geq 1}\inf_{y\in[-1+2\varepsilon,1-2\varepsilon]}\partial_\xi \Pi^{n}({(\Pi^n)}^{-1}((1+y)/2)) \geq c_\varepsilon$. 

    Now take $t\in [m^{-2},T]$.
    Then, $m\sqrt{t}\geq 1$ and, for every $y \in K$, 
    $$\partial_x\pi_t^m(x_t^m(y))=(1/\sqrt{t})\partial_\xi \Pi^{m\sqrt{t}}\left({\left(\Pi^{m\sqrt{t}}\right)}^{-1}((1+y)/2)\right)\geq c_\varepsilon/\sqrt{t}.$$
    Since $\sigma(t,y,m)=2\partial_x\pi_t^m(x_t^m(y))$, then, for $t\in [m^{-2},T]$, $\inf_{y\in K}\sigma(t,y,m)\geq 2 c_\varepsilon/\sqrt{t}$.
    Therefore,
    $$\lim_{m\to \infty}\int_0^T\inf_{y\in K}\sigma(t,y,m)^2\diff t\geq \lim_{m\to \infty} 4 c_\varepsilon^2\ln(Tm^2)=\infty,$$ 
    proving the claim.
\end{proof}

\FloatBarrier \newpage

\section{Related Models}
\label{appendix:variants}

\subsection{Uncertain Preference Intensity}
\label{appendix:variants:FSS}

In our baseline model, the decision-maker chooses between two alternatives, $a$ and $b$, whose payoffs depend on whether $\theta>0$ or $\theta<0$, and, prior to choosing, they can learn about $\theta$ by observing $X_t=\theta \kappa^{-1} t + B_t$, where $B_t$ is a Brownian motion independent of $\theta$.

In the closely related model of \citet{DrugowitschMoreno-BoteChurchlandShadlenPouget2012JNeuro} and \citet{FudenbergStrackStrzalecki2018AER}, the decision-maker is not just uncertain about which alternative is optimal, but also about the relative preference intensity.
Specifically, they consider a case which is equivalent to having $\theta$ is normally distributed with mean $m_0$ and variance $v_0$ and payoffs given by $u(a,\theta)=\theta$ and $u(b,\theta)=-\theta$.
In this case, posterior beliefs about $\theta$ are at any point in time normally distributed with mean $m_t:=\mathbb E[\theta\mid \mathcal F^X_t]=v_t (m_0/v_0)+v_t \kappa^{-1} X_t$ and $v_t:=\operatorname{Var}(\theta\mid \mathcal F^X_t)=(v_0^{-1}+\kappa^{-2}t)^{-1}$. 
The problem in \citet{FudenbergStrackStrzalecki2018AER} is then to choose a stopping time $\tau$ adapted to $\mathcal F^X$ to maximise the expected payoff net of the time cost, i.e.,
\begin{align}
    \sup_\tau \mathbb E[|m_\tau|-c\tau].
    \label{equation:FSS-objective}
\end{align}

This setting ties-in both uncertainty about preferences and uncertainty about problem complexity, since the realised preference intensity is given by $|\theta|$ and the realised problem complexity is determined by $\kappa |\theta|^{-1}$.
The double role of $\theta$, affecting both the learning rate and preferences, differentiates this case from both our baseline setting in the main text as well as the extensions we consider in \hyref{appendix:general}.
Nevertheless, we show that the qualitative implications of \hyref{theorem:exogenous-stopping:speed-accuracy}[Theorems] and \ref{theorem:optimal-stopping:speed-accuracy} remain the same.

\citet[Theorem 4, Proposition 3]{FudenbergStrackStrzalecki2018AER} show that the optimal stopping time is given by $\tau^\kappa:=\inf\{t>0||m_t|\geq\overline m_t^\kappa\}$ for some Lipschitz continuous $\overline m^\kappa:\mathbb R_+\to \mathbb R_{++}$ that implicitly depends on $v_0$, $\kappa$, and $c$ and satisfies $\overline m_t^\kappa/\sqrt{v_t}$ being decreasing in $t$.
Letting $p_t:=\mathbb P(\theta>0|\mathcal F^X_t)=\Phi(m_t v_t^{-1/2})$, 
and $\overline p_t^\kappa = \Phi(\overline m_t^\kappa v_t^{-1/2})$, we then obtain $\tau^\kappa:=\inf\{t>0|p_t \notin C_t^\kappa\}$, where $C_t^\kappa:=(1-\overline p_t^\kappa,\overline p_t^\kappa)$ with $\overline p:\mathbb R_+\to (1/2,1)$, Lipschitz continuous and decreasing in $t$.

We note that, holding fixed $\kappa$, with greater realised problem complexity $\kappa |\theta|^{-1}$, the expected stopping time increases and accuracy decreases. 
Indeed, considering the case in their Proposition 3, with $m_0=0$, \hyref{corollary:exogenous-stopping:symmetric-thresholds:speed}[Corollary] applies, and $\tau^\kappa$ increases in the first-order stochastic dominance sense in $|\theta|^{-1}$.
In contrast, accuracy, which is here given by $\mathbb P(\alpha_{\tau^\kappa}\in \argmax_\alpha u(\alpha,\theta))=\overline p_{\tau^\kappa}^\kappa$, decreases in $|\theta|^{-1}$, owing to the fact that $\overline p_t$ decreases in $t$ and $\tau^\kappa$ increases in the first-order stochastic dominance sense in $|\theta|^{-1}$.

We show that the expected optimal stopping time for this problem is also non-monotone in $\kappa$.
This, as we mentioned in \hyref{section:optimal-stopping:related-models}[Section], corresponds to a \emph{recognised} distributional change in effective problem complexity.
\begin{proposition}
    Let $\tau^\kappa$ be the earliest optimal stopping time that solves \eqref{equation:FSS-objective}. 
    Then, 
    $\lim_{\kappa \to 0}\mathbb E[\tau^\kappa]=\lim_{\kappa\to \infty}\mathbb E[\tau^\kappa]=0$.
\end{proposition}
\begin{proof}
    First, note that 
    $\mathbb E[|m_{\tau^\kappa}|-c\tau^\kappa]\geq |m_0| \iff 
    c \mathbb E[\tau^\kappa] \leq 
    \mathbb E[|m_{\tau^\kappa}|-|m_0|]\leq \mathbb E[|m_{\tau^\kappa}-m_0|]\leq \sqrt{\mathbb E[(m_{\tau^\kappa}-m_0)^2]}$. 
    As, by the law of total variance, $\mathbb E[(m_t-m_0)^2]=\operatorname{Var}(m_t)=\operatorname{Var}(\mathbb E[\theta\mid \mathcal F_t^X])=\operatorname{Var}(\theta)-\mathbb E[\operatorname{Var}(\theta)\mid \mathcal F_t^X]=v_0-v_t$,
    and $v_0-v_t=v_0 -{(v_0^{-1}+\kappa^{-2}t)}^{-1}=\frac{v_0^2\kappa^{-2}t}{(1+v_0\kappa^{-2}t)}\leq v_0^2\kappa^{-2}t$.
    Hence, 
    $c \mathbb E[\tau^\kappa] \leq \sqrt{\mathbb E[(m_{\tau^\kappa}-m_0)^2]}=\sqrt{\mathbb E[v_0-v_{\tau^\kappa}]}\leq v_0\kappa^{-1}\sqrt{\mathbb E[\tau^{\kappa}]} \iff \mathbb E[\tau^{\kappa}]\leq v_0^2\kappa^{-2} c^{-2} \to 0$ as $\kappa\to \infty$.

    To show that $\lim_{\kappa\to 0}\mathbb E[\tau^{\kappa}]=0$ fix a deterministic time $T^\kappa$ such that the decision-maker stops at $t=\kappa$.
    Since $\mathbb E[(\theta-m_{T^\kappa})^2]=v_{T^\kappa}={(v_0^{-1}+\kappa^{-1})}^{-1}\to 0$ as $\kappa\to 0$, $m_{T^\kappa}\to \theta$ in $L^2$, and, therefore, also in $L^1$.
    Consequently, $\mathbb E[|\theta|-|m_{T^\kappa}|]\leq \mathbb E[|\theta-m_{T^\kappa}|]\to 0$ as $\kappa \to 0$.
    Since this stopping time is suboptimal, 
    $\mathbb E[|m_{T^\kappa}|]-c \kappa=\mathbb E[|m_{T^\kappa}|-c T^\kappa]\leq \mathbb E[|m_{\tau^\kappa}|-c \tau^\kappa]$.
    Moreover, perfect information is better than the information obtained upon stopping optimally, i.e., $\mathbb E[|m_{\tau^\kappa}|-c \tau^\kappa]\leq \mathbb E[|\theta|]- c\mathbb E[\tau^\kappa]$.
    We then have 
    $c\mathbb E[\tau^\kappa]\leq \mathbb E[|\theta|]- \mathbb E[|m_{T^\kappa}|]+c \kappa\to 0$ as $\kappa \to 0$.
\end{proof}
\begin{proposition}
    Let $\tau^\kappa$ be the earliest optimal stopping time that solves \eqref{equation:FSS-objective}. 
    Then, $\mathbb E[\overline p^\kappa_{\tau^\kappa}]$ decreases in $\kappa$.
\end{proposition}

\begin{proof}
    We compare problems in information time.
    For each $\kappa>0$, let $s:=\kappa^{-2}t$ and define
    $
        Z_s^\kappa:=\kappa^{-1}X_{\kappa^2 s}.
    $
    Since $X_t=\theta\kappa^{-1}t+B_t$, we have
    $
        Z_s^\kappa
        =
        \theta s+\widetilde B_s,
    $
    where $\widetilde B_s:=\kappa^{-1}B_{\kappa^2s}$ is a Brownian motion.
    Thus, after the time-change $t=\kappa^2s$, the posterior belief process is independent of $\kappa$.
    In particular,
    $
        v_s:=\operatorname{Var}(\theta\mid \mathcal F^Z_s)
        =
        (v_0^{-1}+s)^{-1}
    $
    and 
    $
        p_s:=\mathbb P(\theta>0\mid \mathcal F^Z_s)
    $ 
    do not depend on $\kappa$.
    The only effect of $\kappa$ is to change the cost per unit of information time from $c$ to $c\kappa^2$, since
    $
        c \diff t=c\kappa^2 \diff s.
    $

    By \citet[Theorem 4]{FudenbergStrackStrzalecki2018AER}, the optimal stopping time is the first exit time from a symmetric continuation region.
    By \citet[Theorem 4, Proposition 3]{FudenbergStrackStrzalecki2018AER}, the associated posterior-mean boundary is decreasing in the cost parameter.
    Hence, writing $C:=c\kappa^2$ for the cost per unit of information time, there exists a boundary
    $
        \overline p^C_s
        :=
        \Phi\left(
            \frac{\overline m^C_s}{\sqrt{v_s}}
        \right)
        \in(1/2,1)
    $
    such that the earliest optimal stopping time in information time is
    $
        S^C
        :=
        \inf\{s\geq0:p_s\notin(1-\overline p^C_s,\overline p^C_s)\},
    $
    and such that, whenever $C_1<C_2$,
    $
        \overline m^{C_2}_s\leq \overline m^{C_1}_s
        $ for all $s\geq0.
    $
    Therefore,
    $
        \overline p^{C_2}_s\leq \overline p^{C_1}_s
        $ for all $s\geq0.
    $
    It follows that the continuation region under $C_2$ is contained in the continuation region under $C_1$, and hence
    $S^{C_2}\leq S^{C_1}$.

    Now $p_s=\mathbb P(\theta>0\mid\mathcal F^Z_s)$ is a bounded martingale.
    Since $\max\{p,1-p\}=1/2+|p-1/2|$ is convex, the process $\max\{p_s,1-p_s\}$ is a bounded submartingale.
    Thus, by optional sampling, if $C_1<C_2$, then
    $
        \mathbb E[\max\{p_{S^{C_2}},1-p_{S^{C_2}}\}]
        \leq
        \mathbb E[\max\{p_{S^{C_1}},1-p_{S^{C_1}}\}].
    $
    At the stopping time $S^C$, the posterior belief hits the boundary of the continuation region, so $\max\{p_{S^{C}},1-p_{S^{C}}\}=\overline p^C_{S^C}$.
    Therefore,
    $
        \mathbb E[\overline p^{C_2}_{S^{C_2}}]
        \leq
        \mathbb E[\overline p^{C_1}_{S^{C_1}}].
    $

    Finally, returning to calendar time, the problem with parameter $\kappa$ corresponds to the information-time cost
    $
        C^\kappa:=c\kappa^2,
    $
    and
    $
        S^{C^\kappa}=\kappa^{-2}\tau^\kappa,
        $ $
        \overline p^{C^\kappa}_{S^{C^\kappa}}
        =
        \overline p^\kappa_{\tau^\kappa}.
    $
    Hence, if $0<\kappa_1<\kappa_2$, then $C^{\kappa_1}<C^{\kappa_2}$, and the preceding inequality gives
    $
        \mathbb E[\overline p^{\kappa_2}_{\tau^{\kappa_2}}]
        \leq
        \mathbb E[\overline p^{\kappa_1}_{\tau^{\kappa_1}}].
    $
    Thus $\kappa\mapsto\mathbb E[\overline p^\kappa_{\tau^\kappa}]$ is decreasing.
\end{proof}

\subsection{Optimal Stopping with Discounting}
\label{appendix:variants:discounting}

We now consider a variation of the problem in \hyref{section:setup}[Section] in which the decision-maker, instead of bearing an explicit cost of time, exponentially discounts payoffs at a rate $r>0$. 
Formally, the decision-maker chooses a stopping time $\tau^*$ to maximise their expected discounted payoff, $$\sup_\tau \mathbb E[\exp(-r\tau )\max_\alpha\mathbb E[u(\alpha,\theta)\mid \mathcal F_\tau^X]].$$
This is a special case of the problem examined in \hyref{appendix:general:optimal}[Appendix], where $r>0=c$ and $\diff p_t = 2 \kappa^{-1} p_t(1-p_t)\diff \tilde B_t$.
Therefore, from \hyref{proposition:optimal:speed-accuracy:general}[Proposition], we know that $\mathbb P(\alpha_{\tau^*}\in \argmax_\alpha u(\alpha,\theta))$ decreases in $\kappa$ and that $\mathbb E[\tau^*]$ is non-monotone in $\kappa$.

We make the simplifying assumption that $u(\alpha,\theta)=1_{\alpha = \theta}$.
Since beliefs evolve according to a time-homogeneous stochastic differential equation, by \hyref{proposition:optimal:stopping-thresholds:general}[Proposition][(1) and (5)], the optimal stopping time is characterised by two fixed stopping thresholds, $\tau^*:=\inf\{t>0|p_t\notin (\underline p, \overline p)\}$.
As $v(p):=\max_\alpha u(\alpha,p)$ is symmetric about 1/2, it follows from \hyref{proposition:optimal:stopping-thresholds:general}[Proposition][(6)] that $\overline p = 1-\underline p$.
Since $\tau^*=\inf\{t>0|p_t\notin (1-\overline p,\overline p)\}=\inf\{t>0|2\kappa^{-1} X_t^\kappa  \notin (-\logit(\overline p),\logit(\overline p))\}$, from \citet[p. 309, 3.0.5(b)]{BorodinSalminen2002Book} we obtain 
\begin{align*}
    &\sup_\tau \mathbb E[\exp(-r\tau )\max_\alpha\mathbb E[u(\alpha,\theta)\mid \mathcal F_\tau^X]]
    =
    \mathbb E[\exp(-r\tau^*)|\theta =1]\overline p
    =
    \frac{\exp(\logit(\overline p)/2)/2}{\cosh(1/2 \sqrt{2 r + \kappa^{-2}} \kappa \logit(\overline p))}.
\end{align*}
Taking first-order conditions, we obtain
$\overline p = \logit^{-1}(2 \tilde \kappa^{-1} \tanh^{-1}(\tilde \kappa^{-1}))$, where $\tilde \kappa:=\sqrt{2 r \kappa^2+1}$.
From \citet[p. 309, 3.0.1]{BorodinSalminen2002Book}, the expected stopping time is given by $$\mathbb E[\tau^*]=\kappa^{2}\tanh(\logit(\overline p)/2)\logit(\overline p)/2$$ and, substituting, we obtain 
$$\mathbb E[\tau^*]=\frac{\tilde \kappa^2-1}{r\tilde \kappa} \tanh^{-1}(\tilde \kappa^{-1})\tanh(\tilde \kappa^{-1} \tanh^{-1}(\tilde \kappa^{-1})),$$ which is quasi-concave in $\tilde\kappa\geq 1$ and, hence, quasi-concave in $\kappa\geq 0$.

\subsection{Static Costly Information Acquisition}
\label{appendix:variants:cia}

In our baseline model, the decision-maker chooses a time at which to stop acquiring information.
By choosing a stopping time $\tau$, the decision-maker effectively chooses a distribution over posterior beliefs $\pi^\tau$, where $p_\tau \sim \pi^\tau$, such that $\mathbb E_{p_\tau\sim \pi^\tau}[p_\tau]=p_0$, which entails a benefit $B(\pi^\tau)=\mathbb E[\max_{\alpha}u(\alpha,p_\tau)]$ and a cost $C(\pi^\tau)=c\mathbb E[\tau]$.

In this section, we turn to the connection between the non-monotonic relation between speed and accuracy and the cost of information acquisition.
Specifically, we consider a decision-maker facing uncertainty about $\theta\in \Theta$ and can acquire a Blackwell experiment $\pi$ to learn about $\theta$.
For simplicity, we focus on the case in which $\Theta$ is finite. 
Following existing literature \citep[e.g.,][]{KamenicaGentzkow2011AER,CaplinDeanLeahy2022JPE,LipnowskiRavid2023WP}, we model Blackwell experiments as a distribution over posterior beliefs, that is, $\pi\in \Pi:=\{\tilde\pi\in \Delta(\Delta(\Theta)) | \mathbb E_\pi[p]=p_0\}$.
We denote by $C:\Pi \to \overline{\mathbb R}_+$
the cost of the experiment, which we assume is strictly increasing with respect to the Blackwell order, continuous with respect to the L\'evy-Prokhorov metric, strictly convex, and normalised such that a fully uninformative experiment bears zero cost, 
$C(\delta_{p_0})=0$, where $\delta_{p_0}$ denotes probability distribution that assigns probability one to $p_0$.
We model the benefit of information as $B:\Pi\to \mathbb R$, which we assume is bounded, linear, and strictly increasing in the Blackwell order. 
This arises as the decision-maker's expected payoff from maximising their expected payoff given posterior beliefs $p\sim \pi$, that is, $B(\pi)=\mathbb E_{p\sim \pi}[\max_\alpha \mathbb E_p[u(\alpha,\theta)]]$.

The decision-maker chooses an experiment to maximise their value of information, $V(\pi;K):=B(\pi)-K C(\pi)$, where $K\geq 0$ is a scaling parameter.
We write $V(K):=\sup_{\pi\in \Pi}V(\pi;K)$, $\pi^*(K):=\argmax_{\pi\in \Pi}V(\pi;K)$, $B^*(K):=B(\pi^*(K))$, and $C^*(K):=C(\pi^*(K))$.

To relate this setting to our baseline model, we define $T^*(K)$ as a strictly monotone function of total cost $K C^*(K)$.
We interpret $T^*(K)$ as the expected time, with the scaling $K$ factor as a measure of problem complexity.
In the case of our baseline model, where the decision-maker is choosing a stopping time, we have $K=\kappa^2$ and $C(\pi)=(c/2)\left(\mathbb E_{p\sim \pi}[\psi(p)]-\psi(p_0)\right)$,
where $\psi(p):=p\logit(p)+(1-p)\logit((1-p))$, \citep[Proposition 3]{MorrisStrack2019WP}.

A simple application of Berge's maximum theorem proves the following:
\begin{proposition}
    $T^*(K)$ is non-monotone in $K$.
\end{proposition}
\begin{proof}
    Let $\pi_I$ denote a fully informative experiment, assigning probability $p_0(\theta)$ to $\mathbf 1_\theta$, and $\pi_U$ a fully uninformative experiment.
    As $\Pi$ is compact with respect to the L\'evy-Prokhorov metric, and $V$ is continuous in $(\pi,K)$, then by Berge's maximum theorem, $V(K)$ is continuous in $K$.
    Hence, $\lim_{K\to 0}V(K)=V(0)=B(\pi_I)$ and 
    $0=\lim_{K\to 0}V(0)-V(K)=\lim_{K\to 0}B(\pi_I)-B^*(K)+K C^*(K)\geq \lim_{K\to 0}K C^*(K)\geq 0$.
    Moreover, $0\leq \lim_{K\to \infty} V(K)/K=\lim_{K\to \infty} B^*(K)/K-C^*(K)\leq \lim_{K\to \infty} B(\pi_I)/K-C^*(K)=\lim_{K\to \infty} -C^*(K)\leq 0 \implies C(\pi^*(K))\to 0\implies \pi^*(K)\to \pi_U$.
    Then, $B^*(K)\to B(\pi_U)$ and, by optimality, $B(\pi_U)\leq B^*(K)-K C^*(K)\implies \lim_{K\to \infty}K C^*(K)\leq \lim_{K\to \infty}B^*(K)-B(\pi_U)=0$. 
    To see that it is strictly positive somewhere, note that for some close to fully informative experiment $\hat\pi$, $B(\hat \pi)>B(\pi^U)>0$ and $0=C(\pi^U)<C(\hat \pi)<\infty$ as $C$ and $B$ are strictly increasing in Blackwell order.
    Then, 
    $V(K)\geq B(\hat\pi)-K C(\hat\pi)$ which is strictly positive for small but strictly positive $K$. This immediately implies that $K C^*(K)>0$.
\end{proof}
Quasi-concavity in the general case requires $-K V''(K)/V'(K)$ crosses one at most once.\footnote{
    Note that, since $V$ is convex, by Alexandrov's theorem, $V'$ and $V''$ exist almost everywhere. Then, by an application of the envelope theorem, $K C^*(K)=-K V'(K)$. Since $K C^*(K)$ is non-monotone, $-K V''(K)/V'(K)$ must cross one at least once; crossing it exactly once is sufficient and necessary for quasi-concavity.
} 
It is, however, easy to construct counter-examples, even if $C$ is in the class of uniformly posterior separable cost functions \citep{CaplinDeanLeahy2022JPE}.

\subsection{Optimal Control with Discounting}
\label{appendix:variants:discounting-ability}

We consider a variant of the model in \hyref{section:ability}[Section] in line with \citet{MoscariniSmith2001Ecta}, accomodating for discounting in addition to the flow cost.
We require that $c>0,c'>0,c''>0$, and $c'''\geq 0$ as in our baseline.
This guarantees that $\lim_{e\to \infty} e c'(e)-c(e)=\infty$, condition which is directly assumed in \citet{MoscariniSmith2001Ecta}.
We fix $\lambda,\kappa>0$. 

The decision-maker then maximises 
$$\mathbb E\left[\max_\alpha\mathbb E[\exp(-r\tau)u(\alpha,\theta)\mid \mathcal F_\tau^X]-\int_0^\tau \exp(-r t)c(e_t/\lambda)\diff t\right].$$
Following \citet{MoscariniSmith2001Ecta}, let $g(e):=e c'(e)-c(e)$, which $g'>0$ on $e>0$, given $c''>0$.
Then, optimal effort level solves
$e = \lambda g^{-1}(r V^{\kappa,\lambda}(p))$, where $$V^{\kappa,\lambda}(p):=\sup_{\tau,(e_t)}\mathbb E\left[\max_\alpha\mathbb E[\exp(-r\tau)u(\alpha,\theta)\mid \mathcal F_\tau^X]-\int_0^\tau \exp(-r t)c(e_t/\lambda)\diff t\right].$$
From \citet[Lemma 2]{MoscariniSmith2001Ecta}, $\exists \overline p^{\kappa,\lambda},\underline p^{\kappa,\lambda}: 0<\underline p^{\kappa,\lambda}<\overline p^{\kappa,\lambda}<1$ such that the optimal stopping time $\tau^{\kappa,\lambda}:=\inf\{t>0|p_t^{\kappa,\lambda}\notin C^{\kappa,\lambda}\}$, with $C^{\kappa,\lambda}:=(\underline p^{\kappa,\lambda},\overline p^{\kappa,\lambda})$ and $p_t^{\kappa,\lambda}$ solves $\diff p_t^{\kappa,\lambda} = \kappa^{-1} \lambda g^{-1}(r V^{\kappa,\lambda}(p_t)) p_t(1-p_t)\diff B_t$ with $p_0^{\kappa,\lambda}=p_0$.
Moreover, $V^{\kappa,\lambda}$ is $\mathcal C^1$ and $\mathcal C^2$ everywhere but at $\{\underline p^{\kappa,\lambda},\overline p^{\kappa,\lambda}\}$.

Let $c_{\lambda}(e):=c(e/\lambda)$ and $\gamma<1$.
Then, 
(1)
$c_{\lambda}(e)=c_{\gamma\lambda}(\gamma e)<c_{\gamma\lambda}(e)$;
(2) 
${c_{\lambda}}'(e)=c'(e/\lambda)/\lambda = \gamma c'(\gamma e/(\gamma\lambda))/(\gamma \lambda)=\gamma {c_{\gamma\lambda}}'(\gamma e)<{c_{\gamma\lambda}}'(e)$; 
and (3)
${c_{\lambda}}''(e)=c''(e/\lambda)/\lambda^2 = \gamma^2 c''(\gamma e/(\gamma\lambda))/(\gamma \lambda)^2=\gamma^2 {c_{\gamma\lambda}}''(\gamma e)<{c_{\gamma\lambda}}''(e)$, which follows from $c'''\geq 0$.
From \citet[Proposition 5(c)-(d)]{MoscariniSmith2001Ecta}, $V^{\kappa,\lambda}$ increases in $\lambda$ and decreases in $\kappa$ and $C^{\kappa,\lambda}$ increases in the subset order in $\lambda$ and decreases in $\kappa$.
Given that $\diff p_t^{\kappa,\lambda} = \kappa^{-1} \lambda g^{-1}(r V^{\kappa,\lambda}(p_t)) p_t(1-p_t)\diff B_t$, then, \hyref{proposition:exogenous:speed:shrinking:general}[Proposition] and \hyref{proposition:exogenous:speed-accuracy:nested:general} apply and accuracy increases in $\lambda$ and decreases in $\kappa$.
As $0<\max_\alpha u(\alpha,p_0)\leq V^{\kappa,\lambda}(p_t)\leq \mathbb E[\max_\alpha u(\alpha,\theta)]<\infty$, then, from \hyref{proposition:optimal:speed-accuracy:general}[Proposition], 
$
\lim_{\kappa\to 0}\mathbb E[\tau^{\kappa,\lambda}]
=
\lim_{\kappa\to \infty}\mathbb E[\tau^{\kappa,\lambda}]
=
\lim_{\lambda\to 0}\mathbb E[\tau^{\kappa,\lambda}]
=
\lim_{\lambda\to \infty}\mathbb E[\tau^{\kappa,\lambda}]
=0
$.

\FloatBarrier \newpage

\section{Technical Lemmas}
\label{appendix:technical-lemmas}

\begin{lemma}
    \label{lemma:convexity-prior-general}
    Let $\Theta\subseteq \mathbb R$ be a standard Borel space, 
    $\mathcal P(\Theta)$ denote a convex set of probability measures on
    $\Theta$, and $(\Omega,\mathcal F)$ be a measurable space carrying the observable experiment. 
    Let $\Pi$ be a class of admissible non-anticipative policies,
    and assume that $\Pi$ does not depend on the prior
    $\nu\in\mathcal P(\Theta)$.

    For every $\pi\in\Pi$, assume that $\theta\mapsto \mathbb P_\theta^\pi$ is a probability kernel from $\Theta$ to $(\Omega,\mathcal F)$, where $\mathbb P_\theta^\pi$ denotes the law of the observable experiment under state $\theta$ and policy $\pi$. 
    For every prior $\nu\in\mathcal P(\Theta)$, define the joint law
    $\mathbb P_\nu^\pi$ on $\Theta\times\Omega$ by
    $
        \mathbb P_\nu^\pi(d\theta,d\omega)
        :=
        \nu(d\theta)\mathbb P_\theta^\pi(d\omega).
    $
    Let $u^\pi:\Theta\times\Omega\to\mathbb R$ be the total payoff generated by policy $\pi$. 
    Assume that $u^\pi$ is measurable and integrable under $\mathbb P_\nu^\pi$ for all $\nu\in\mathcal P(\Theta)$ and all $\pi\in\Pi$.

    Define
    $$
        V(\nu):=
        \sup_{\pi\in\Pi}
        \mathbb E_\nu^\pi\left[u^\pi(\theta,\omega)\right].
    $$
    Then $V$ is convex on $\mathcal P(\Theta)$.
\end{lemma}

\begin{proof}
    Fix a policy $\pi\in\Pi$. Define
    $
        U(\nu,\pi):=
        \mathbb E_\nu^\pi\left[u^\pi(\theta,\omega)\right].
    $
    By the definition of $\mathbb P_\nu^\pi$,
    $$
        U(\nu,\pi)
        =
        \int_\Theta
        \left(
            \int_\Omega
            u^\pi(\theta,\omega)\,
            \mathbb P_\theta^\pi(d\omega)
        \right)
        \nu(d\theta).
    $$
    Thus, for fixed $\pi$, the map $\nu\mapsto U(\nu,\pi)$ is affine.

    Let $\nu_0,\nu_1\in\mathcal P(\Theta)$ and let
    $
        \nu_\lambda:=\lambda\nu_1+(1-\lambda)\nu_0,
    $ 
    for 
    $
        \lambda\in[0,1].
    $
    Since $\mathcal P(\Theta)$ is convex, $\nu_\lambda\in\mathcal P(\Theta)$.
    For every $\pi\in\Pi$, 
    $
        U(\nu_\lambda,\pi)
        =
        \lambda U(\nu_1,\pi)+(1-\lambda)U(\nu_0,\pi).
    $
    Therefore
    \begin{align*}
        V(\nu_\lambda)
        &=
        \sup_{\pi\in\Pi}
        U(\nu_\lambda,\pi)\\
        &=
        \sup_{\pi\in\Pi}
        \left[
            \lambda U(\nu_1,\pi)
            +(1-\lambda)U(\nu_0,\pi)
        \right]\\
        &\le
        \lambda\sup_{\pi\in\Pi}U(\nu_1,\pi)
        +
        (1-\lambda)\sup_{\pi\in\Pi}U(\nu_0,\pi)\\
        &=
        \lambda V(\nu_1)+(1-\lambda)V(\nu_0).
    \end{align*}
    Hence $V$ is convex.
\end{proof}

We note that \hyref{lemma:convexity-prior-general}[Lemma] applies to general
information processes, controlled or uncontrolled. 
The policy class $\Pi$ may include, for example, a control process, a stopping rule, and a terminal action rule. 
The only substantive requirements are that the admissible policy class is
prior-independent and that, conditional on each state $\theta$, the law of the
observable experiment under a fixed policy is well defined.

\begin{lemma}
    \label{lemma:quasi-submodularity}
    Let $X, Y \subseteq \mathbb R$, 
    $f:X\times Y\to \mathbb R$ be strictly increasing with respect to the natural product order and $f(X\times Y)$ be convex.
    If $h:\mathbb R \to \mathbb R$ is strictly quasi-concave on $f(X\times Y)$, then, $h\circ f:X\times Y \to \mathbb R$ is strictly quasi-submodular.
\end{lemma}
\begin{proof}
    Take any $w=(x,y)$ and $w'=(x',y')$ such that $w\vee w'\ne w,w'$ and $w,w'\ne w\wedge w'$, and assume the premise, $h(f(w\wedge w'))\geq h(f(w))$. 
    We want to show that $h(f(w')) > h(f(w\vee w'))$.

    First note that, if $h$ is strictly increasing or strictly decreasing on $f(X\times Y)$, the claim is either vacuously or trivially satisfied. 
    Suppose then that $h$ is non-monotone. 
    Then, as $h$ is strictly quasi-concave, there is $z^*\in Z:=f(X\times Y)$ such that $h$ is strictly increasing on $Z\cap(-\infty, z^*]$ and strictly decreasing on $Z\cap[z^*,\infty)$.

    As $f$ is strictly increasing, $f(w\wedge w')<f(w)$ and so $f(w)> z^*$, as otherwise both $f(w\wedge w')$ and $f(w)$ would belong to the region in which $h$ is strictly increasing contradicting the premise.

    There are then two cases: $f(w')\geq z^*$ and $f(w')<z^*$. 
    If $f(w')\geq z^*$, then, as $f$ is strictly increasing, $f(w\vee w')> f(w')\geq z^*$ and $h(f(w\vee w'))< h(f(w'))$, as both $f(w\vee w')$ and $f(w')$ are in the region in which $h$ is strictly decreasing.
    Suppose then that $f(w')<z^*$.
    Then, $f(w\wedge w')<f(w')<z^*<f(w)$.
    Given that $h$ is quasi-concave and $f(w')\in (f(w\wedge w'),f(w))$, then $h(f(w'))\geq \min\{h(f(w\wedge w')),h(f(w))\}=h(f(w))$.
    Since $z^* <f(w)<f(w\vee w')$, $h(f(w))>h(f(w\vee w'))$, and we conclude $h(f(w'))>h(f(w\vee w'))$.
\end{proof}

\FloatBarrier \newpage

\end{document}